\documentclass[twocolumn]{aastex61}
\usepackage{comment}
\usepackage{ifthen}
\usepackage{amsmath}
\usepackage{makecell}
\newcommand{\forloop}[5][1]%
{%
\setcounter{#2}{#3}%
\ifthenelse{#4}%
	{%
	#5%
	\addtocounter{#2}{#1}%
	\forloop[#1]{#2}{\value{#2}}{#4}{#5}%
	}%
	{%
	}%
}% 

\newcommand{\ctbd}[1]{}

\newcommand{\loopand}{\ifnum\value{planetcounter}=2 and \else\fi}
\newcommand{\loopcomma}{\ifnum\value{planetcounter}<2 ,\else. \fi}
\newcommand{\loopcommanoperiod}{\ifnum\value{planetcounter}<2 ,\else \space\fi}
\newcommand{\loopcommanospace}{\ifnum\value{planetcounter}<2 ,\else \fi}

\newcommand{\nlctotal}{{83,717,159}}
\newcommand{\nstarstotal}{{56,401,549}}

\newcommand{\nlctotalshort}{{\ensuremath{8.4\times10^7}}}
\newcommand{\nstarstotalshort}{{\ensuremath{5.6\times10^7}}}

\newcounter{planetcounter}

\newboolean{emulateapj}
\setboolean{emulateapj}{true}

\newboolean{rvtablelong}
\setboolean{rvtablelong}{true}

\newboolean{astroph}
\setboolean{astroph}{true}
\shortauthors{Hartman et al.}
\shorttitle{The T16 Project}
\ifthenelse{\boolean{emulateapj}}{
    
}{
    
}

\begin{document}

%% Titlepage
\title{%%
The T16 Project: Image Subtraction Light Curves from TESS Cycle 1 Full-Frame Images for Stars with $T < 16$
}

\correspondingauthor{Joel Hartman}
\email{jhartman@astro.princeton.edu}

\author[0000-0001-8732-6166]{Joel~D.~Hartman}
\affil{Department of Astrophysical Sciences, Princeton University, NJ 08544, USA}

\author[0000-0001-7204-6727]{G\'asp\'ar~\'A.~Bakos}
\altaffiliation{Packard Fellow}
\affil{Department of Astrophysical Sciences, Princeton University, NJ 08544, USA}
\affil{MTA Distinguished Guest Fellow, Konkoly Observatory, Hungary}

\author[0000-0002-0514-5538]{Luke~G.~Bouma}
\affil{Observatories of the Carnegie Institution for Science, 813 Santa Barbara Street, Pasadena, CA 91101, USA}
\affil{Department of Astronomy, California Institute of Technology, Pasadena, CA 91125, USA}

\author{Zoltan Csubry}
\affil{Department of Astrophysical Sciences, Princeton University, NJ 08544, USA}

%% EOF authors

% #####################################################################
%% abstract
\begin{abstract}
%++++++++++++++++++++++++++++++++++++++++++++++++++++++++++++++++++++++
%++++++++++++++++++++++++++++++++++++++++++++++++++++++++++++++++++++++

\setcounter{footnote}{10}
We present \nlctotal\ light curves for \nstarstotal\ stars with
$T < 16$\,mag observed in the Full-Frame Images (FFIs) of
Cycle 1 of the NASA {\em TESS} mission. These light curves were
extracted from subtracted images produced by the Cluster Difference
Imaging Survey \citep[CDIPS;][]{bouma:2019:cdips}. We make public the
raw image subtraction light curves, together with light curves
de-trended against instrumental systematics. We compare the light
curves to other publicly available light curves from the {\em TESS}
FFIs, finding that for a substantial fraction of stars with $T < 16$, the T16
project provides the highest precision FFI
light curves available. We demonstrate that the detrended T16 light
curves are generally as good as, or better than, than the light curves
from other projects for the known TOIs. We also show that the
un-detrended light curves can be used to study high amplitude variable
stars. The light curves are being made available through the NASA
Mikulski Archive for Space Telescopes (MAST). Light curve production
is underway for additional {\em TESS} Cycles.
\setcounter{footnote}{0}
\end{abstract}

% #####################################################################
%% keywords
\keywords{
%%
%%
%
% TBD
    light curves, photometry
}

%% EOF keywords
%% EOF titlepage

% #####################################################################
%% Introduction
\section{Introduction}
\label{sec:introduction}
%++++++++++++++++++++++++++++++++++++++++++++++++++++++++++++++++++++++
%++++++++++++++++++++++++++++++++++++++++++++++++++++++++++++++++++++++
%% EOF introduction

The NASA {\em Transiting Exoplanet Survey Satellite} ({\em TESS}; \citealp{ricker:2015}) mission has conducted an unprecedented,
nearly all-sky, space-based survey for small transiting planets around
bright stars. During the primary mission
approximately 307,000 stars with $T \lesssim 12$\,mag were observed
every two minutes at high photometric precision for a duration of
$\sim 1$\,month near the ecliptic plane, and up to $1$\,year of
continuous observations near the ecliptic poles. Light curves for
these stars were produced by the {\em TESS} Science Processing
Operations Center \citep[SPOC;][]{jenkins:2016} at NASA Ames and have been searched for
transit signals and made publicly available on the NASA Mikulski
Archive for Space Telescopes (MAST).

{\em TESS} has been approved for two extended missions so far. During
the first two year extended mission (from 2020 July through 2022
September) the Southern ecliptic hemisphere was re-observed, while the
pointings for the Northern ecliptic hemisphere observations were
adjusted to include a set of observations along the ecliptic plane,
and a set of observations concentrated on the ecliptic pole. During
the ongoing second extended mission {\em TESS} is
observing fields both in the Northern and Southern ecliptic
hemispheres.

In addition to the 2 minute cadence targets, stacked full-frame images
(FFIs) were downloaded from the spacecraft. These were obtained at a
cadence of 30\,minutes, 10\,minutes, and 200\,seconds for the primary
mission, first extended mission, and second extended mission,
respectively. Included in the {\em TESS} FFIs are more than 100 million
stars with $T < 16$\,mag for which {\em TESS} should produce light
curves with a 1\,hr photometric precision better than 1\%.

\begin{deluxetable*}{llllrrr}
\tablewidth{0pc}
\tabletypesize{\scriptsize}
\tablecaption{
    Summary of projects that have released FFI light curves as HLSPs on MAST
    \label{tab:ffiprojects}
}
\tablehead{
    \multicolumn{1}{c}{Project Name} &
    \multicolumn{1}{c}{Sectors\tablenotemark{a}} &
    \multicolumn{1}{c}{Target} &
    \multicolumn{1}{c}{Photometric} &
    \multicolumn{1}{c}{Number of} &
    \multicolumn{1}{c}{Number of} &
    \multicolumn{1}{c}{References} \\
    \multicolumn{1}{c}{} &
    \multicolumn{1}{c}{Covered} &
    \multicolumn{1}{c}{Selection} &
    \multicolumn{1}{c}{Method} &
    \multicolumn{1}{c}{Stars} &
    \multicolumn{1}{c}{Light Curves} &
    \multicolumn{1}{c}{}
}
\startdata
~~~~TGLC & 1--41 & $T < 16$ & Aperture and PSF Photometry & $9.4 \times 10^7$ & $2.4 \times 10^8$ & (13) \\
~~~~GSFC-ELEANOR-LITE & 1--26 & $T < 16$ & Aperture Photometry & $9.3 \times 10^7$ & $1.5 \times 10^8$ & (14) \\
~~~~QLP & 1--80 & $T < 13.5$ & Simple Aperture Photometry & $2.0 \times 10^7$ & $7.5 \times 10^7$ & (1) (2) (3) (4) \\
~~~~TESS-SPOC & 1--72 & $N_{\rm targets} < 1.6 \times 10^5$ per sector & Simple Aperture Photometry & $3.8 \times 10^6$ & $1.1 \times 10^7$ & (5) \\
~~~~CDIPS & 1--55 & Star cluster members or young stars & Image Subtraction & $1.2 \times 10^6$ & $2.4 \times 10^6$ & (6) \\
~~~~PATHOS & 1--26 & Star cluster members & PSF Photometry & $2.7 \times 10^5$ & $3.9 \times 10^5$ & (9) (10) (11) (12) \\
~~~~DIAMANTE & 1--26 & Transiting Planet Candidates & Image Subtraction & 1556 & 1556 & (7) (8) \\
~~~~T16 & 1--13 & $T < 16$ & Image Subtraction & \nstarstotalshort\ & \nlctotalshort\ & this work \\
\enddata
\tablerefs{(1) \citet{huang:2020a}; (2) \citet{huang:2020b}; (3) \citet{kunimoto:2021}; (4) \citet{kunimoto:2022}; (5) \citet{caldwell:2020}; (6) \citet{bouma:2019:cdips}; (7) \citet{montalto:2020}; (8) \citet{montalto:2023}; (9) \citet{nardiello:2019}; (10) \citet{nardiello:2020a}; (11) \citet{nardiello:2020b}; (12) \citet{nardiello:2021}; (13) \citet{han:2023}; (14) \citet{powell:2022};}
\tablenotetext{a}{As of 2024 Dec 2.}
\end{deluxetable*}

At the time of writing, there are seven efforts, besides that
presented here, that have released large collections of {\em TESS} FFI
light curves as High Level Science Products (HLSPs) on MAST. These are
summarized in Table~\ref{tab:ffiprojects}.

The two projects that have covered the most sectors to date are the MIT
Quick-Look Pipeline \citep[QLP;][]{huang:2020a} project, and the TESS
Science Processing Operations Center Full Frame Image \citep[TESS-SPOC
  FFI, hereafter TESS-SPOC;][]{caldwell:2020} project at NASA
Ames. The former has released light curves for all stars with $T <
13.5$, the latter has released light curves for a more selected sample
of up to 160,000 stars per sector. The TESS-SPOC targets have either
$H < 10$\,mag, or $T < 13.5$\,mag, or a distance within 100\,pc. Both of
these pipelines utilize aperture photometry to extract the photometric
measurements, followed by post-processing of the light curves to
correct for instrumental systematic variations in the measurements.

Two other efforts to produce light curves for all stars down to $T <
16$ include the {\em TESS} {\em Gaia} Light Curve project
\citep[TGLC;][]{han:2023}, and the Goddard Space Flight Center \texttt{eleanor
lite} project \citep[GSFC-ELEANOR-LITE, hereafter GSFC;][]{powell:2022}. The former
uses aperture and PSF-fitting photometry, and utilizes the {\em Gaia}
catalog \citep{gaiadr2} to correct for flux from neighboring stars, while the
GSFC project uses ``simple aperture'' photometry through the
\texttt{eleanor} tool \citep{feinstein:2019} to produce light curves. Both of
these projects have released light curves for the first two {\em TESS} Cycles, while TGLC has released light curves for the third {\em TESS} Cycle as well. Both projects greatly exceed QLP and TESS-SPOC in terms of the total number of light curves made public.

Finally three projects have published light curves for particular
classes of stars. This includes the DIAMANTE project \citep{montalto:2020, montalto:2023}, which uses the
image subtraction method of photometry \citep[e.g.,][]{alard:1998}, and has made light
curves public for transiting planet candidates, and the PATHOS \citep{nardiello:2019,nardiello:2020a,nardiello:2020b,nardiello:2021} and
Cluster Difference Imaging Photometric Survey
\citep[CDIPS;][]{bouma:2019:cdips} projects which use PSF photometry,
and image subtraction photometry, respectively, and have made public
light curves for young stars and stars that are members of clusters.

In this paper, we build on the prior work of the CDIPS project to
release image-subtraction-derived light curves for {\em all} stars
with $T < 16$\,mag in the {\em TESS} Cycle 1 FFI observations.
%  As the
%{\em Gaia} DR2 catalog forms the basis for our point source catalog,
%we choose to employ a cut on the {\em Gaia} $G$ magnitude, rather than
%the estimated {\em TESS} $T$ magnitude for simplicity in selecting
%stars to produce light curves for. Among other considerations, a
%simple selection on $G$ allows stars with missing $BP$ and $RP$
%magnitudes to be included.

In the following section we briefly describe the {\em TESS} observations. In Section~\ref{sec:red} we discuss the methods used to produce the T16 light curves. We describe the resulting dataset, its precision, and a comparison to other sets of light curves in Section~\ref{sec:results}. We conclude the paper in Section~\ref{sec:conclusion}.

% #####################################################################
\section{TESS Observations}
\label{sec:tess}
%++++++++++++++++++++++++++++++++++++++++++++++++++++++++++++++++++++++
%++++++++++++++++++++++++++++++++++++++++++++++++++++++++++++++++++++++

Here we present light curves from the first year of {\em TESS}
observations, dubbed Cycle I. This cycle was divided into 13 observing
campaigns, called sectors. Each sector consisted of two consecutive
13.7\,day elliptical orbits of the {\em TESS} spacecraft. Every
perigee the spacecraft paused observations for approximately 3\,hr to
downlink the data to Earth. Momentum dumps, where the speed of the
spacecraft momentum wheels were reset, were performed every
2.5\,days. The pointing of the spacecraft is destabilized at these
periods, causing systematic variations in the light curves that are
associated with these events.

The {\em TESS} spacecraft uses four identical $f/1.4$ lenses of pupil
diameter 10.5\,cm, and a field of view of $24^{\circ} \times
24^{\circ}$. The lenses are aligned so that the full {\em TESS} field
of view is a $24^{\circ} \times 96^{\circ}$ stripe on the sky. Each
lens images onto a set of four $2048 \times 2048$ back-illuminated
MIT/Lincoln Laboratory CCID-80 detectors at a pixel scale of
$21$\arcsec\,{\rm pixel}$^{-1}$. The effective bandpass of the {\em
  TESS} instrument spans from 600\,nm to 1000\,nm, and is referred to
as the $T$-band. The CCDs are read out at a cadence of 2\,sec. In order to mitigate against cosmic rays, ten consecutive exposures are collected for each pixel, the brightest and faintest exposures in the collection are rejected, and the remaining eight pixels are summed. This procedure is repeated 90 times, co-adding the results, to produce the FFIs with an effective cadence of 30\,min and exposure time of 24\,min.

Altogether there are 16 different detectors used for a single {\em
  TESS} sector, and we reduce the observations from each sector and
detector independently.

% #####################################################################
\section{Data Reduction Methods}
\label{sec:red}
%++++++++++++++++++++++++++++++++++++++++++++++++++++++++++++++++++++++
%++++++++++++++++++++++++++++++++++++++++++++++++++++++++++++++++++++++

\subsection{Difference Image Photometry}
\label{sec:phot}

We make use of the difference image reduction of the {\em TESS} FFIs
carried out by the CDIPS project, and published in
\citet{bouma:2019:cdips}.  Here we provide a brief summary of the
reduction process, referring the reader to this prior work for
details. We follow this summary with a more detailed description of
the methods used to carry out the photometric measurements from the
difference images that form the new contribution of the present work.

The CDIPS processing began with the calibrated {\em TESS} FFIs
summarized in Section~\ref{sec:tess}. An estimate of the large-scale
background is first subtracted from the images, where the estimate is
based on a median box-fillter followed by a Gaussian
blurring. Saturated and other problematic pixels are masked and the
WCS solutions included in the FFI headers, obtained from MAST, are
verified for each image to exclude images with problematic
astrometry. An astrometric reference image is then selected from the
set of observations for a given combination of sector/camera/CCD. The
background-subtracted science images are registered to the reference
image using the existing WCS solutions and the tools within the {\sc
  fitsh} package \citep{pal:2012}. A photometric reference image
is then produced for the sector/camera/CCD by aligning and matching
the flux, background and PSF of a selection of $\sim 50$ science
frames using the {\sc fitsh} tools. The reference image is then
convolved to each science frame using the {\sc ficonv} tool from the
{\sc fitsh} package, and the resulting difference images are the
starting point for the new work that we describe in the present paper.

We then followed the process of \citet{bouma:2019:cdips} for the
photometry, but here we made use of a much larger list of
stars. Whereas the CDIPS project focused on making light curves for
stars identified as potential members of open clusters, moving groups,
stellar associations, or that were identified as candidate young
stars, here we make light curves for {\em }all stars with $T < 16$\,mag in
the Gaia DR2 catalog. We estimate $T$ for each source from the {\em Gaia} DR2 photometry using eq.~1 from \citet{stassun:2019}:
\begin{eqnarray}\label{eqn:tmag}
T & = & G \\
 & & - 0.00522555(BP-RP)^3 \notag \\
 & & + 0.0891337(BP-RP)^2 \notag \\
 & & - 0.633923(BP-RP) \notag \\
 & & + 0.0324473, \notag
\end{eqnarray}
where $BP$, $RP$ and $G$ are the magnitudes from the three different
pass-bands in the {\em Gaia} catalog, and where we assume $BP-RP = 0$
for sources where either measurement is unavailable.

We measure aperture photometry on the difference images using the WCS
solutions and the fixed astrometric positions from Gaia to center the
aperture for each source. Here we use the same set of three apertures
as used by the CDIPS project, namely circular apertures of radii 1.0,
1.5 and 2.25 pixels. Differential flux photometry is measured using the
{\sc fiphot} program, included in {\sc fitsh}. This process also
subtracts from each measurement of the differential flux any residual
positive or negative background on the difference images that are
estimated using annuli of inner radius 7.0 pixels and outer radius 13 pixels.

To convert the differential flux measurements to total flux
measurements we provide as input to {\sc fiphot} an estimate of the
flux on the photometric reference image of each source through each of
the three apertures. Because stars on the reference image may be
highly blended, rather than directly measuring this flux for each
source, we make use of the $G$, $BP$ and $RP$ photometry in the Gaia
DR2 catalog. We determine a transformation from these magnitudes to
fluxes on the photometric reference images through each of the
apertures by fitting a relation to the observations for a set of
uncrowded stars. We find that the scatter on this relation is
typically $\sim 0.1$\,mag. Any errors in this process for an
individual star will lead to an overall error on the amplitude of any
variations in the resulting light curve, but will otherwise not affect
the shape of the light curve or the signal-to-noise ratio of any
variations that are present. Special caution should be taken for high
amplitude variable stars, where the actual flux on the reference image
may differ significantly from the expected flux based on the {\em
  Gaia} catalog photometry. This can lead to significant errors in the
amplitude of variability in the derived light curves for these stars.

Note that any non-varying flux is automatically excluded from the difference images. Because we use the {\em Gaia} catalog photometry to determine the reference flux of each source, the resulting light curves that we produce will exclude any non-variable contamination from blended neighbors. The resulting light curve is effectively corrected for dilution from any sources that are unresolved by {\em TESS}, but resolved by {\em Gaia}. {\em However, if any star has an unresolved neighbor that is itself a variable, the variable flux from that neighbor may be sampled by the aperture used to measure the differential flux of the target star in question.} The varying flux from the variable neighbor will then be combined with the estimated reference flux of the target star based on {\em Gaia} to produce a magnitude light curve that includes a contaminated variable signal. The amplitude of this signal, in magnitudes, can be exceedingly large in the case of a faint star that is blended with a bright variable star. In fact negative net fluxes, yielding NaN values for the magnitude, are possible if the differential flux from the variable neighbor is negative and has an absolute value that exceeds the estimated reference flux of the target star.

%The result from this process is a set of files, one per image, that
%contain the photometric fluxes for each of the stars in a given image
%with $T < 16$\,mag. We transform these photometry files into light
%curves (i.e., one file per star, each containing all of the
%photometric fluxes measured for that stars) using the {\sc grcollect}
%program included in {\sc fitsh}.

\subsection{Light Curve Time Stamps}
\label{sec:lctimes}

The {\em TESS} FFIs contain in the image headers the {\em TESS}
Barycentric Julian Dates (TBJD) at the beginning and end of each
observation. These are given on the Barycentric Dynamical Time (TDB)
system, which differs from the UTC system by a discrete value that
increments whenever leap-seconds are held in the UTC system. The
barycentric corrections are applied for the sky coordinates of a
specified pixel in the image. Because each source observed in the
image has a different sky position, separate barycentric corrections
need to be applied for each light curve. To do this we first compute
the mid BJD time for each observation from the image headers, and
subtract the applied barycentric correction to determine the JD at the
time of observation. We then use the {\sc VARTOOLS} program
\citep{hartman:2016:vartools} to convert the JD to BJD for each
source. Here we use the {\em Gaia} DR3 coordinates for each source,
accounting for proper motion, and we determine the position of the
{\em TESS} observatory with respect to the center of the Earth at the
time of each observation using the JPL-Horizons Telnet interface
\citep{horizons}. We confirm that this method reproduces the
barycentric corrections listed in the FFI image headers for the sky
coordinates corresponding to the listed image positions to within
0.15\,s for all observations, with a mean difference between the
barycentric corrections that we measure, and those given in the
headers, of 0.002\,s, and a standard deviation of 0.039\,s. Note that
if we do not account for the orbit of {\em TESS} then the barycentric
corrections may differ by as much as 1.3\,s.

\subsection{Light Curve Detrending}
\label{sec:detrending}

To facilitate the detection of short time-scale variability it is
necessary to apply trend filtering methods to the light curves. The
detrending process that we apply differs from what was done by the
CDIPS project, which separately applied two different detrending methods: Principle Component Analysis
\citep[PCA; e.g.,][]{ivezic:2014}, and the Trend Filtering
Algorithm \citep[TFA;][]{kovacs:2005:TFA}. Like CDIPS, we also apply
TFA, but we first
apply a Spline-based External Parameter Decorrelation (SEPD), which we
describe more below, and then we run TFA on the SEPD-filtered light curves. Due to the much larger number of light curves that we are processing here compared to what was handled by CDIPS, we do not apply PCA, which, as an iterative process, is computationally slower than SEPD and TFA. The detrending is carried out using the {\sc
  VARTOOLS} program \citep{hartman:2016:vartools}.

The SEPD method that we apply to the raw light curves involves fitting
a linear combination of basis functions to each light curve, and then
subtracting the optimal fit. The detrending is performed against four
variables: time, the X and Y CCD position of the source, and the
temperature of the CCD. Here we use basis splines to remove low
frequency variations from the light curves, which for {\em TESS} tend to be
dominated by instrumental systematics, particularly unmodeled
scattered light.  We use a third order spline with a
knot spacing of 1\,day, and split the fit into independent bases on
any time gaps in the light curve that exceed 0.5\,days. This is done
to handle significant variations that occur just before and after data
downlinks. For the X and Y position and the temperature we use a
simple linear relation for the basis functions.  In performing the
decorrelations we exclude greater than $3\sigma$ outliers from the
fit. This SEPD correction differs from the PCA-based filtering applied
by CDIPS in that here the correction includes an explicit
decorrelation against time to remove long timescale variations and maximize the ability to detect shorter
timescale variations such as eclipses or planetary transits. For
longer term variations, filtering methods that are tailored to the
specific type of longer variability searched for may be needed, and we
therefore choose not to present a generic filtering that is unlikely
to be optimal for any particular class of longer period variables.

Note that we did not attempt to quantify the significance of
  the different detrending factors, nor did we conduct a thorough
  exploration of other potential factors. Such an exercise may be
  fruitful, but is beyond the scope of the present paper, which
  focuses on the delivery of the light curves that have already been
  produced.

The second detrending that is performed on the SEPD-filtered light
curves is an application of the Trend Filtering Algorithm (TFA) of
\citet{kovacs:2005:TFA}. Here, for each Sector/Camera/CCD combination
a representative list of 200 light curves is randomly selected such
that the stars are distributed uniformly across the CCD and uniformly
in magnitude. Variable stars are iteratively identified and removed
from the list. Each light curve is then fit as a linear combination
of the 200 template light curves, and the best-fit model is then
subtracted to yield the TFA-corrected light curve. Here we use the
same lists of template light curves as used by the CDIPS project for
each Sector/Camera/CCD.

{\em We stress again that any variability on timescales longer than 1\,day
  will be filtered from the detrended light curves} (columns labelled
  EPD or TFA in Table~\ref{tab:lccolumns}), and that the pre-detrended
  light curves should instead be used to study longer period
  variations (columns labelled IRM in Table~\ref{tab:lccolumns}). For a demonstration of this, see the discussion on Cepheid Variables in Section~\ref{sec:examples}.

% #####################################################################
\section{Results}
\label{sec:results}
%++++++++++++++++++++++++++++++++++++++++++++++++++++++++++++++++++++++
%++++++++++++++++++++++++++++++++++++++++++++++++++++++++++++++++++++++

\subsection{Summary of Available Data}
\label{sec:datasummary}

\begin{deluxetable*}{lrl}
\tablewidth{0pc}
\tabletypesize{\tiny}
\tablecaption{
    Data columns included in light curve files produced by the T16 project
    \label{tab:lccolumns}
}
\tablehead{
    \multicolumn{1}{c}{Column Name} &
    \multicolumn{1}{c}{FITS Data Type} &
    \multicolumn{1}{c}{Column Description}
}
\startdata
RSTFC & 17A & {\em TESS} image identifier \\
TMID\_UTC & D & UTC-based Julian Date mid exposure time minus 2457000 \\
TMID\_BJD & D & TDB-based Barycentric Julian Date mid exposure time minus 2457000 \\
BGV & D & Background value at the location of the source \\
BGE & D & Background measurement error \\
FSV & D & Measured value of S parameter characterizing the PSF (Eq.~\ref{eqn:sdk}) \\
FDV & D & Measured value of D parameter characterizing the PSF (Eq.~\ref{eqn:sdk}) \\
FKV & D & Measured value of K parameter characterizing the PSF (Eq.~\ref{eqn:sdk}) \\
IFL1 & D & Flux in 1.0 pixel aperture in Analog-to-Digital Units (ADU) \\
IFL2 & D & Flux in 1.5 pixel aperture in ADU \\
IFL3 & D & Flux in 2.25 pixel aperture in ADU \\
IFE1 & D & Flux uncertainty in 1.0 pixel aperture in ADU \\
IFE2 & D & Flux uncertainty in 1.5 pixel aperture in ADU \\
IFE3 & D & Flux uncertainty in 2.25 pixel aperture in ADU \\
IRM1 & D & Instrumental magnitude in 1.0 pixel aperture in magnitude units \\
IRM2 & D & Instrumental magnitude in 1.5 pixel aperture in magnitude units \\
IRM3 & D & Instrumental magnitude in 2.25 pixel aperture in magnitude units \\
IRE1 & D & Instrumental magnitude uncertainty in 1.0 pixel aperture in magnitude units \\
IRE2 & D & Instrumental magnitude uncertainty in 1.5 pixel aperture in magnitude units \\
IRE3 & D & Instrumental magnitude uncertainty in 2.25 pixel aperture in magnitude units \\
EPD1 & D & Spline-EPD detrended magnitude in 1.0 pixel aperture \\
EPD2 & D & Spline-EPD detrended magnitude in 1.5 pixel aperture \\
EPD3 & D & Spline-EPD detrended magnitude in 2.25 pixel aperture \\
TFA1 & D & TFA-detrended magnitude in 1.0 pixel aperture \\
TFA2 & D & TFA-detrended magnitude in 1.5 pixel aperture \\
TFA3 & D & TFA-detrended magnitude in 2.25 pixel aperture \\
XPOS & D & X image position of source on subtracted frame \\
YPOS & D & Y image position of source on subtracted frame \\
TEMP & D & mean CCD temperature in K \\
NTEMPS & D & The number of temperatures averaged to compute TEMP \\
\enddata
\end{deluxetable*}

A total of \nlctotal\ light curves from {\em TESS} Cycle 1 (Sectors 1 through
13) for \nstarstotal\ stars are being made available through the NASA Mikulski
Archive for Space Telescopes (MAST) as a High Level Science Product via \url{http://doi.org/10.17909/8nxx-tw70}\footnote{\url{https://archive.stsci.edu/hlsp/t16/}}. Stars closer to the Southern ecliptic pole have observations from multiple sectors, and will have more than one {\em TESS} Cycle 1 light curve produced by our project. The light
curves are provided in binary FITS table format, with the light curves
organized by Sector, Camera, and CCD. The Gaia DR2 identifier is
included in the name of each light curve. The data columns included in
each light curve file are described in
Table~\ref{tab:lccolumns}. 

Note that the $S$, $D$ and $K$ parameters included in the light curve come from fitting an elliptical Gaussian to the stellar profile. The elliptical Gaussian, $F$, has the form:
\begin{eqnarray}\label{eqn:sdk}
F & = & A \exp{ \left(-\frac{1}{2}B\right) } \\
B & = & S\left( (\Delta x)^2 + (\Delta y)^2 \right) \\
  &   & + D\left( (\Delta x)^2 - (\Delta y)^2 \right) \notag \\
  &   & + 2K\Delta x \Delta y \notag
\end{eqnarray}
where $A$ is the amplitude of the profile, and $\Delta x$ and $\Delta y$ are the $x$ and $y$ distances, in pixels, from the center of the profile.

\subsection{Photometric Precision}
\label{sec:precision}

\begin{figure*}[!ht]
{
\centering
\leavevmode
\includegraphics[width={0.5\linewidth}]{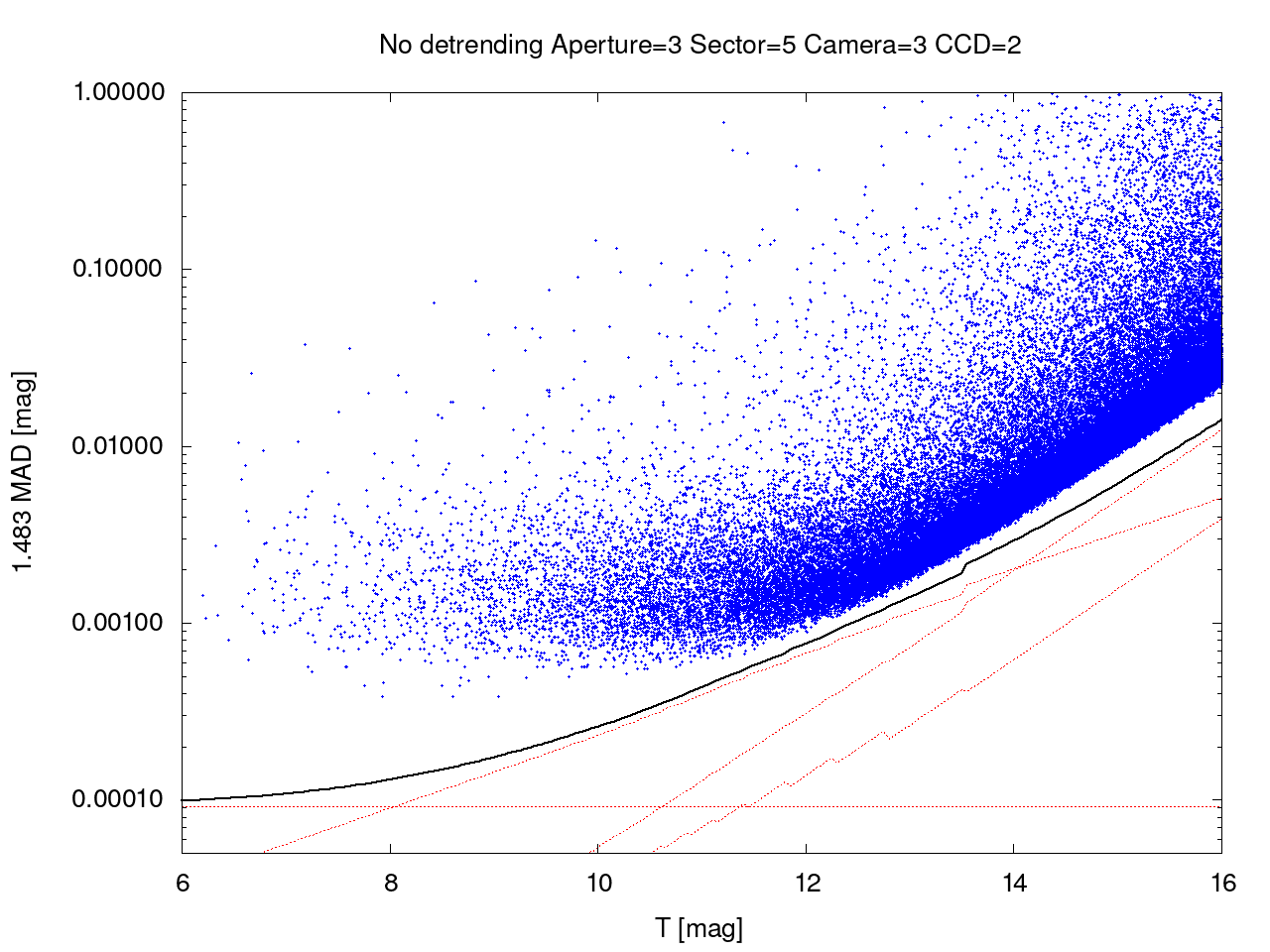}
\hfil
\includegraphics[width={0.5\linewidth}]{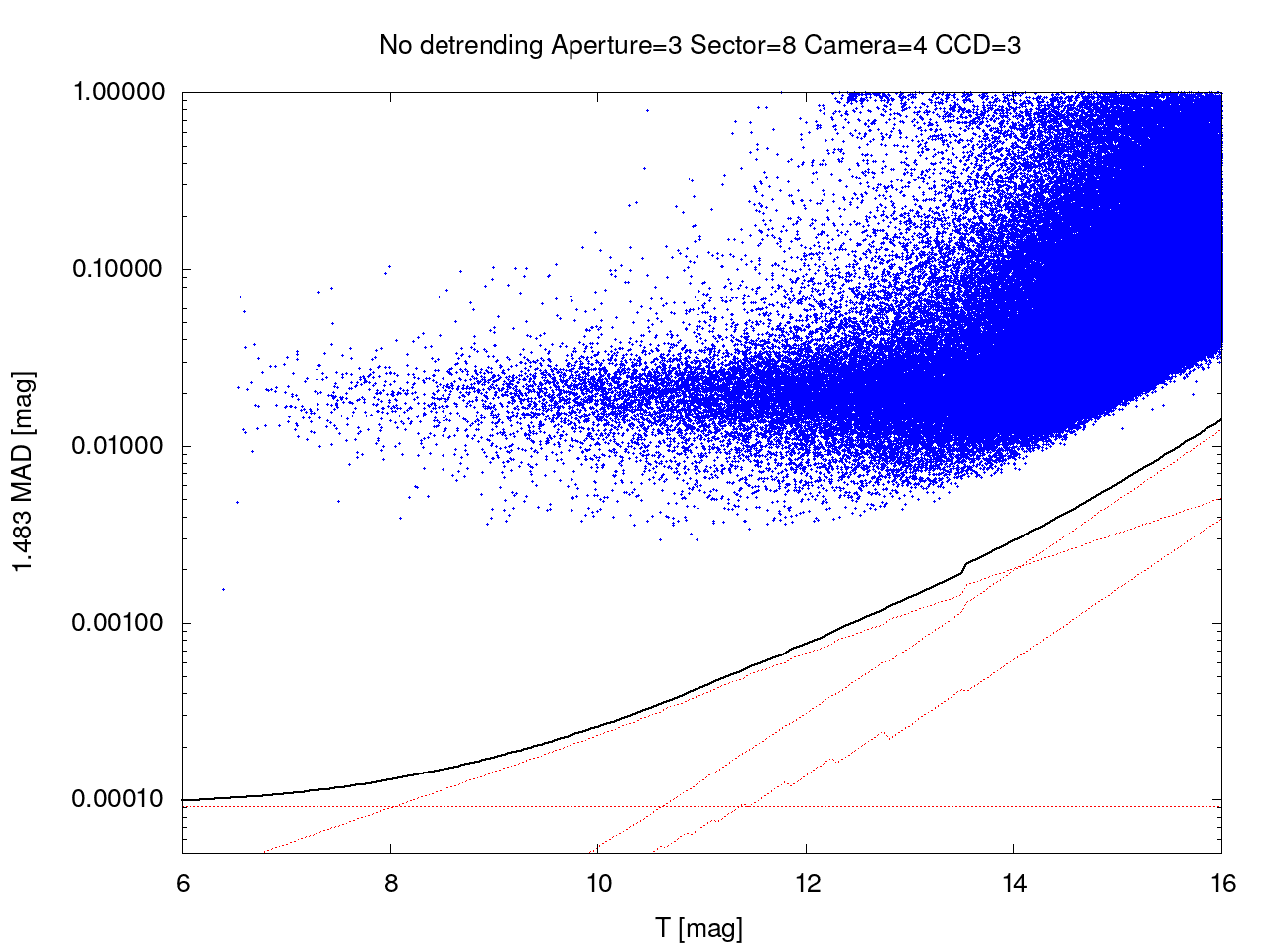}
}
{
\centering
\leavevmode
\includegraphics[width={0.5\linewidth}]{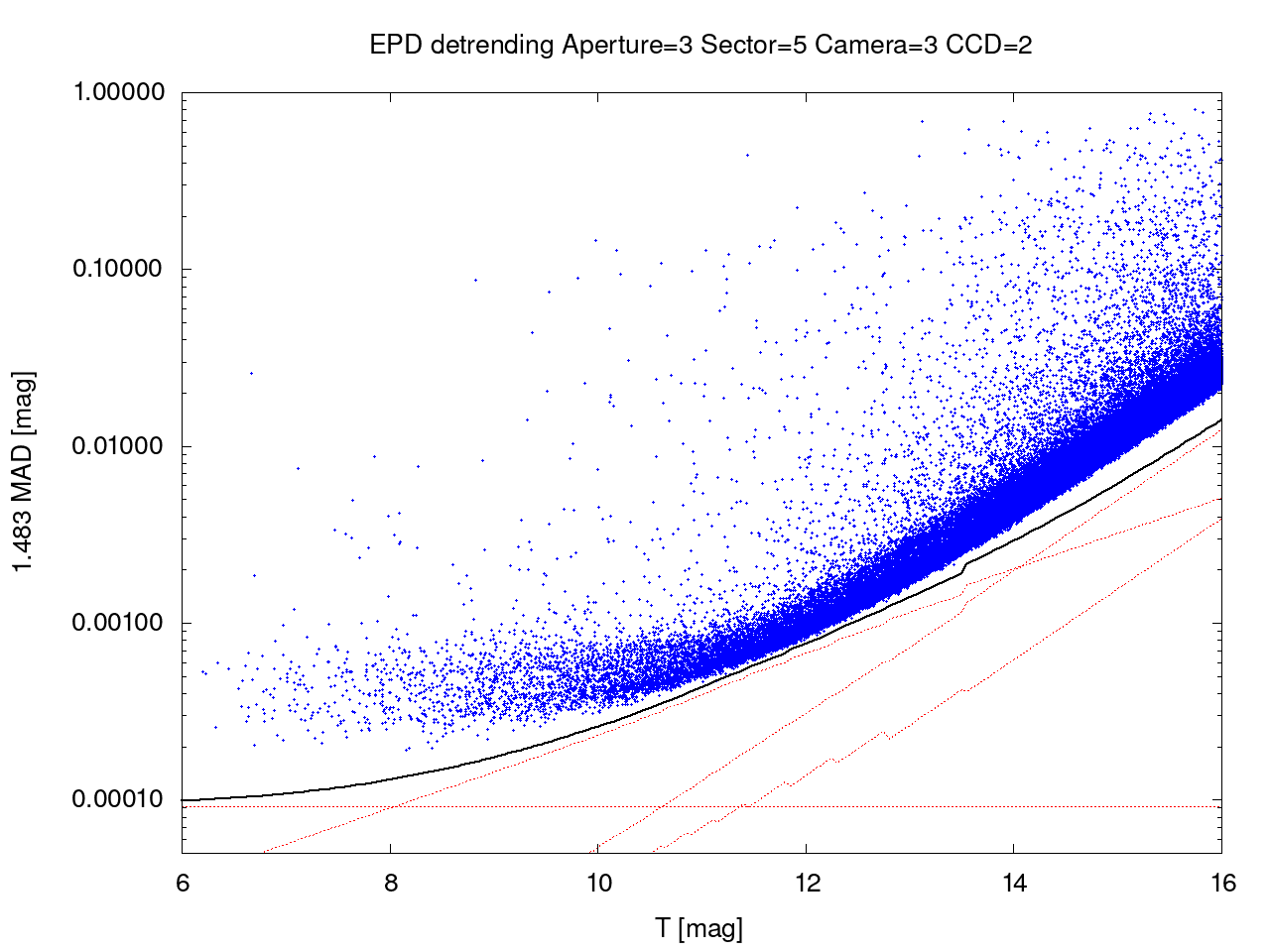}
\hfil
\includegraphics[width={0.5\linewidth}]{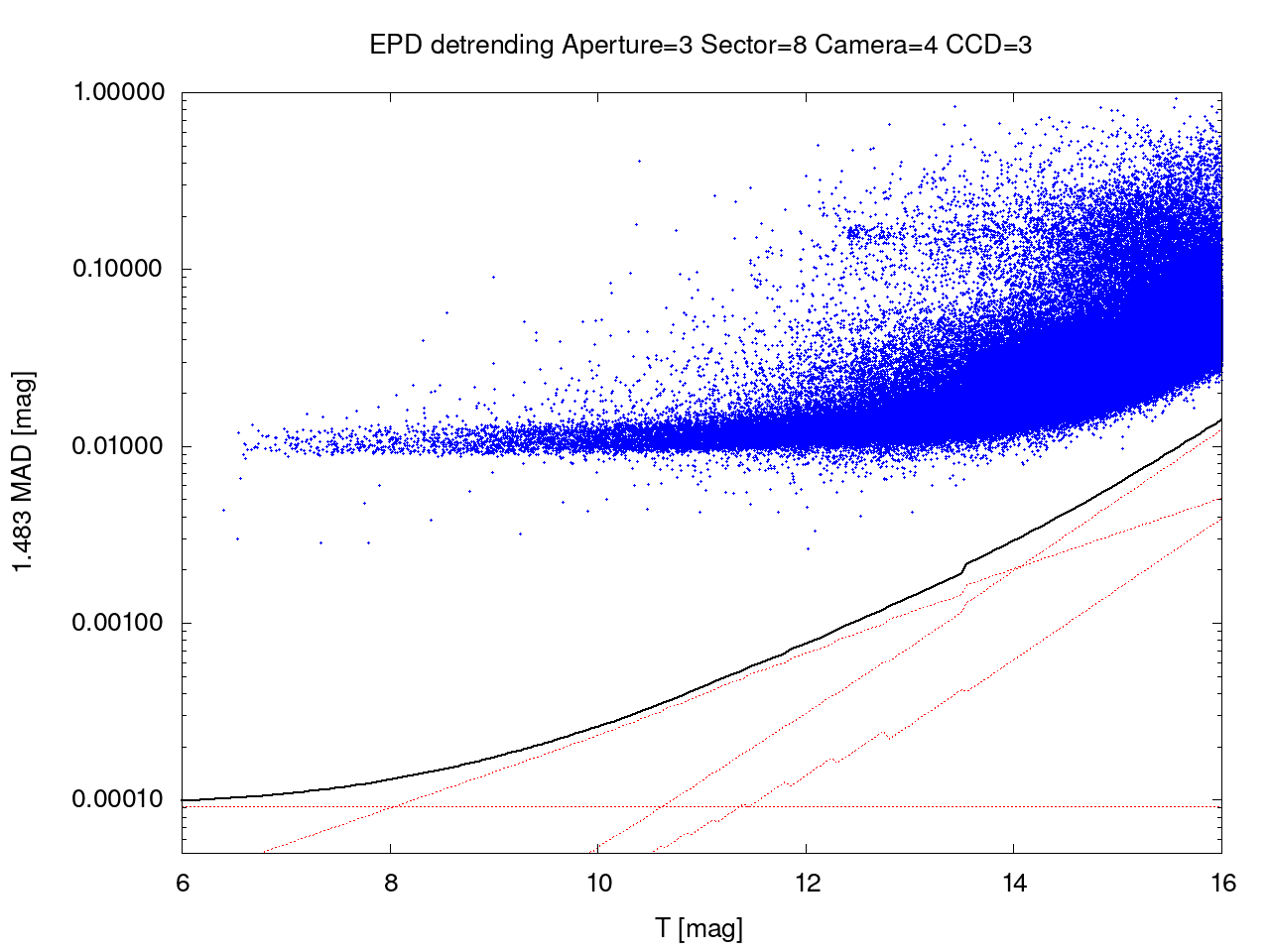}
}
{
\centering
\leavevmode
\includegraphics[width={0.5\linewidth}]{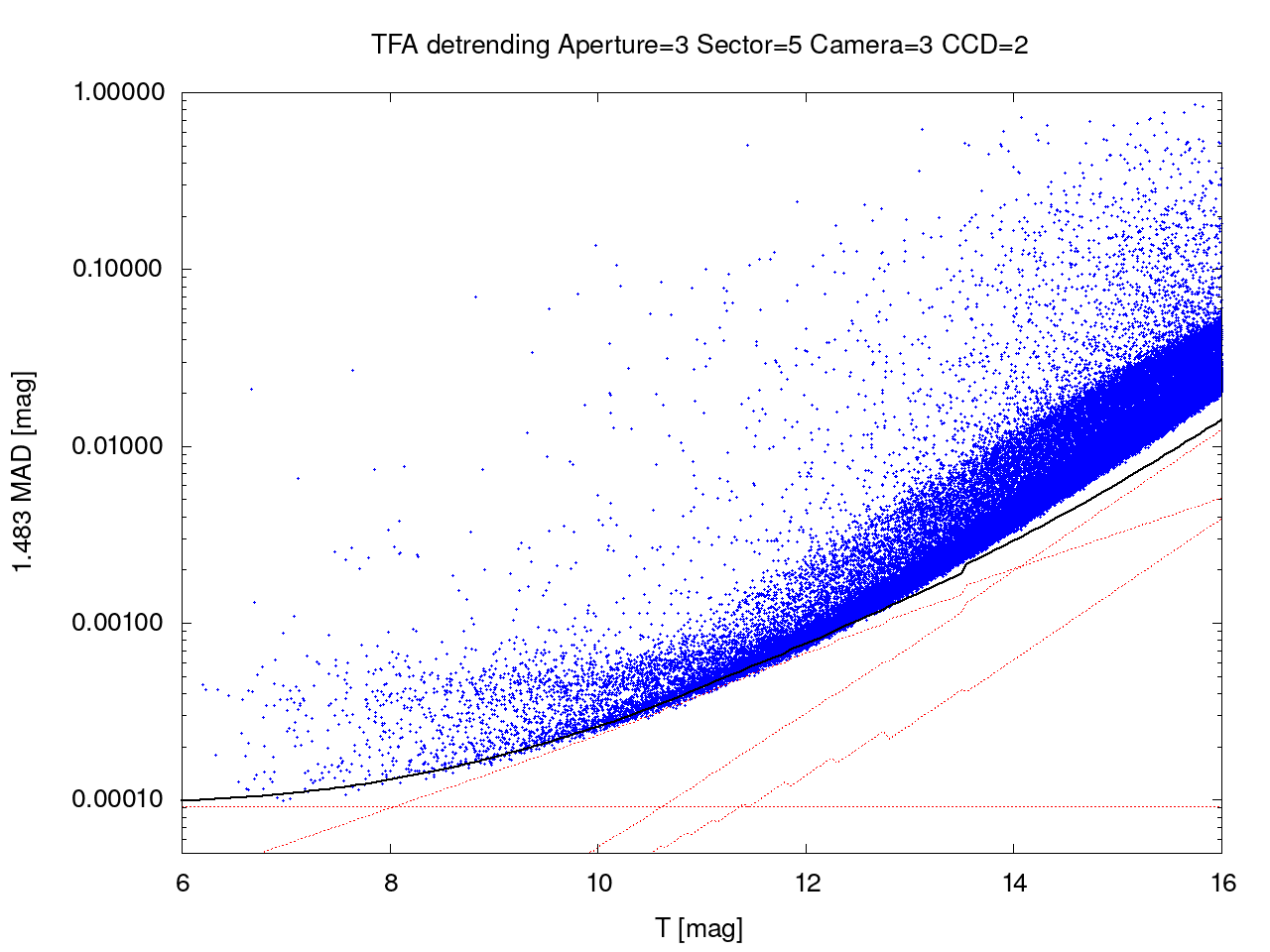}
\hfil
\includegraphics[width={0.5\linewidth}]{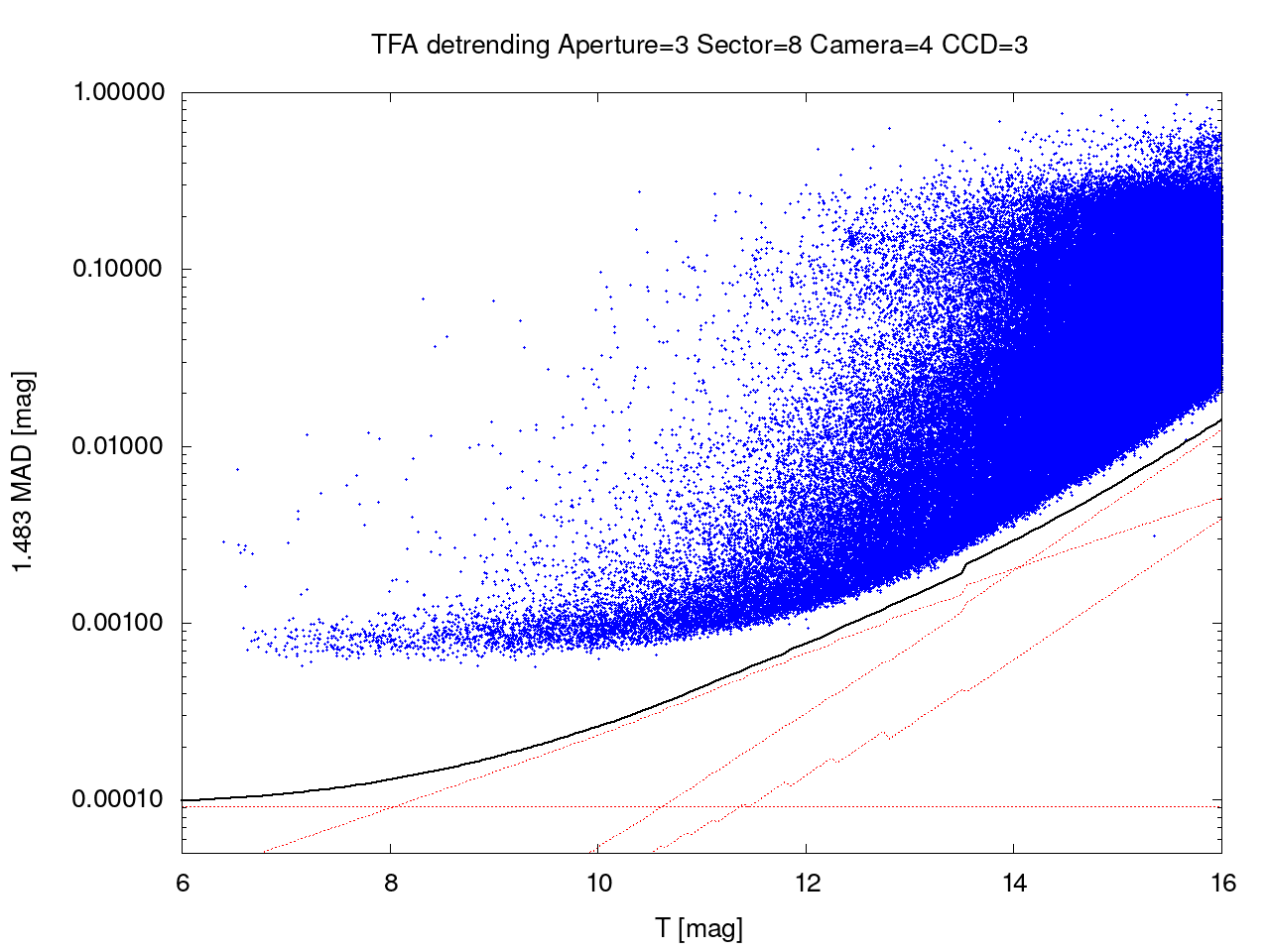}
}
\caption{The T16 aperture 3 light curve MAD plotted vs the source $T$-band magnitude for a Sector/Camera/CCD where the precision approaches the theoretical noise floor after detrending ({\em left}), and for the Sector/Camera/CCD with the poorest realized precision ({\em right}). See Section~\ref{sec:precision} for a discussion of why the systematic noise floor is so high for this particular dataset. The solid line shows the theoretical noise floor from \citet{bouma:2019:cdips}, while the dashed lines show contributions from the source shot noise, background shot noise, and detector read noise. The top row shows the statistics for the un-detrended light curves, the middle row for the SEPD-detrended light curves, and the bottom row for the SEPD+TFA-detrended light curves. Even for Sector/Camera/CCD combinations where the theoretical noise floor is achieved, as shown at left, there are often some light curves in the data with higher levels of systematic errors.
\label{fig:magmadap3}}
\end{figure*}

\begin{figure*}[!ht]
{
\centering
\leavevmode
\includegraphics[width={0.5\linewidth}]{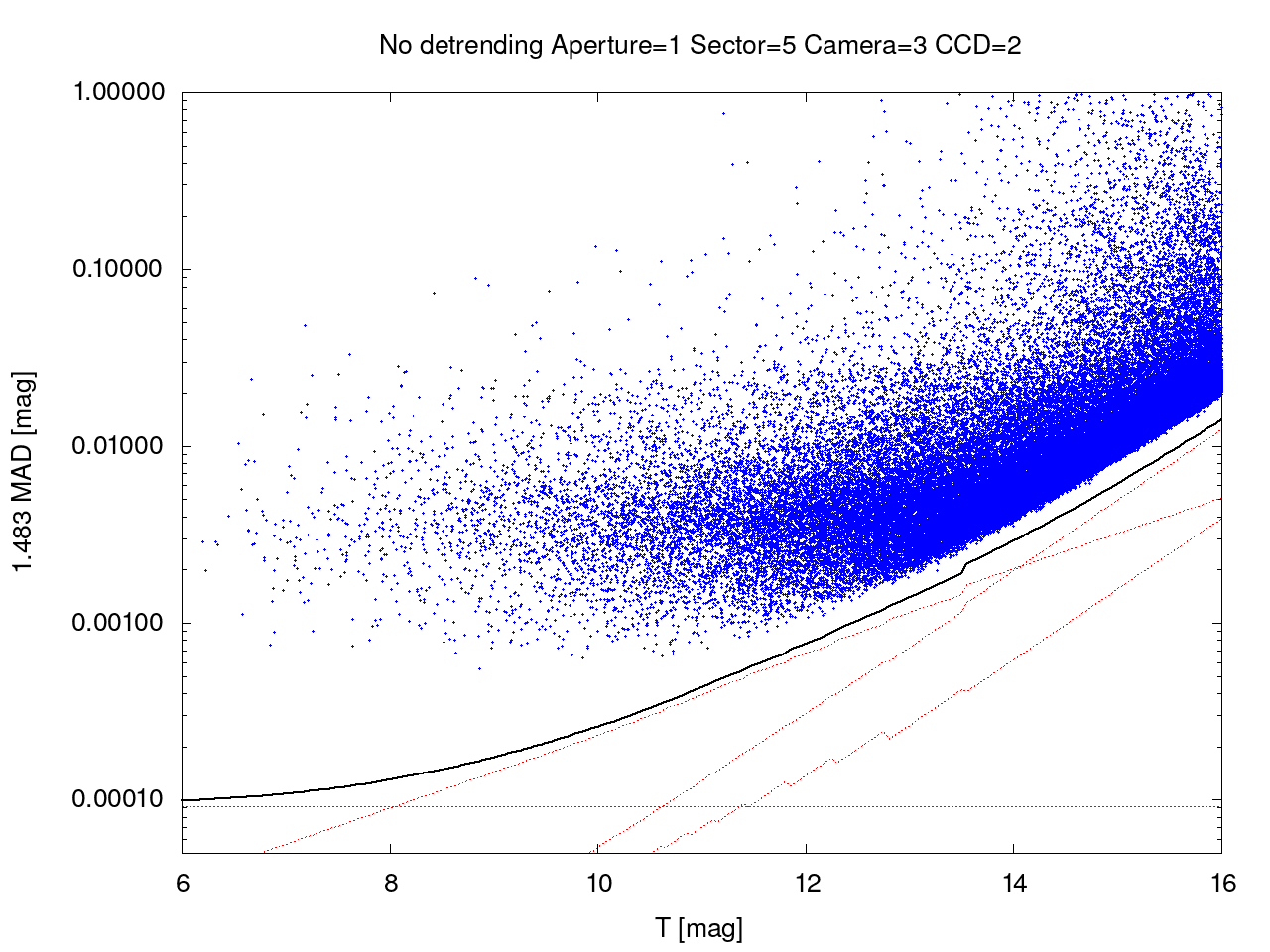}
\hfil
\includegraphics[width={0.5\linewidth}]{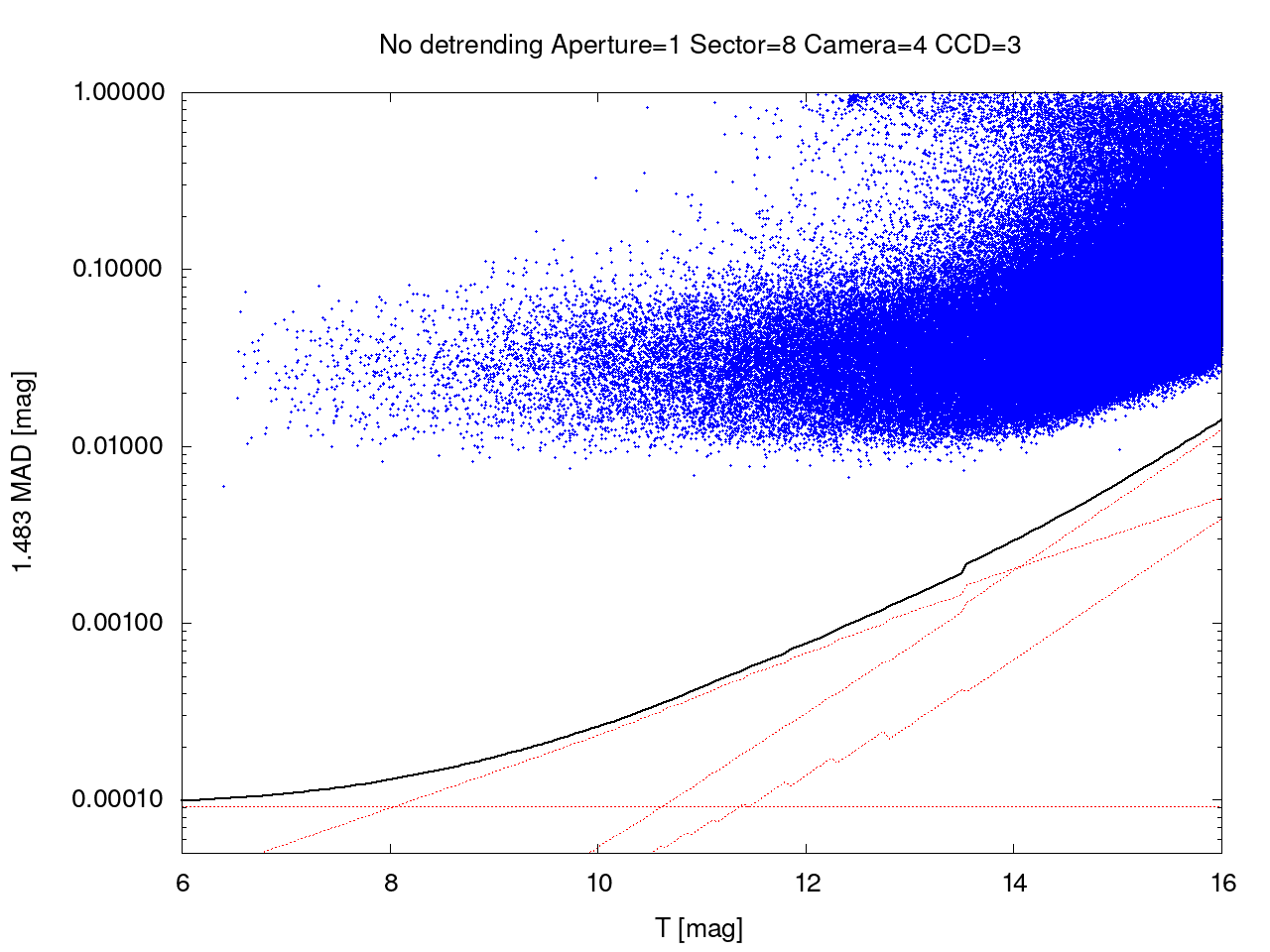}
}
{
\centering
\leavevmode
\includegraphics[width={0.5\linewidth}]{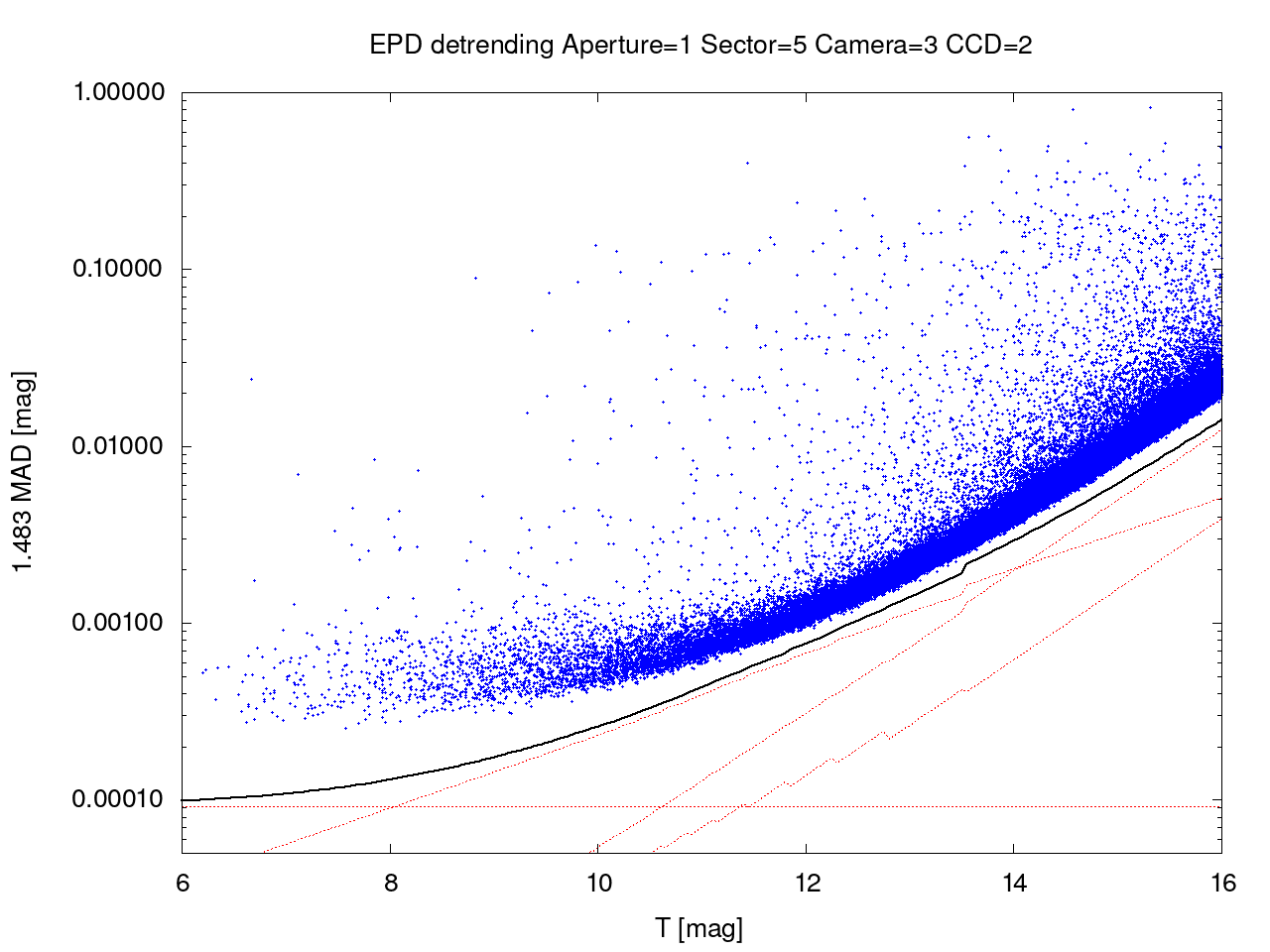}
\hfil
\includegraphics[width={0.5\linewidth}]{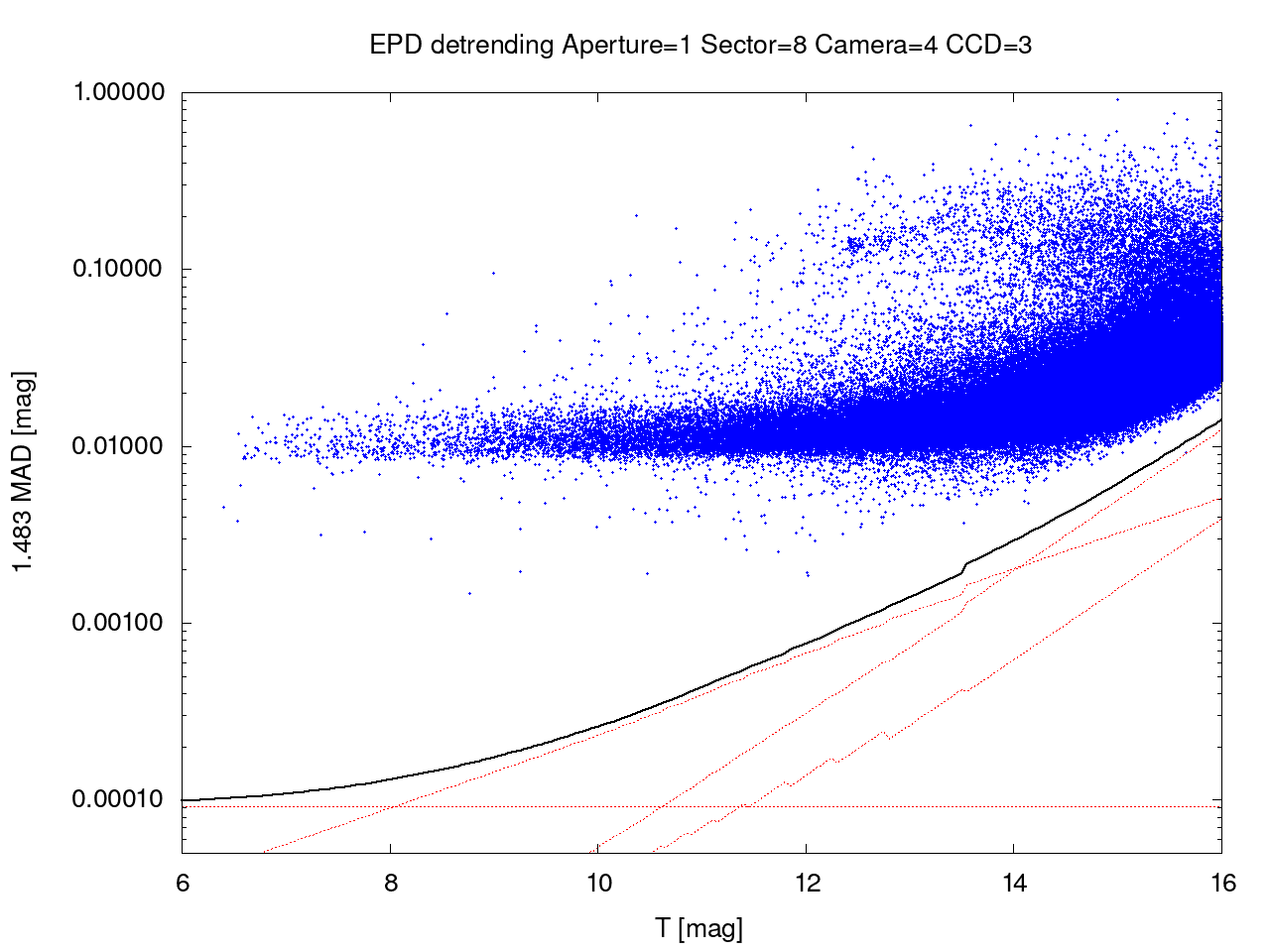}
}
{
\centering
\leavevmode
\includegraphics[width={0.5\linewidth}]{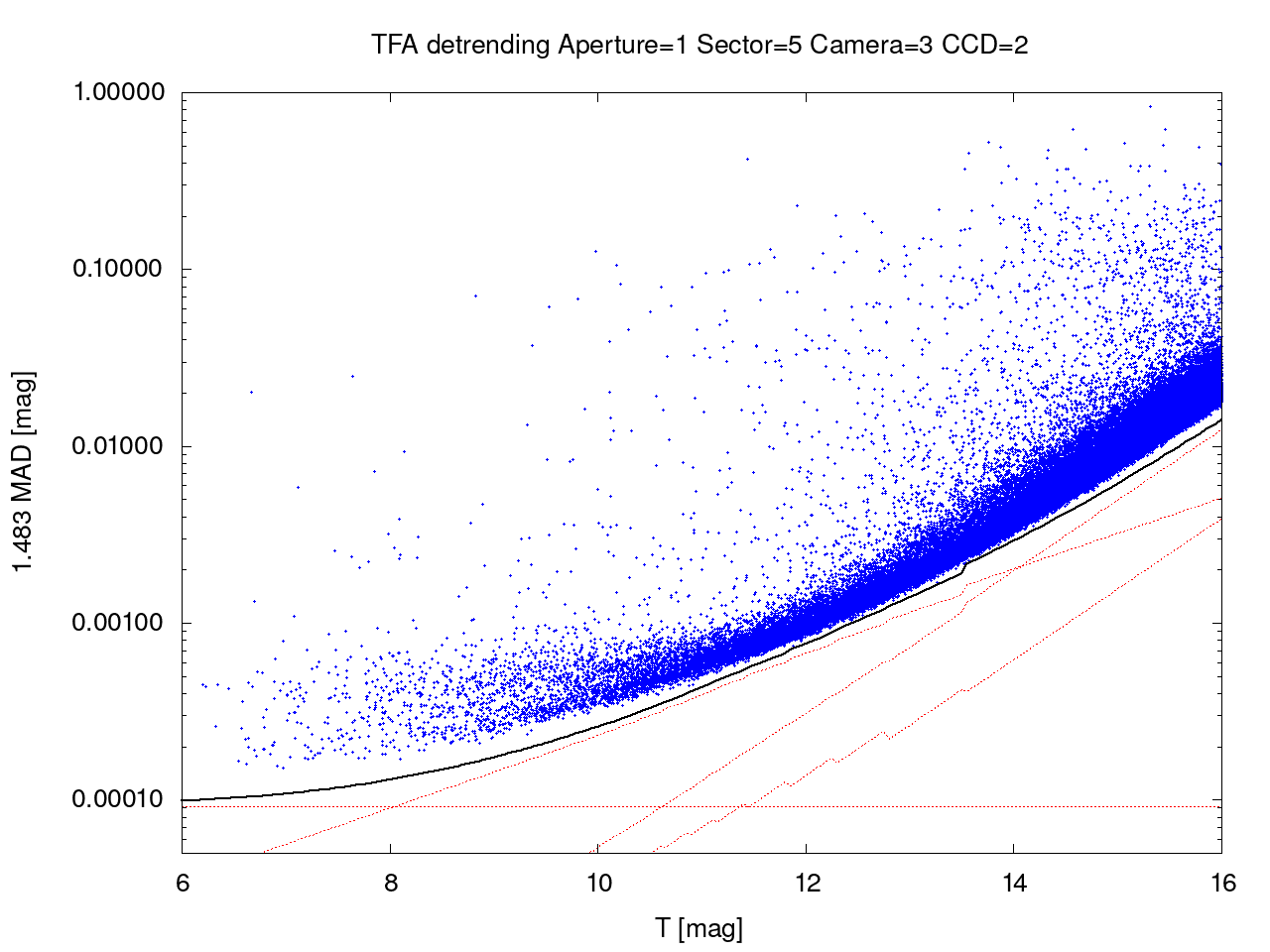}
\hfil
\includegraphics[width={0.5\linewidth}]{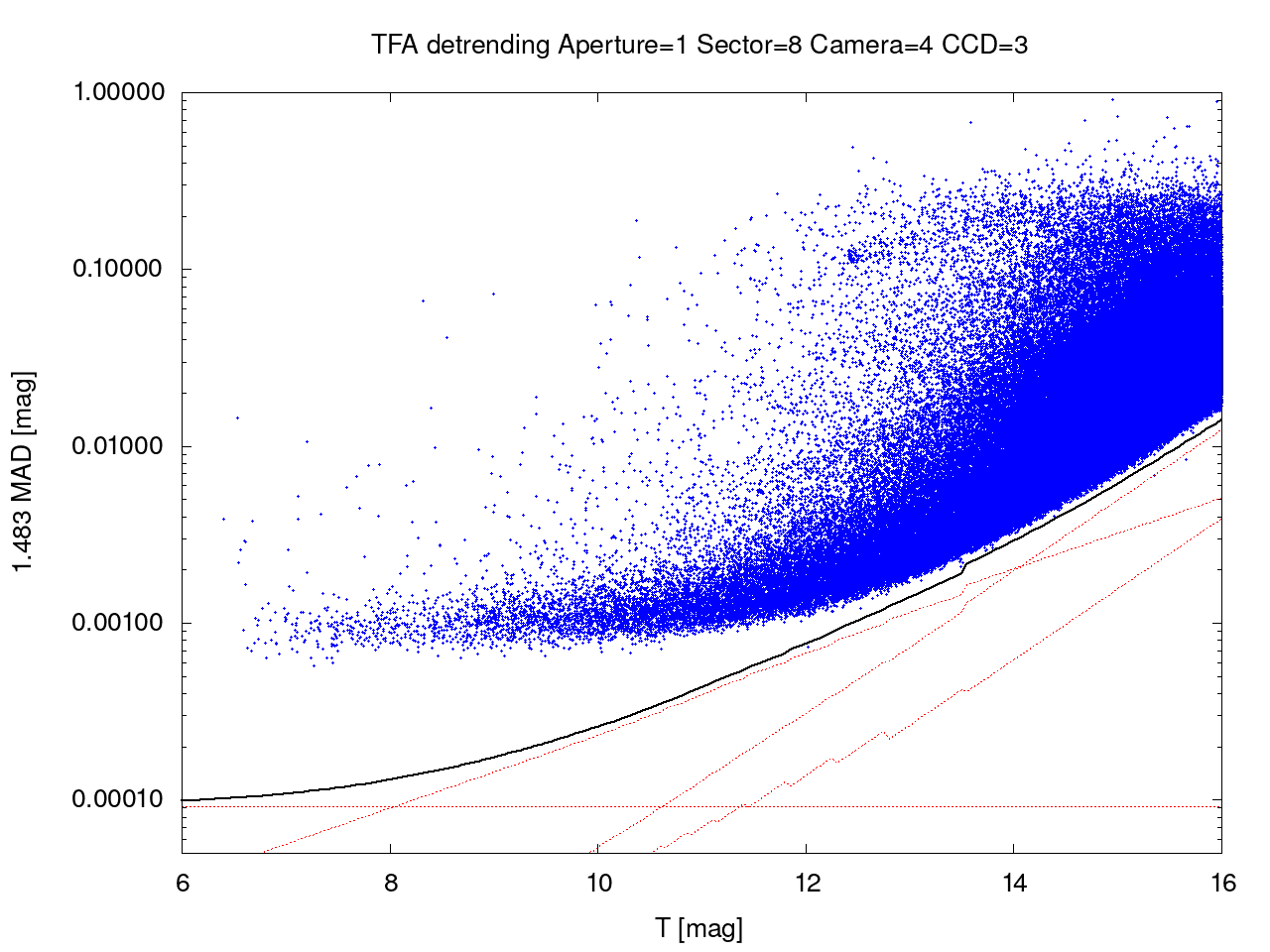}
}
\caption{Similar to Fig.~\ref{fig:magmadap3}, here we show the results for aperture 1. As expected, the precision for fainter stars improves using the smaller aperture, while for brighter stars the larger aperture yields improved precision.
\label{fig:magmadap1}}
\end{figure*}

To characterize the photometric precision attained through this work, we calculate the median absolute deviation (MAD) of each magnitude time-series for each source, where we take
\begin{equation}
{\rm MAD} = {\rm med}_{i}(|mag_{i} - {\rm med}_{j}(mag_{j})|)
\end{equation}
to be the median of the absolute differences from the median.
Figures~\ref{fig:magmadap3} and~\ref{fig:magmadap1} show the $1.483
\times {\rm MAD}$ vs.\ T-band magnitude for light curves from a
selected sample of sector/camera/ccd combinations. Here the scaling
factor of 1.483 is chosen as the expected value $1.483 \times {\rm MAD}$ is
equal to the standard deviation for a Gaussian distribution, in the
large sample limit. A full set of figures showing these relations for
all sector/camera/ccd combinations is available at \url{https://doi.org/10.5281/zenodo.14278698}. Shown in these
figures are estimates for the theoretically expected photometric noise
floor, as calculated in \citet{bouma:2019:cdips}. Individual
contributions from the source shot noise, from the sky background due
to unresolved stars, from read-noise, and a systematic noise
floor, are shown as the various dashed lines in the figure.

\begin{figure}[!ht]
{
\centering
\leavevmode
\includegraphics[width={\linewidth}]{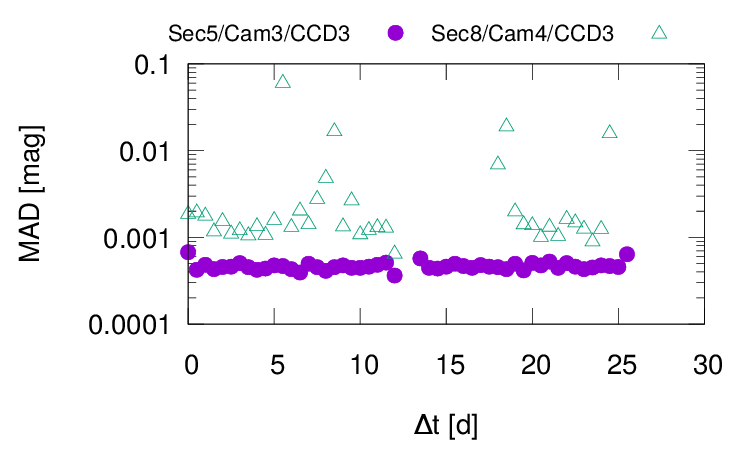}
}
\caption{The median MAD vs.\ time from the first observation in a
  sector for stars with $9 < T < 10$ in the two indicated
  sector/camera/ccd combinations (the same as those shown in
  Fig.~\ref{fig:magmadap3}). The MAD is calculated in 0.5\,day bins
  for each light curve using the undetrended aperture 3
  magnitudes. The MAD is systematically higher at specific times,
  especially at points in the spacecraft orbit with high and complex
  background scattering.
\label{fig:madvstime}}
\end{figure}

In general, without detrending, the scatter in the light curves, as
measured by MAD, is significantly above the theoretical noise floor. This is
due to the large systematic variations present in the light curves,
especially near times in the orbit with high and complex background
scattering. Figure~\ref{fig:madvstime} demonstrates this for the undetrended light curves from the two sectors shown in Fig~\ref{fig:magmadap3}. Here we calculate the median MAD of the aperture 3 light curves for stars within the magnitude range $9 < T < 10$ in 0.5\,d time bins, and show the result as a function of time from the first observation in the sector. For both sector/camera/ccd combinations shown the MAD shows pronounced increases at specific times in the light curves, but the outliers are particularly noticeable for the Sector 8, Camera 4, CCD 3. After applying the SEPD and TFA detrending techniques the noise for
most stars in most sector/camera/ccd combinations approaches the
theoretical noise floor. 

There are several notable exceptions, such as Sector 8, Camera 4, CCD
3 which is shown in Figure~\ref{fig:magmadap3}. The T16 light curves
derived for this sector/camera/ccd exhibit systematic and sudden jumps
to lower brightness that can last for up to 3 days. This may be
related to the presence of a bad column on the CCD and the bright star
$\beta$ Dor which bled into the upper buffer rows of the CCD according
to the {\em TESS} data release notes for Sector 8. One or both of
these features may have impacted the image subtraction process fit for
the change in flux scale between the reference and science images. An
inspection of the subtracted frames shows sudden jumps in the flux
residuals across the entire CCD that suggests a bad fit to the dataset
at these times. We present this here to caution the user that data
anomalies are present in these data, leading in some cases to large
systematic artifacts that are especially pronounced for particular
sector/camera/ccd combinations. In addition to sector 8, camera 4, ccd
3, two other datasets exhibit a systematic noise floor above
0.5\,mmag. These are sector 1, camera 2, ccds 2 and 4. The user is
encouraged to inspect the magnitude-MAD diagrams for any particular
sector/camera/ccd that they wish to use to aid in evaluating the
significance of systematic photometric errors for those data.

\begin{figure*}[!ht]
{
\centering
\leavevmode
\includegraphics[width={0.5\linewidth}]{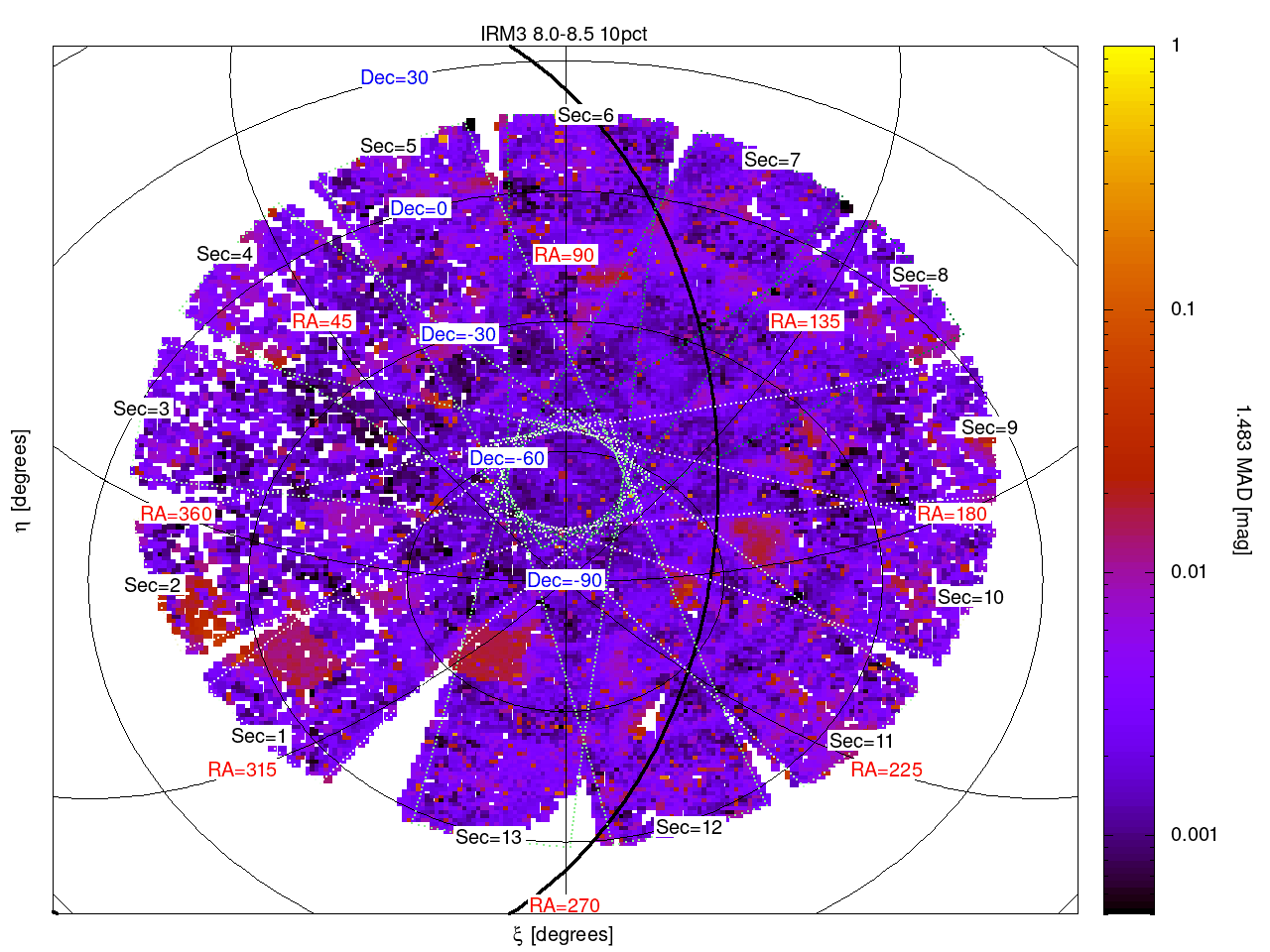}
\hfil
\includegraphics[width={0.5\linewidth}]{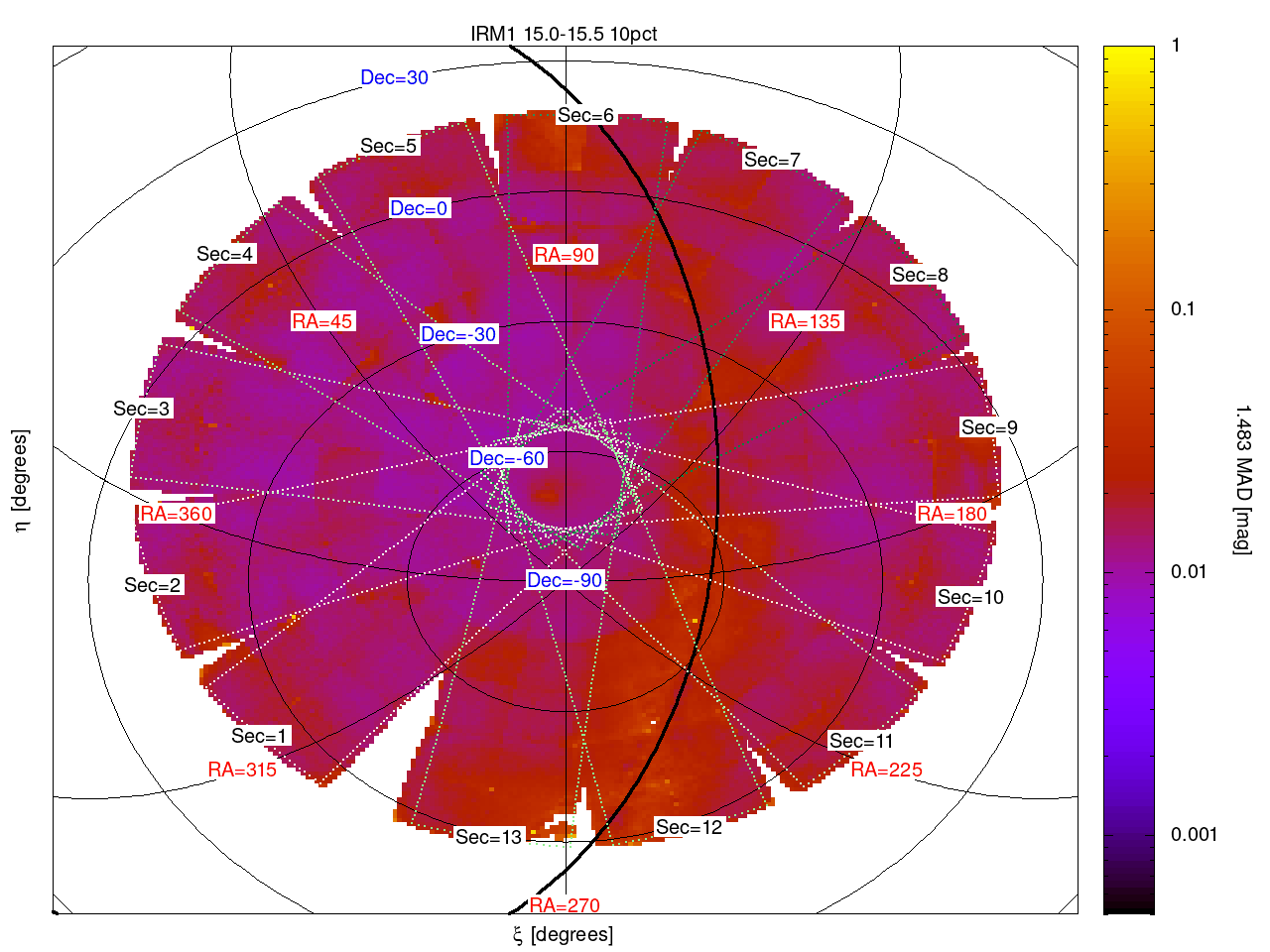}
}
{
\centering
\leavevmode
\includegraphics[width={0.5\linewidth}]{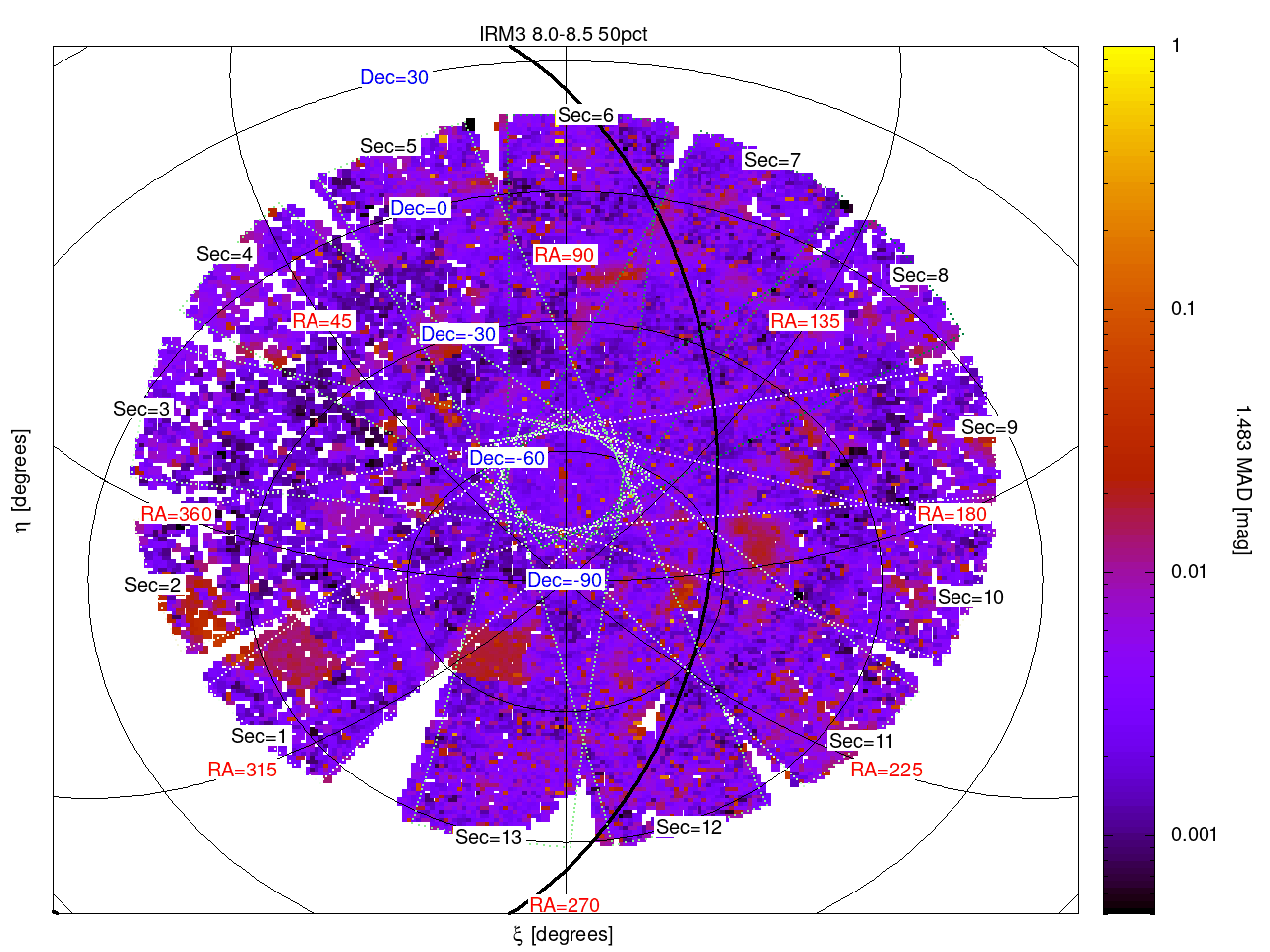}
\hfil
\includegraphics[width={0.5\linewidth}]{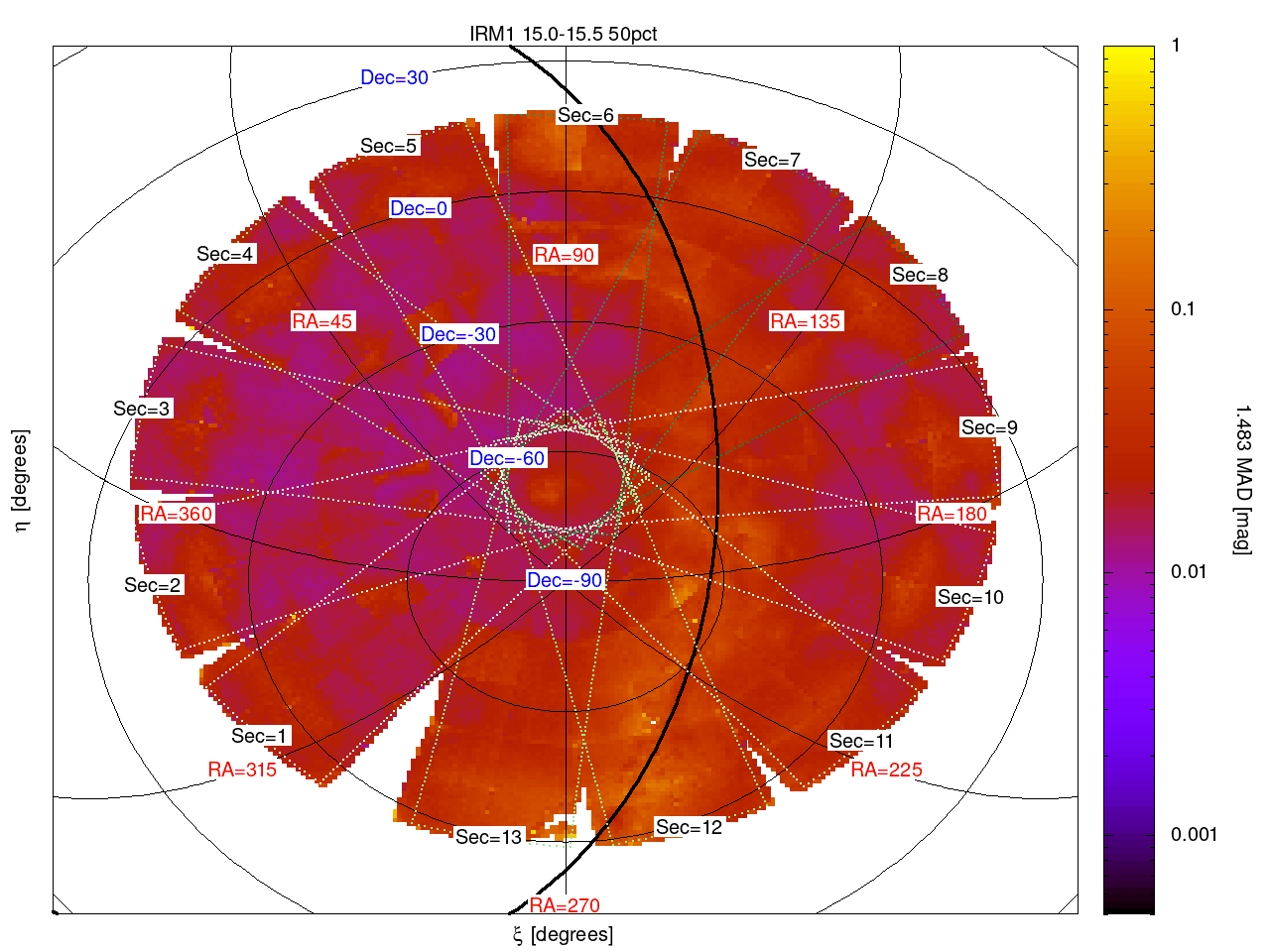}
}
{
\centering
\leavevmode
\includegraphics[width={0.5\linewidth}]{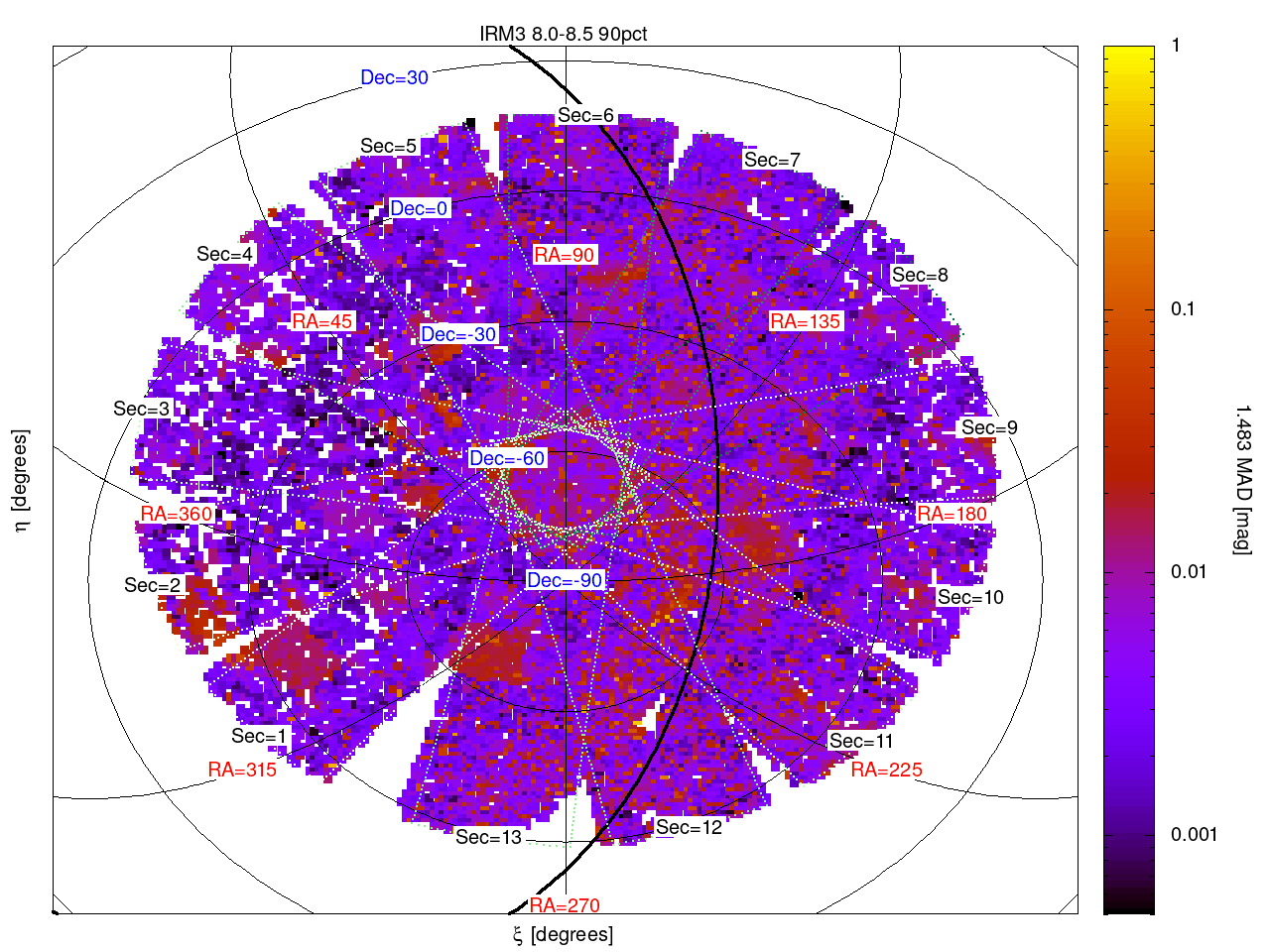}
\hfil
\includegraphics[width={0.5\linewidth}]{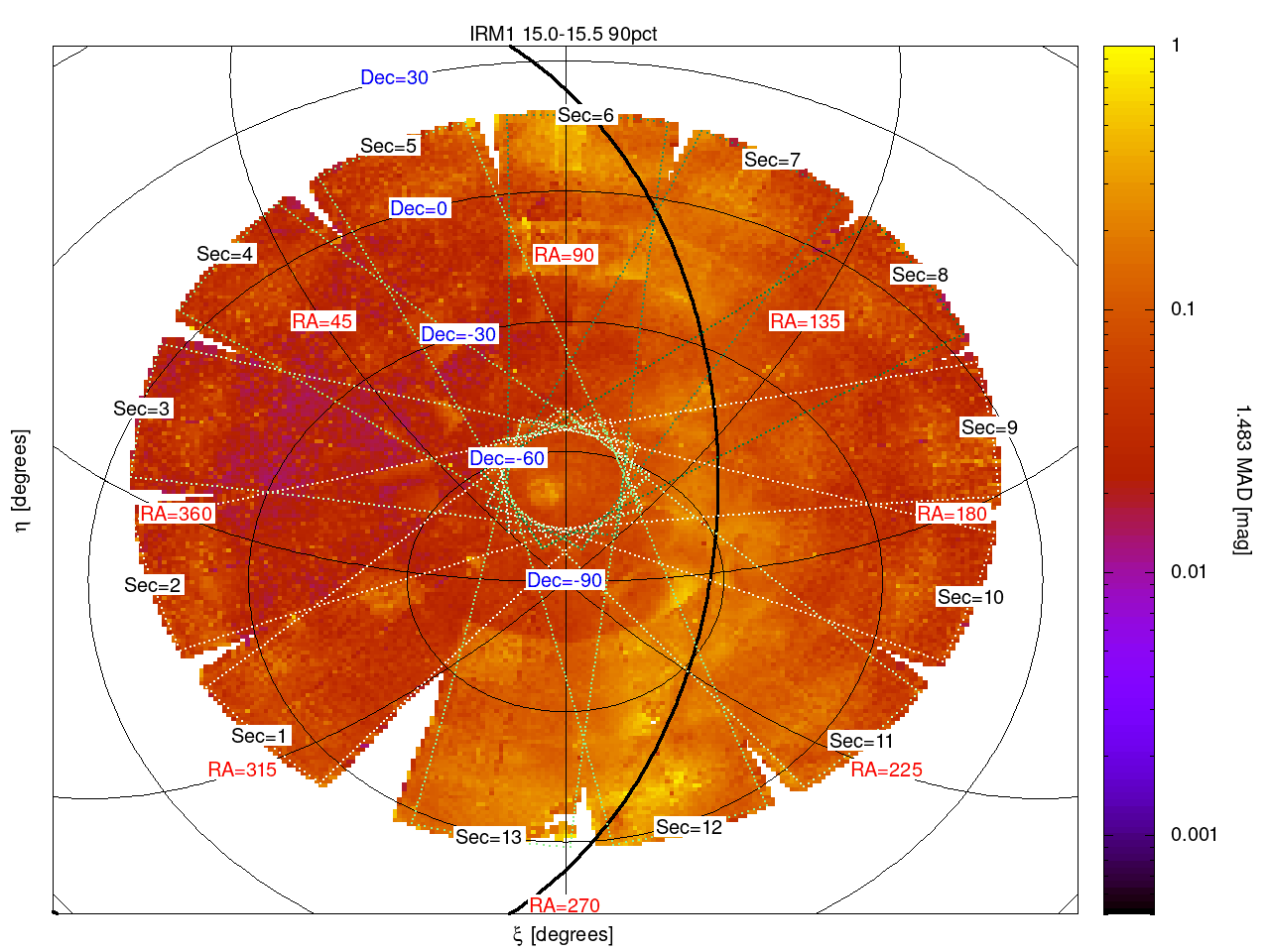}
}
\caption{The 10th percentile (top), median (middle) and 90th percentile (bottom) MAD for pre-detrended light curves calculated for bins in sky position and magnitude. The results are displayed using an arc-projection about the south ecliptic pole. We scale the MAD by a factor of 1.483 so that the expected value is equal to the standard deviation for a Gaussian distribution. Here we show the results for stars with $8 < T < 8.5$ (left) and $15 < T < 15.5$ (right). The sector boundaries are shown and labelled, as are lines of constant right ascension and declination. The dark black line in each panel shows the Galactic plane. For faint stars the MAD is dominated by the background shot noise which is higher along the plane, while for brighter stars the scatter is dominated by instrumental systematics which depend strongly on the Sector/Camera/CCD.
\label{fig:madvspositionirm}}
\end{figure*}

Figure~\ref{fig:madvspositionirm} shows the 10th percentile, median, and 90th percential MAD for the
pre-detrending raw photometry light curves at different $G$-band
magnitudes as a function of position on the sky. For bright stars
(e.g., the $8.0 < T < 8.5$ range shown in the Figure, corresponding to
stars close to saturation) we find that undetrended light curves are
dominated by systematic errors, the level of which varies
significantly among the different Sector/Camera/CCD combinations. For
some combinations of Sector/Camera/CCD the median MAD is as high as
$\sim 0.1$\,mag for these bright stars, while for others the median
MAD is less than $0.001$\,mag. This gives rise to the regions of high
MAD that are localized on the sky, and uncorrelated with Galactic
latitude. For fainter stars (e.g., $15 < T < 15.5$) noise from the sky
background dominates, and the median MAD is highest near the Galactic
plane.

\begin{figure*}[!ht]
{
\centering
\leavevmode
\includegraphics[width={0.5\linewidth}]{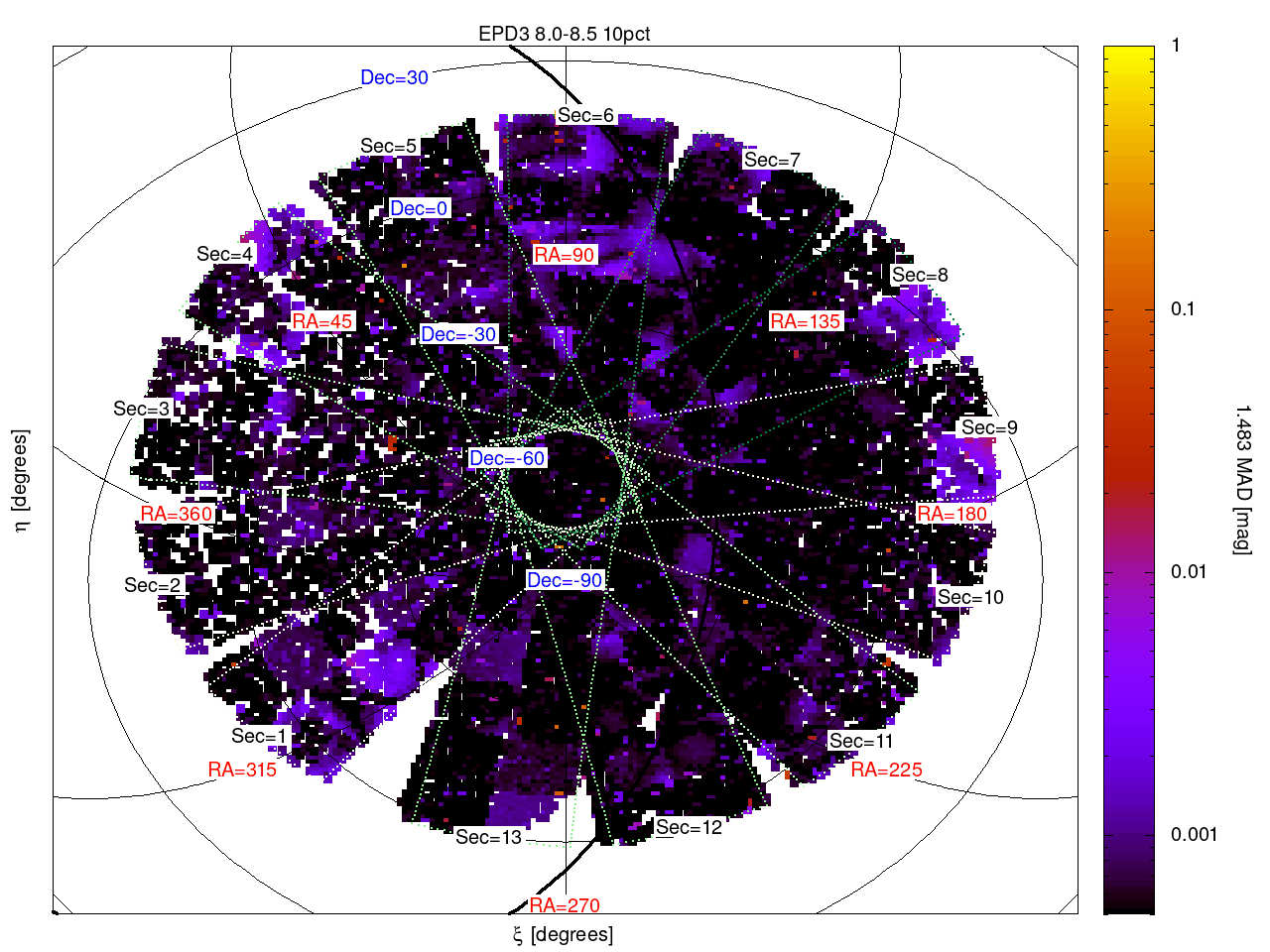}
\hfil
\includegraphics[width={0.5\linewidth}]{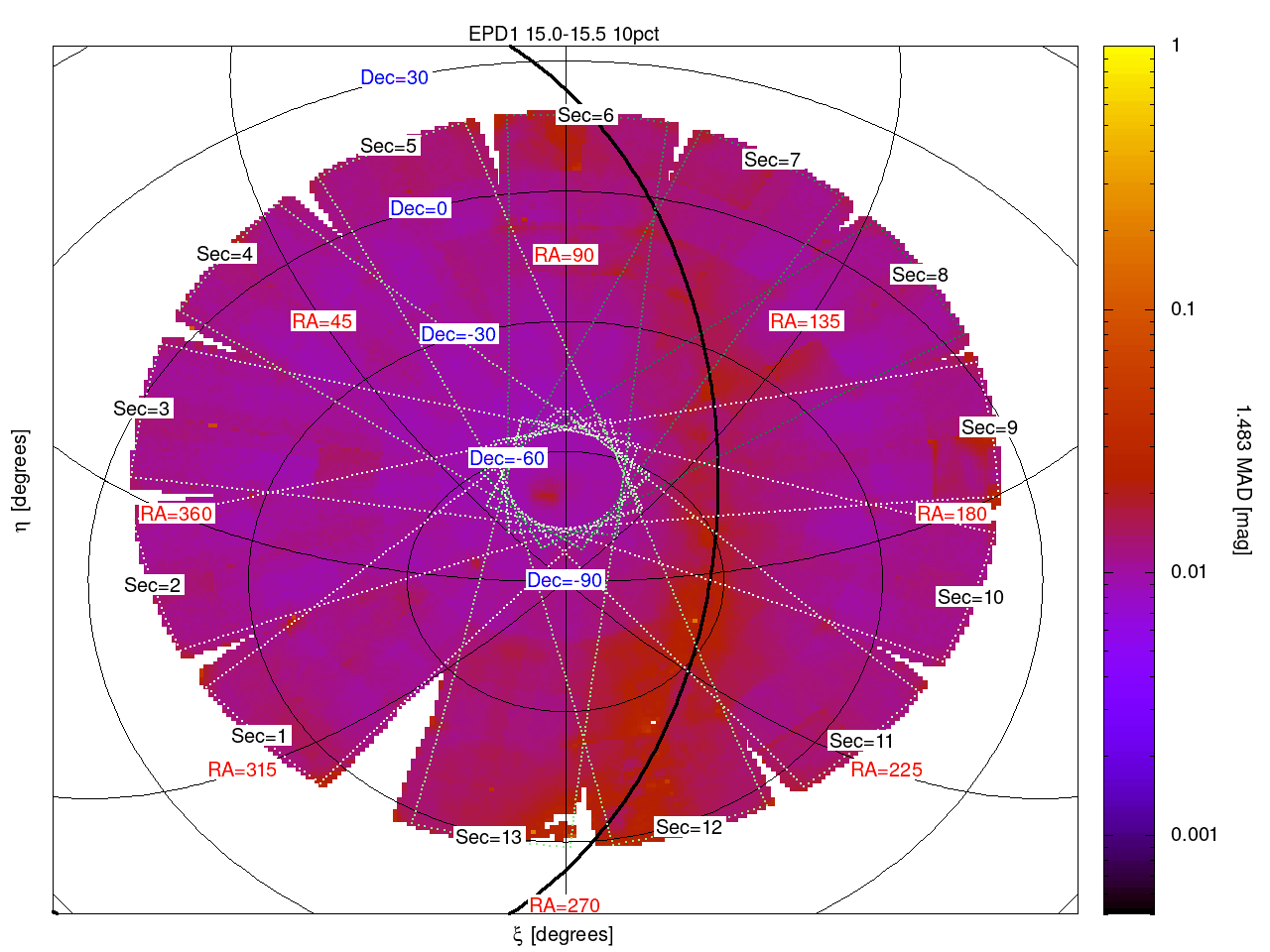}
}
{
\centering
\leavevmode
\includegraphics[width={0.5\linewidth}]{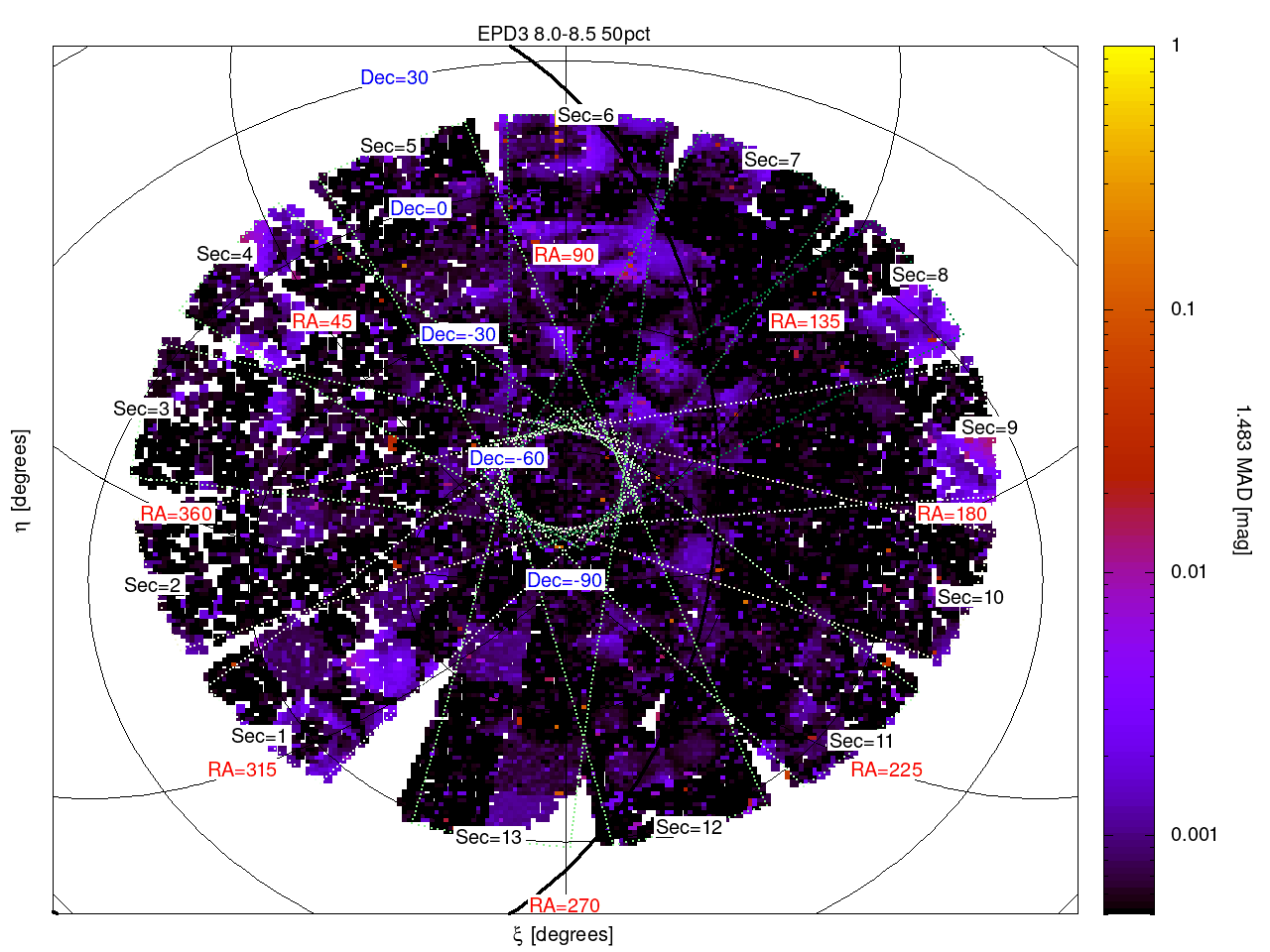}
\hfil
\includegraphics[width={0.5\linewidth}]{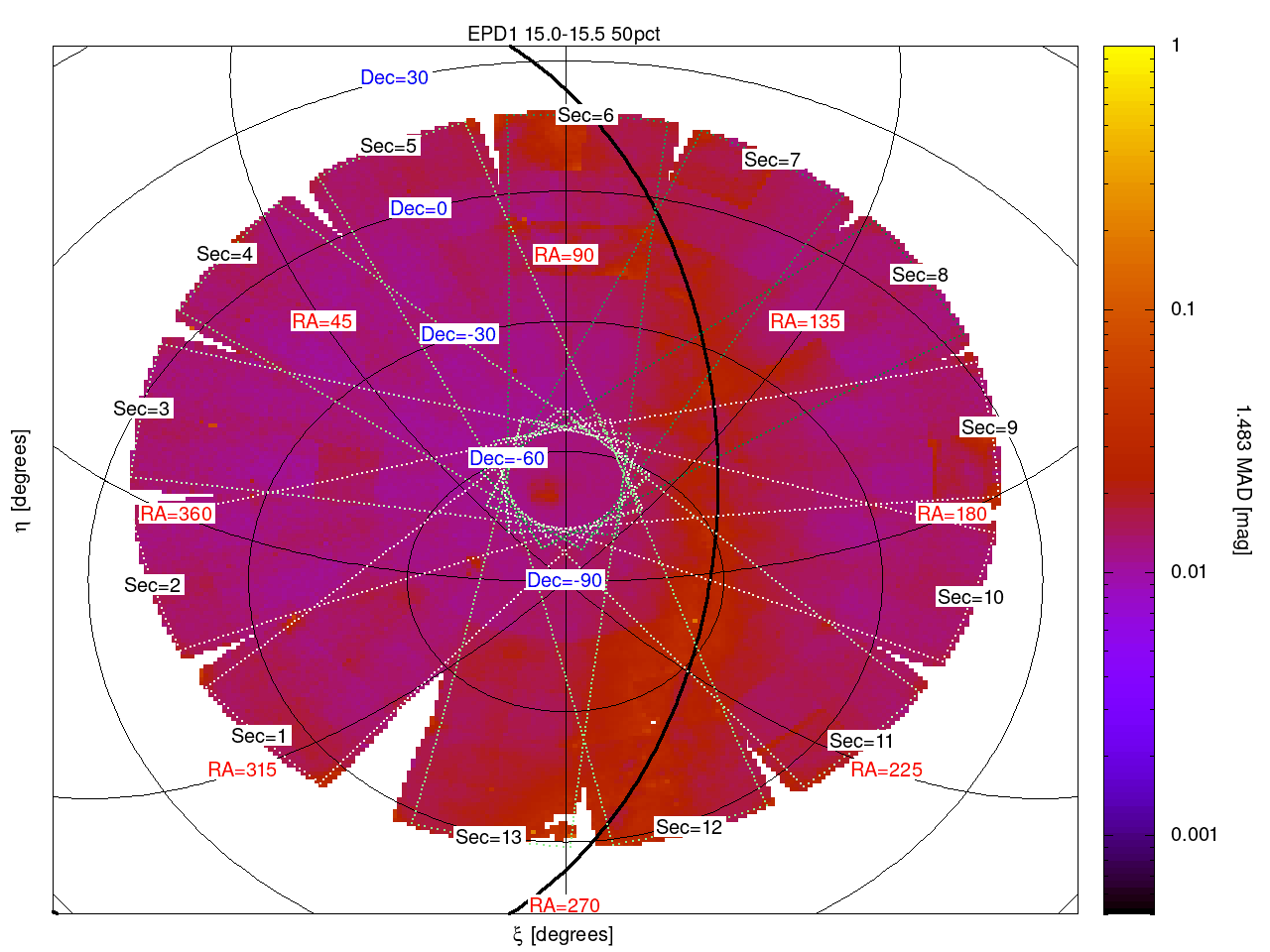}
}
{
\centering
\leavevmode
\includegraphics[width={0.5\linewidth}]{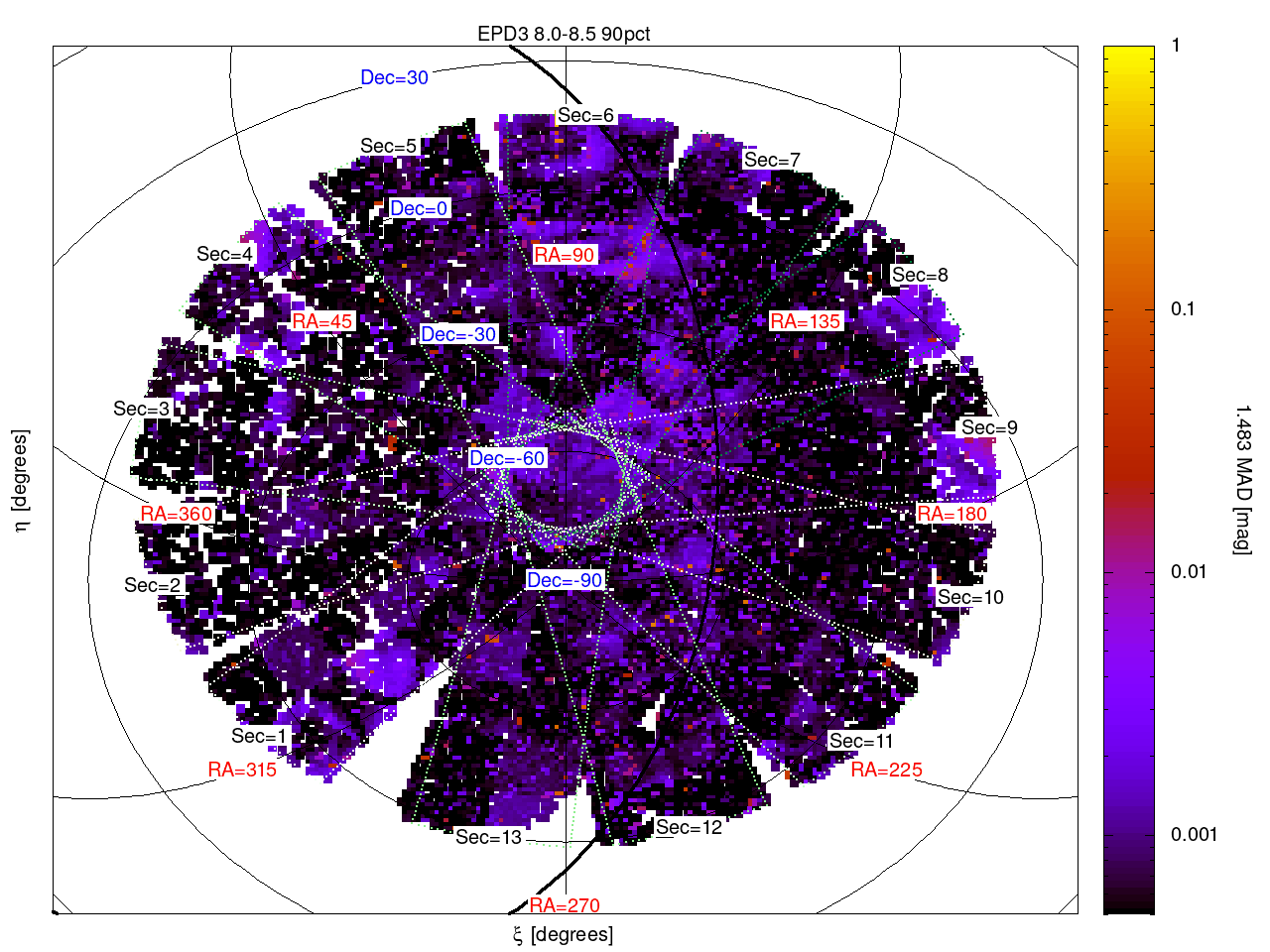}
\hfil
\includegraphics[width={0.5\linewidth}]{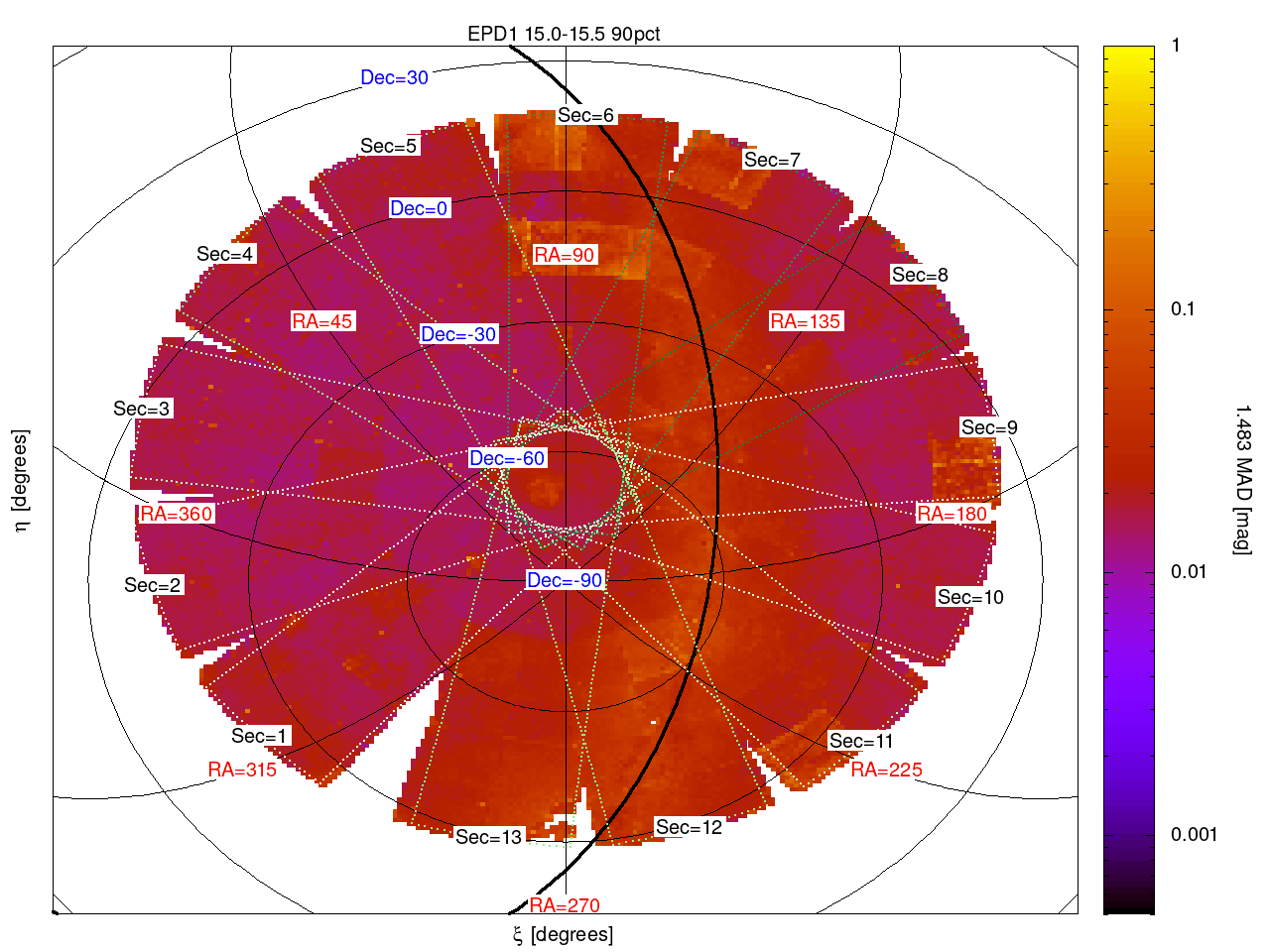}
}
\caption{Same as Figure~\ref{fig:madvspositionirm}, here we show the statistics for SEPD light curves.
\label{fig:madvspositionepd}}
\end{figure*}

\begin{figure*}[!ht]
{
\centering
\leavevmode
\includegraphics[width={0.5\linewidth}]{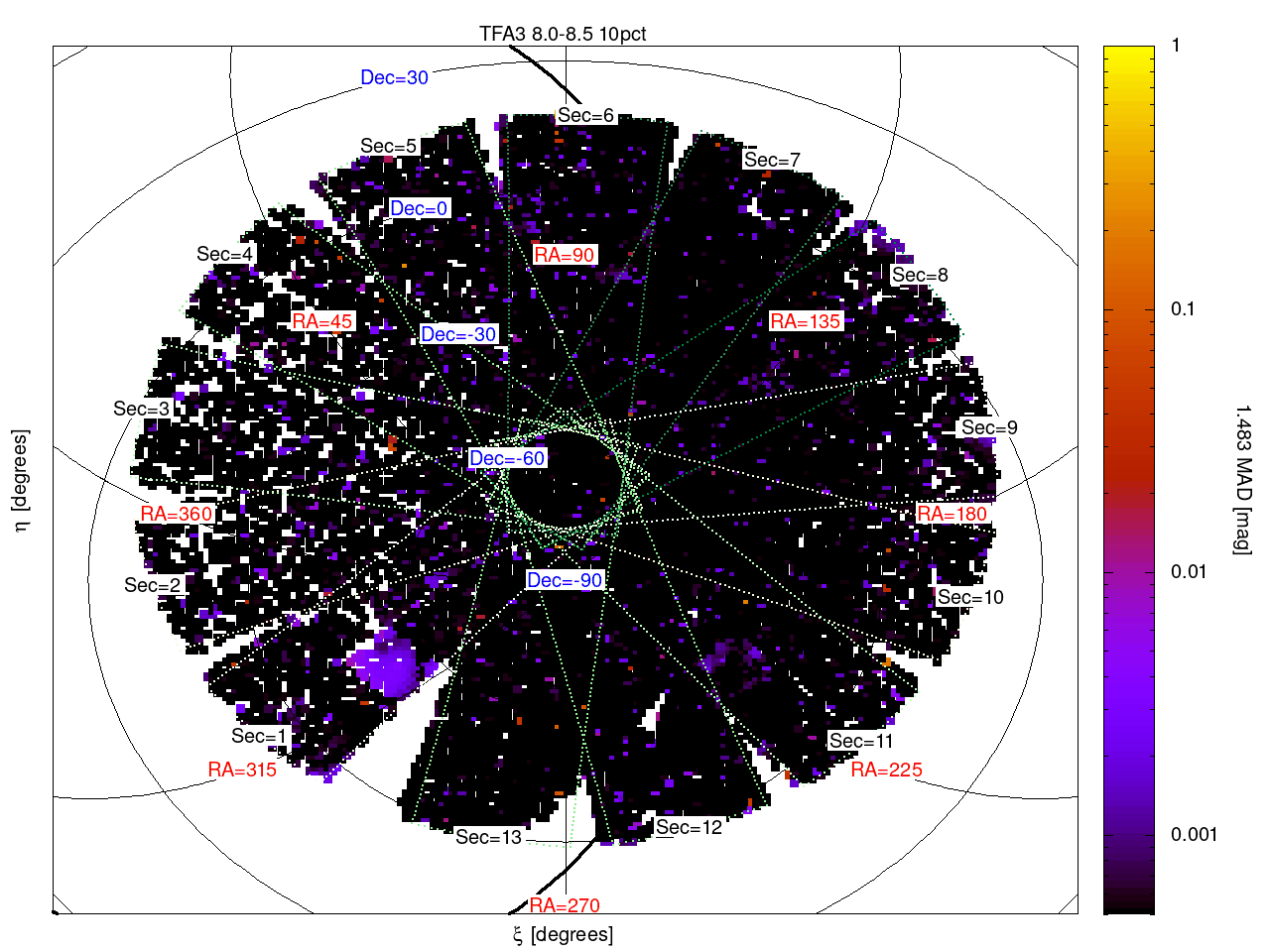}
\hfil
\includegraphics[width={0.5\linewidth}]{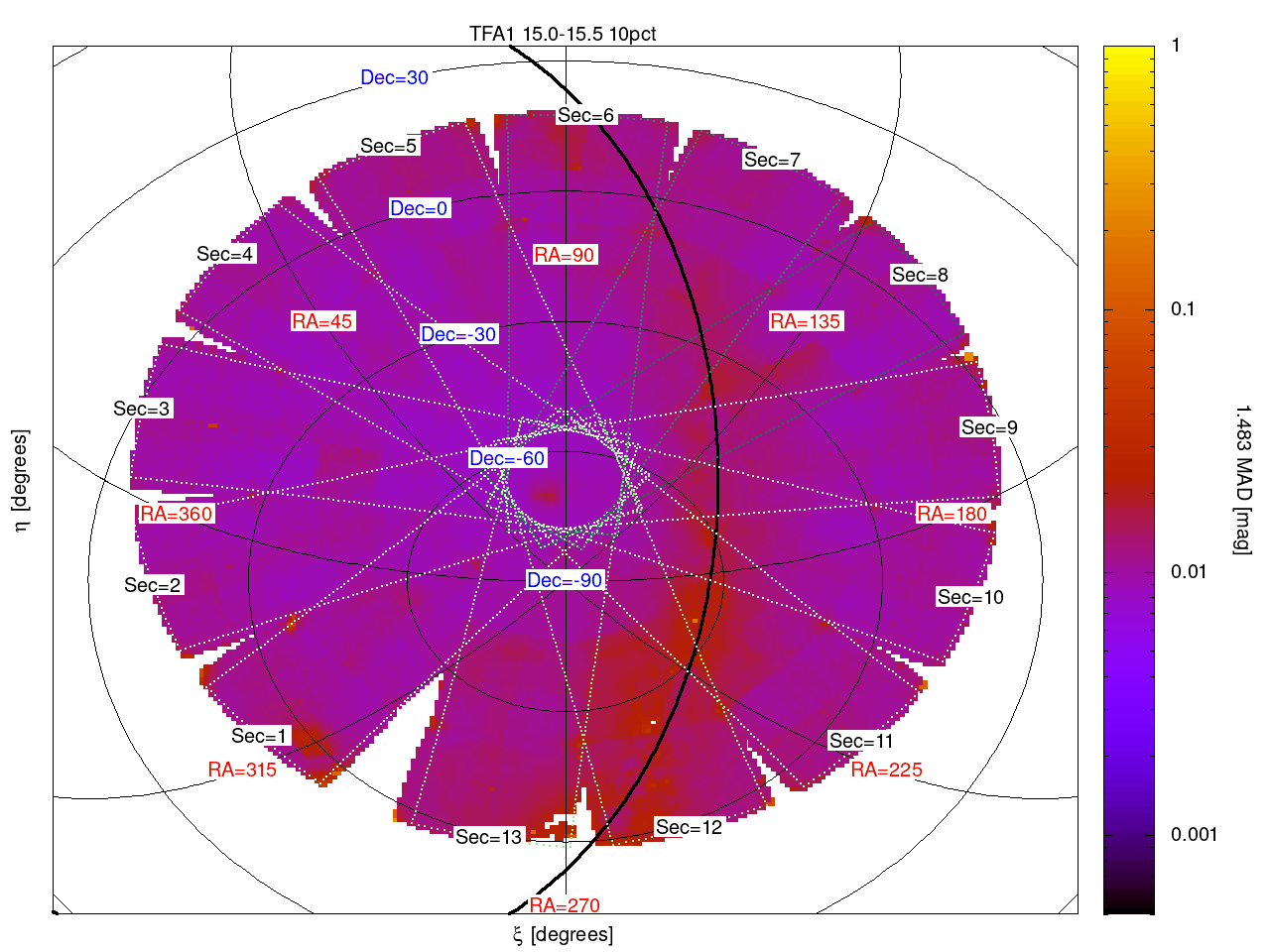}
}
{
\centering
\leavevmode
\includegraphics[width={0.5\linewidth}]{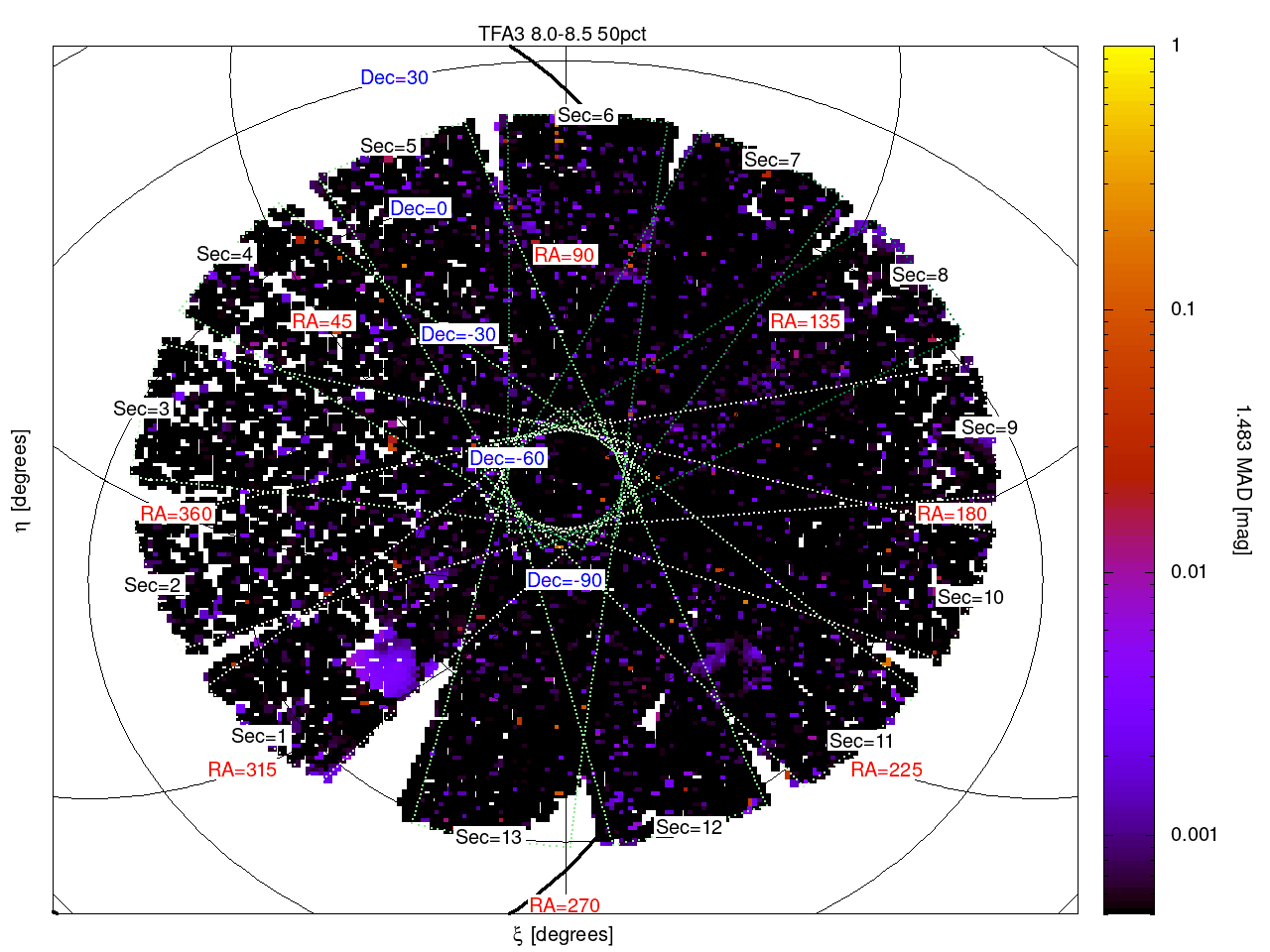}
\hfil
\includegraphics[width={0.5\linewidth}]{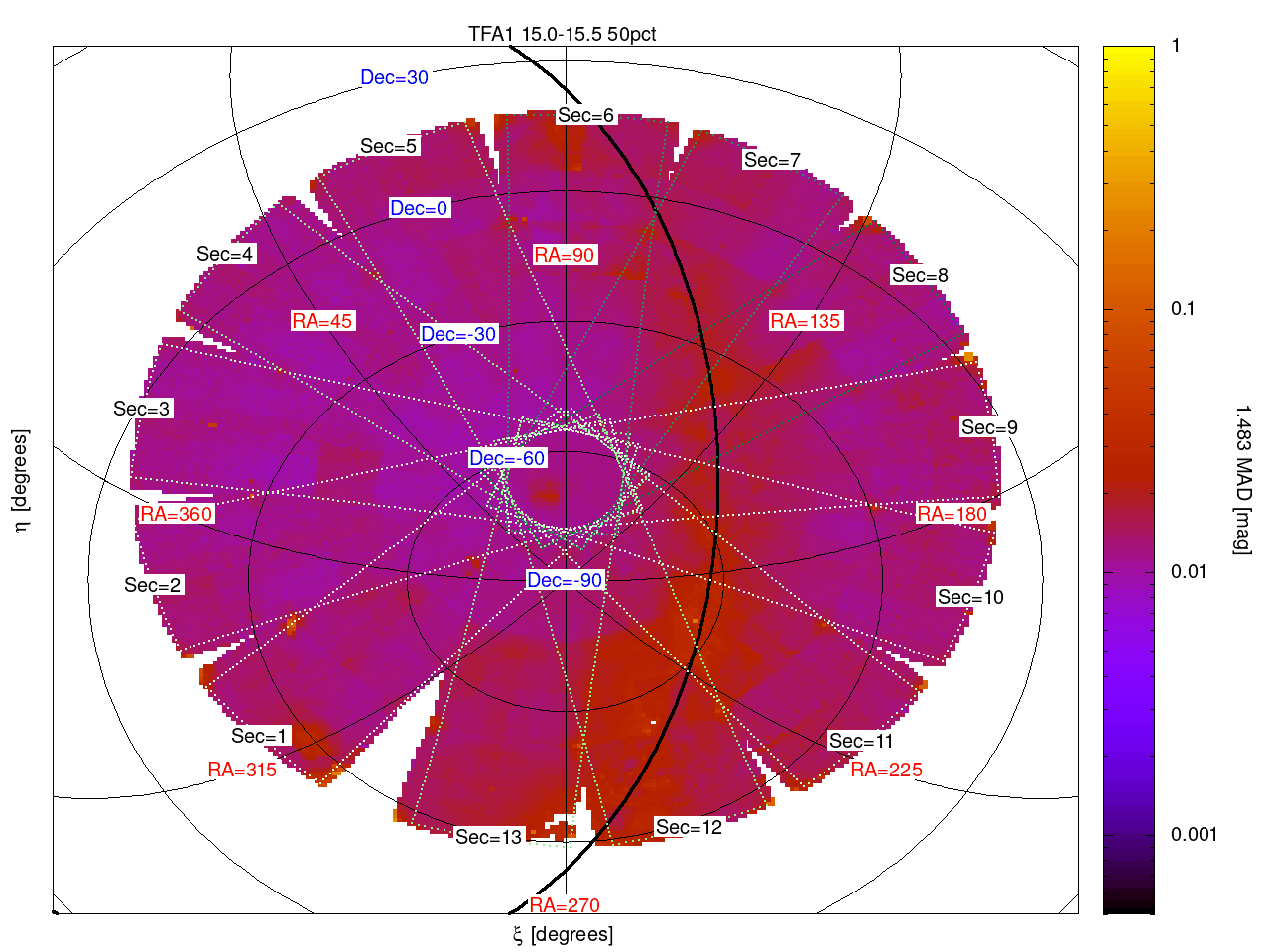}
}
{
\centering
\leavevmode
\includegraphics[width={0.5\linewidth}]{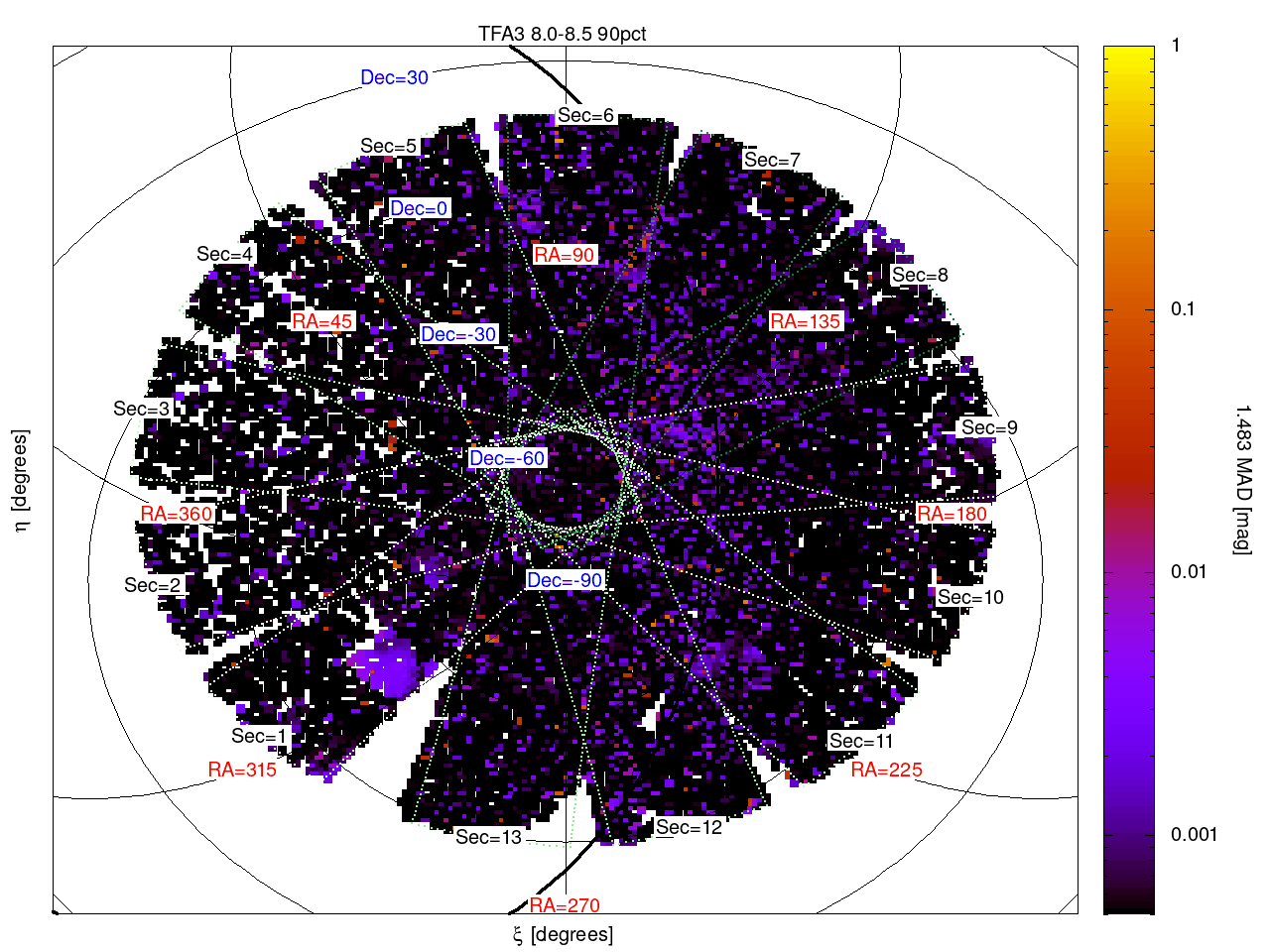}
\hfil
\includegraphics[width={0.5\linewidth}]{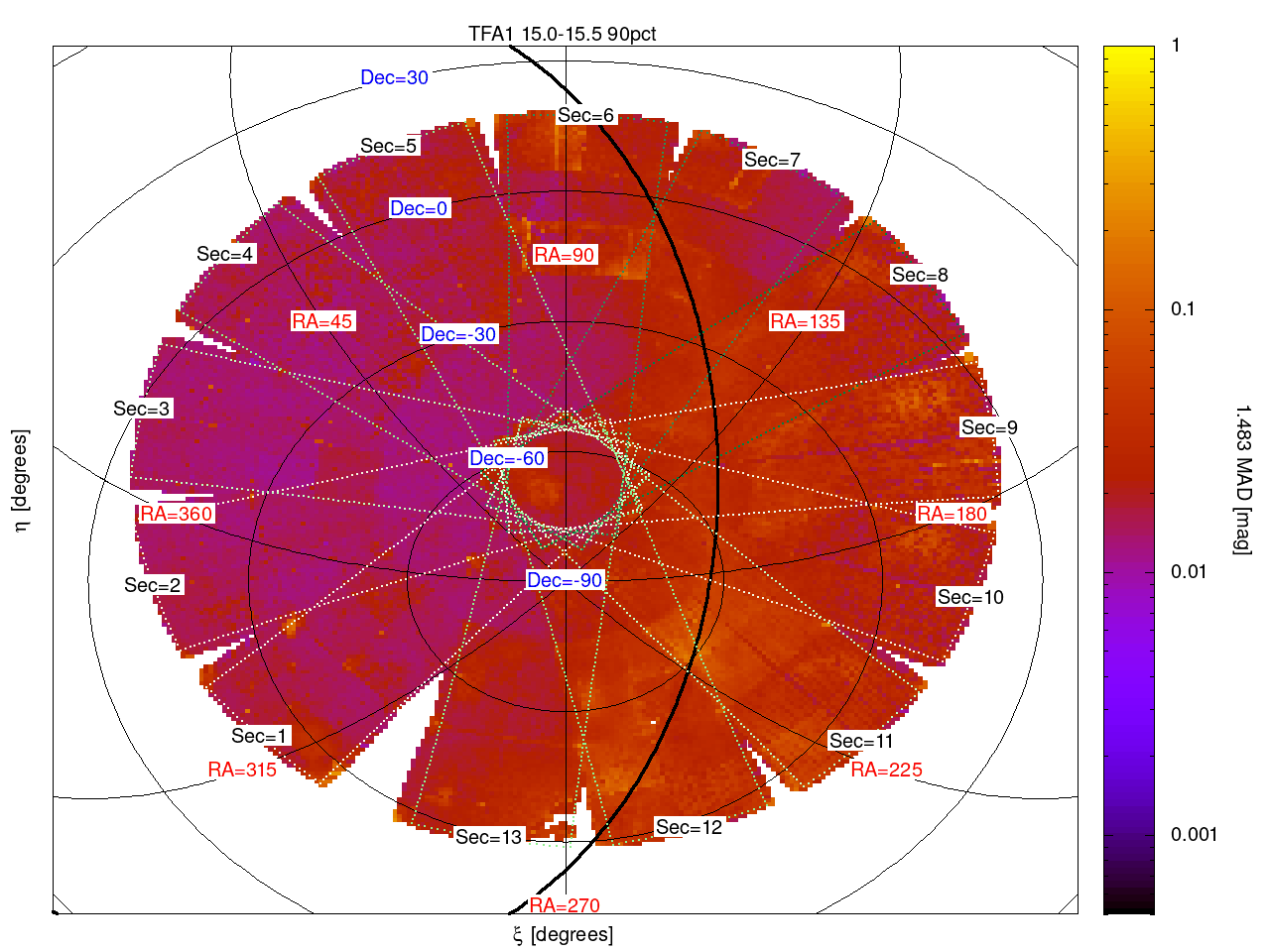}
}
\caption{Same as Figure~\ref{fig:madvspositionirm}, here we show the statistics for TFA light curves.
\label{fig:madvspositiontfa}}
\end{figure*}

Figures~\ref{fig:madvspositionepd} and~\ref{fig:madvspositiontfa} show
similar plots for the SEPD, and TFA-detrended light curves,
respectively. SEPD and TFA progressively reduce the MAD for the bright
stars, with most Sector/Camera/CCD combinations showing median MAD
values below 0.001\,mag after SEPD, and all but one combination
(Sector~1, Camera~2, CCD~4) having a median MAD below this level after
TFA. For the fainter stars the difference between the un-detrended,
SEPD and TFA precision is less pronounced, as expected for white-noise
dominated light curves.

\begin{figure*}[!ht]
{
\centering
\leavevmode
\includegraphics[width={0.5\linewidth}]{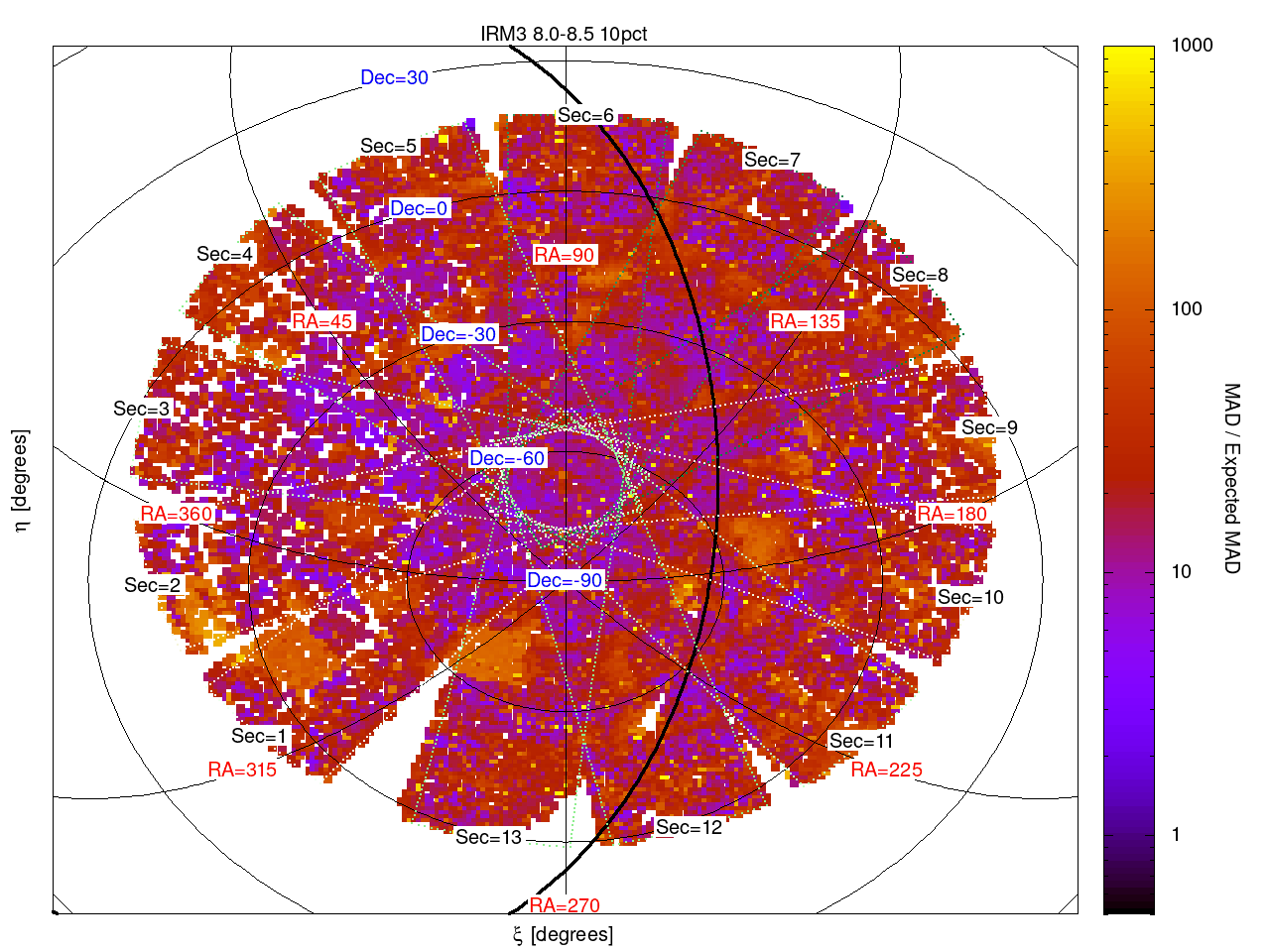}
\hfil
\includegraphics[width={0.5\linewidth}]{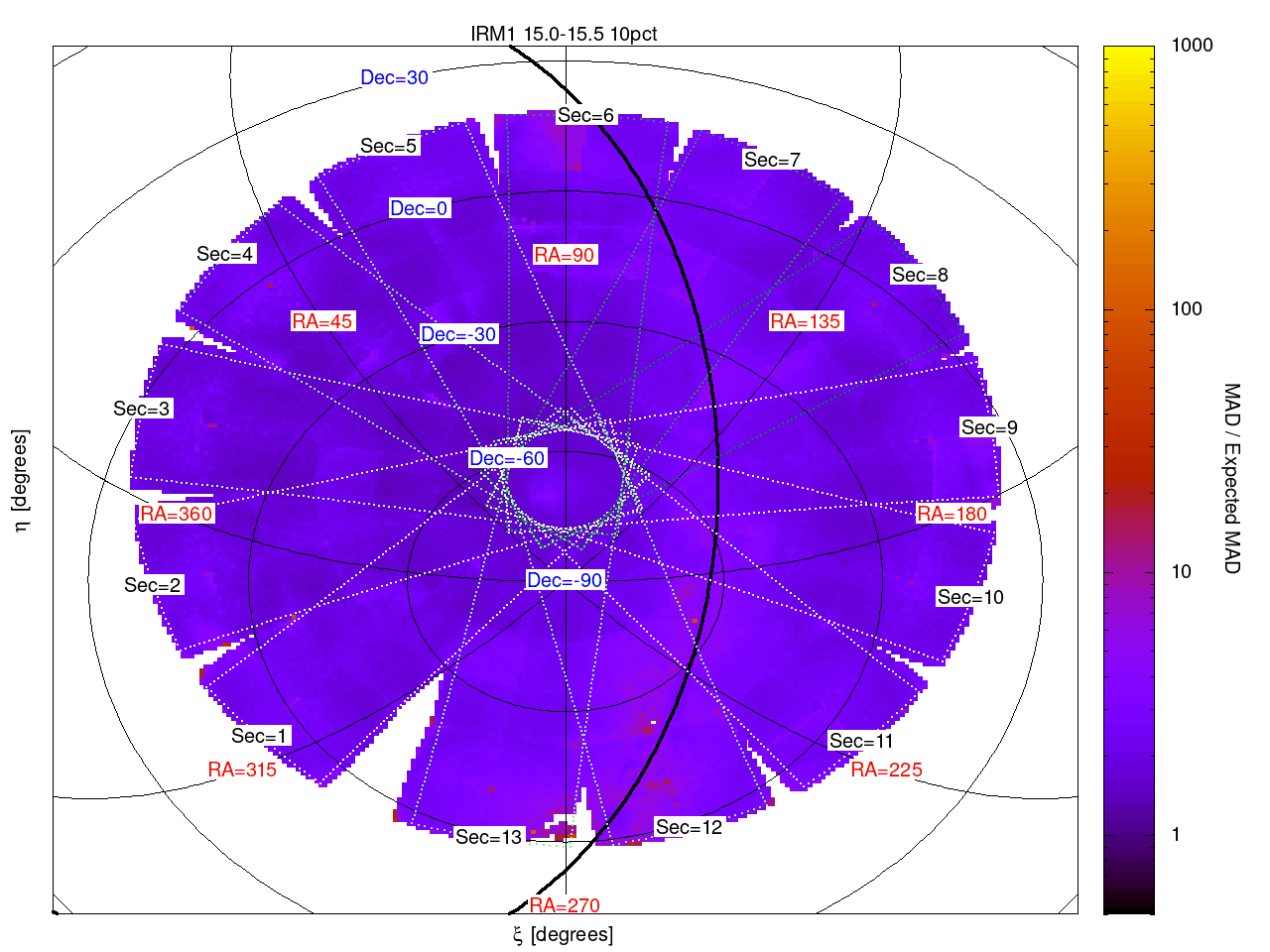}
}
{
\centering
\leavevmode
\includegraphics[width={0.5\linewidth}]{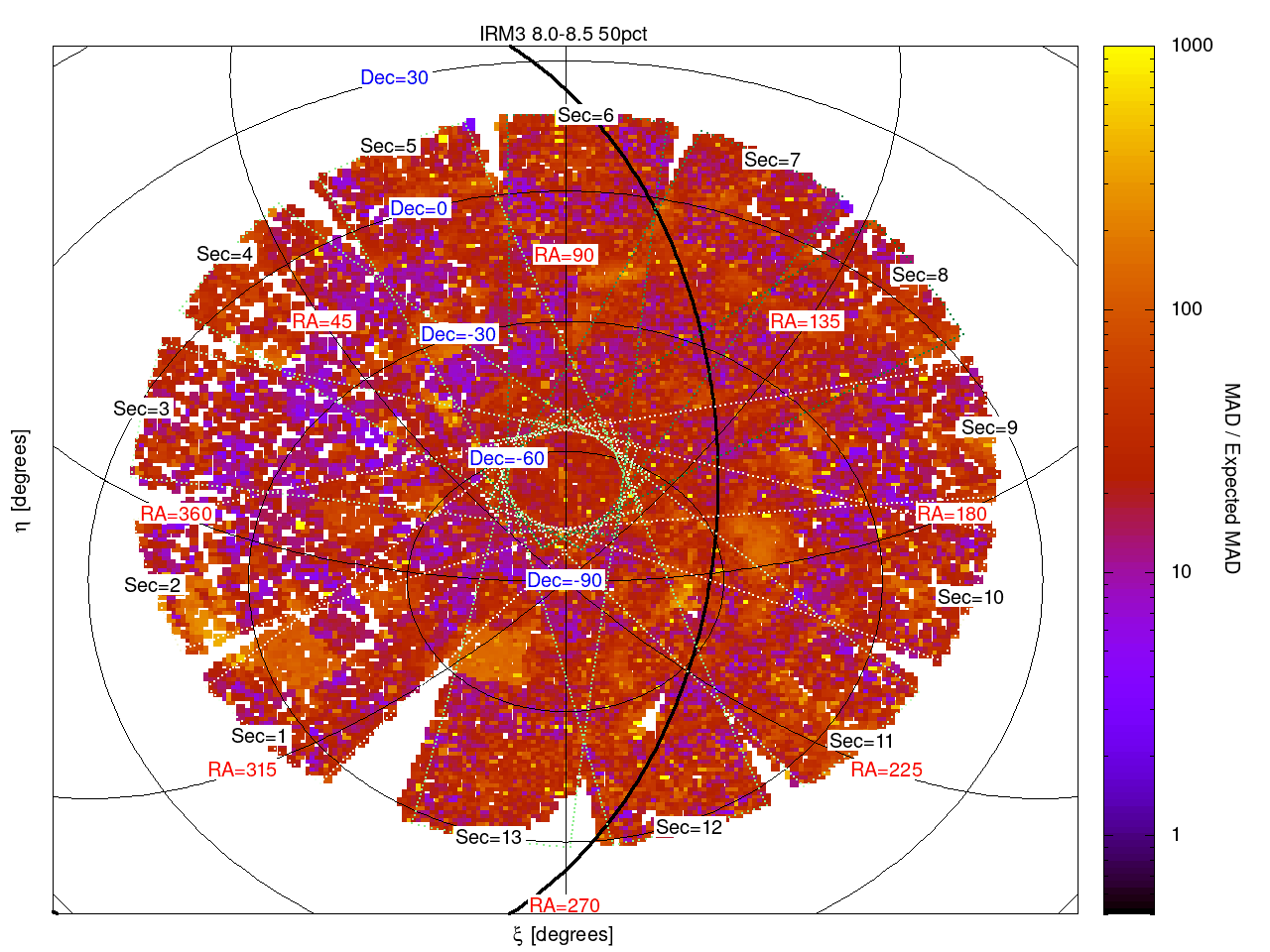}
\hfil
\includegraphics[width={0.5\linewidth}]{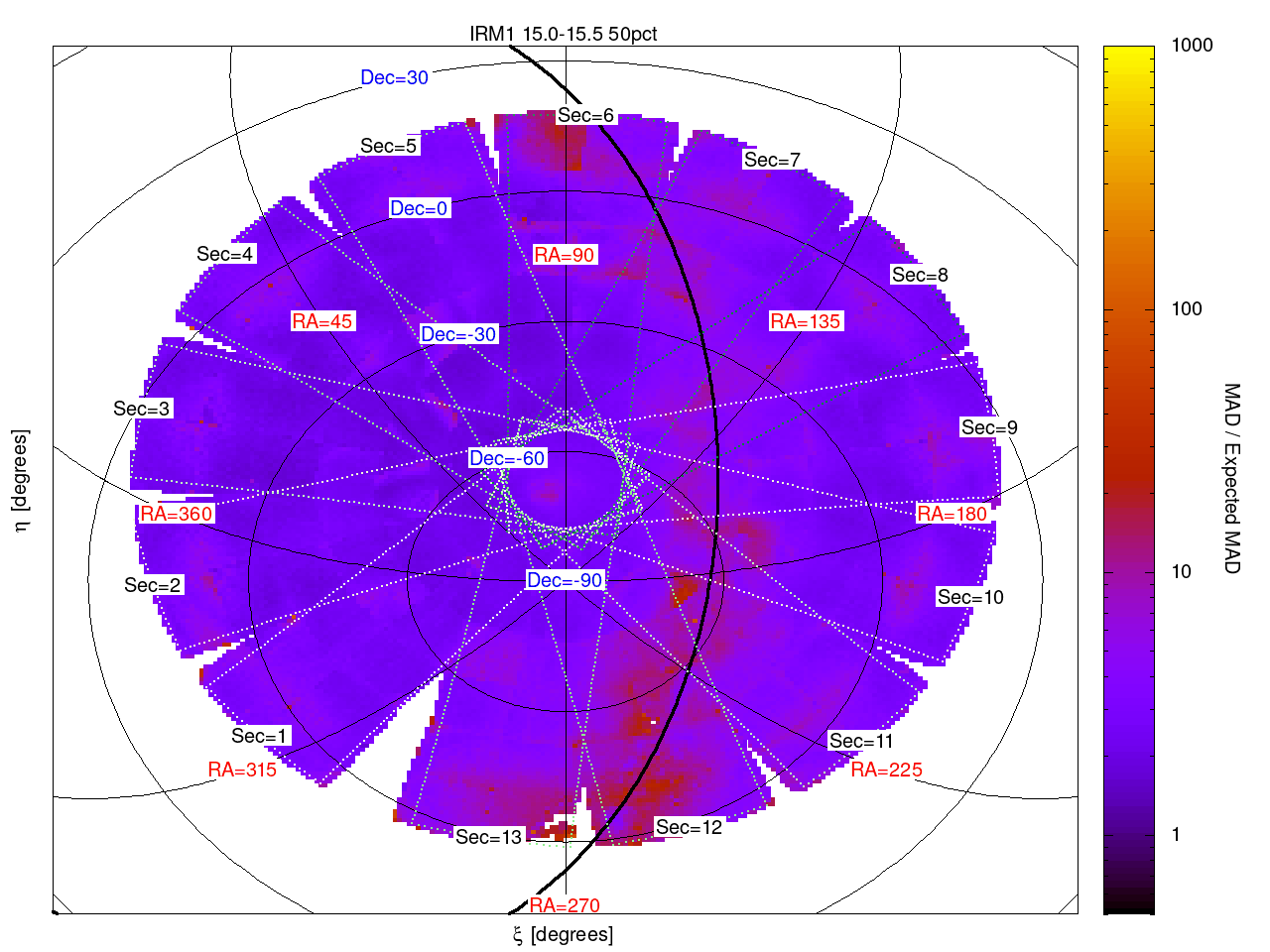}
}
{
\centering
\leavevmode
\includegraphics[width={0.5\linewidth}]{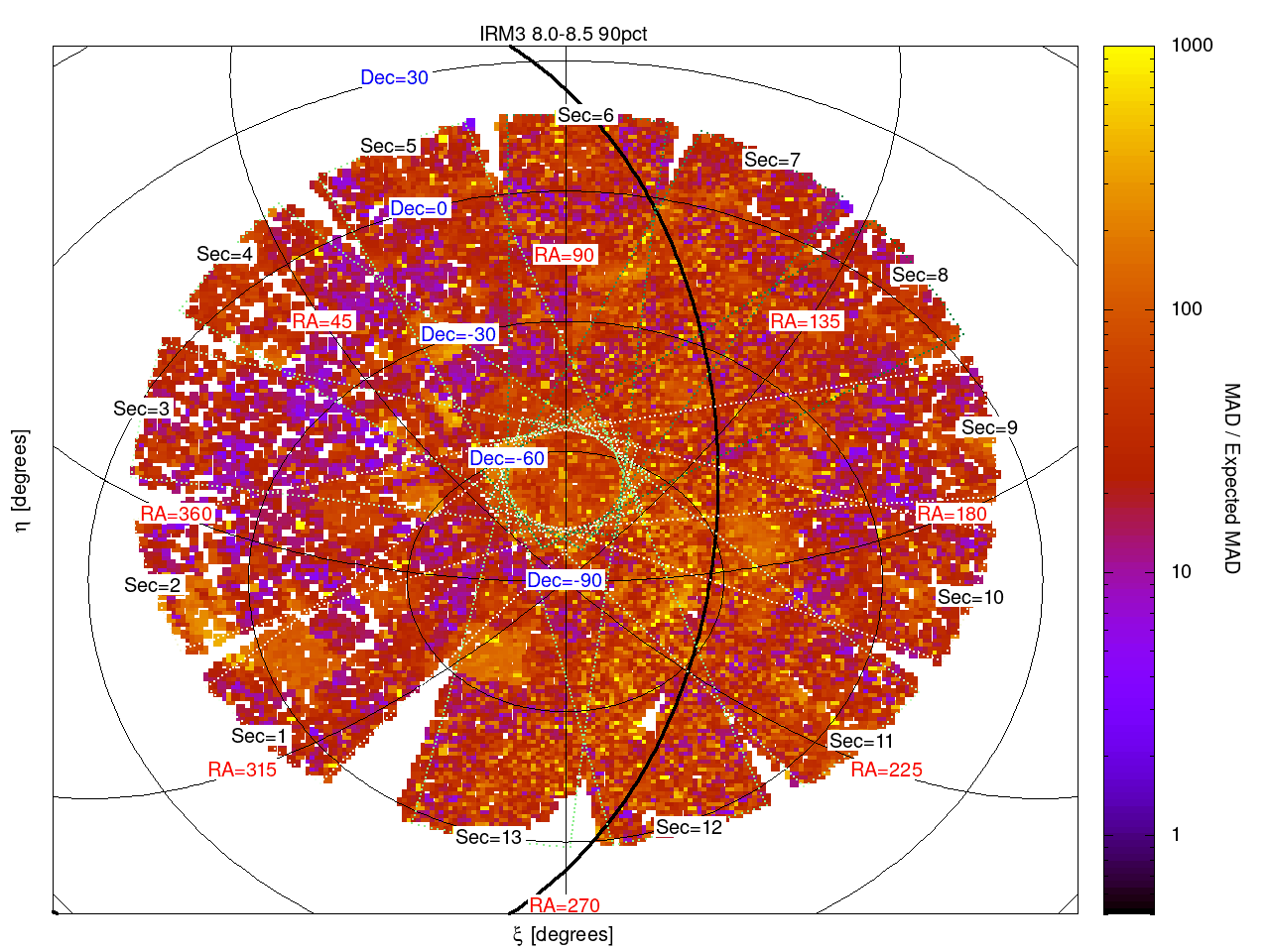}
\hfil
\includegraphics[width={0.5\linewidth}]{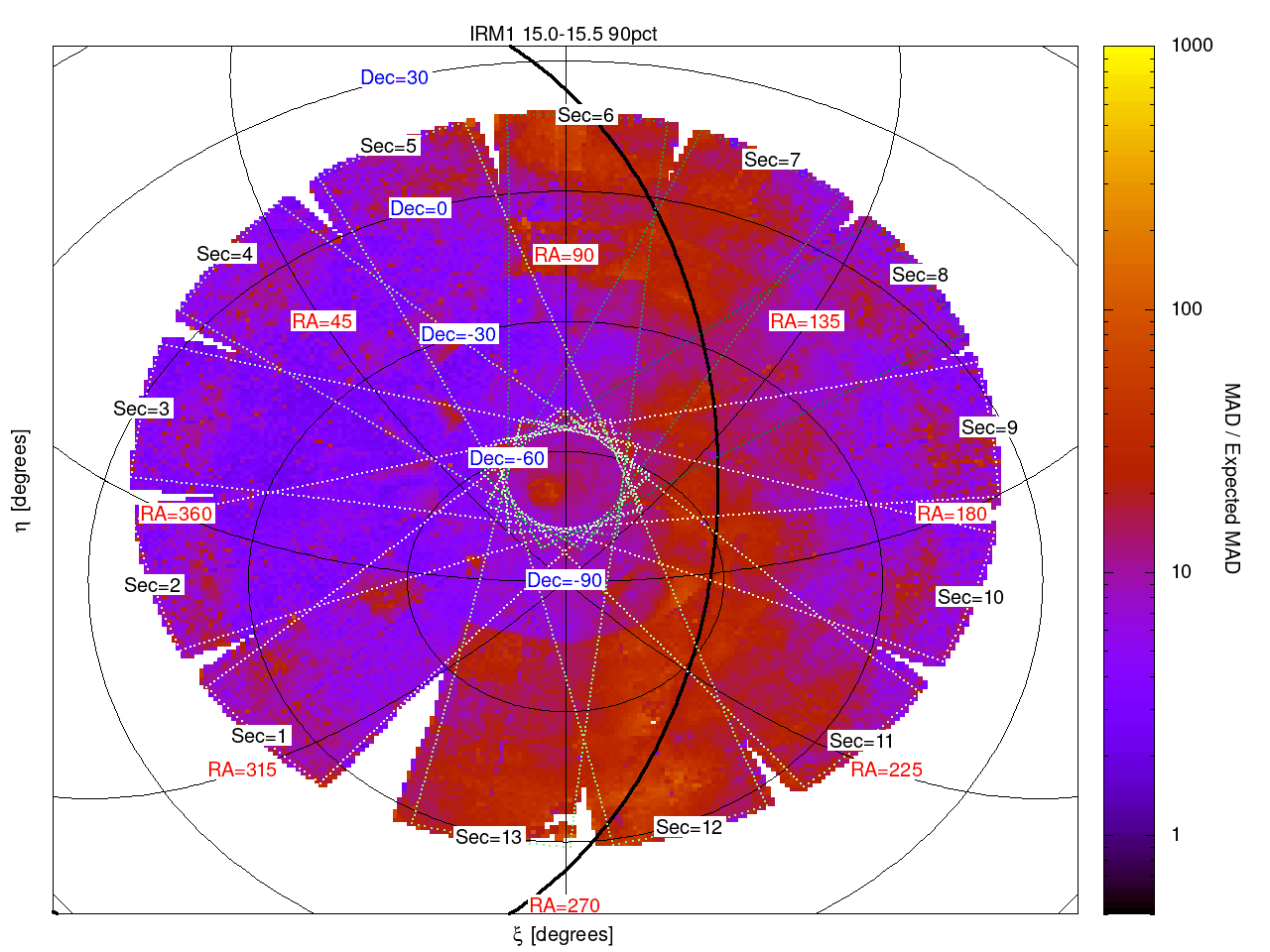}
}
\caption{Same as Figure~\ref{fig:madvspositionirm}, here we show the ratio of the MAD to the expected MAD.
\label{fig:madratiovspositionirm}}
\end{figure*}

\begin{figure*}[!ht]
{
\centering
\leavevmode
\includegraphics[width={0.5\linewidth}]{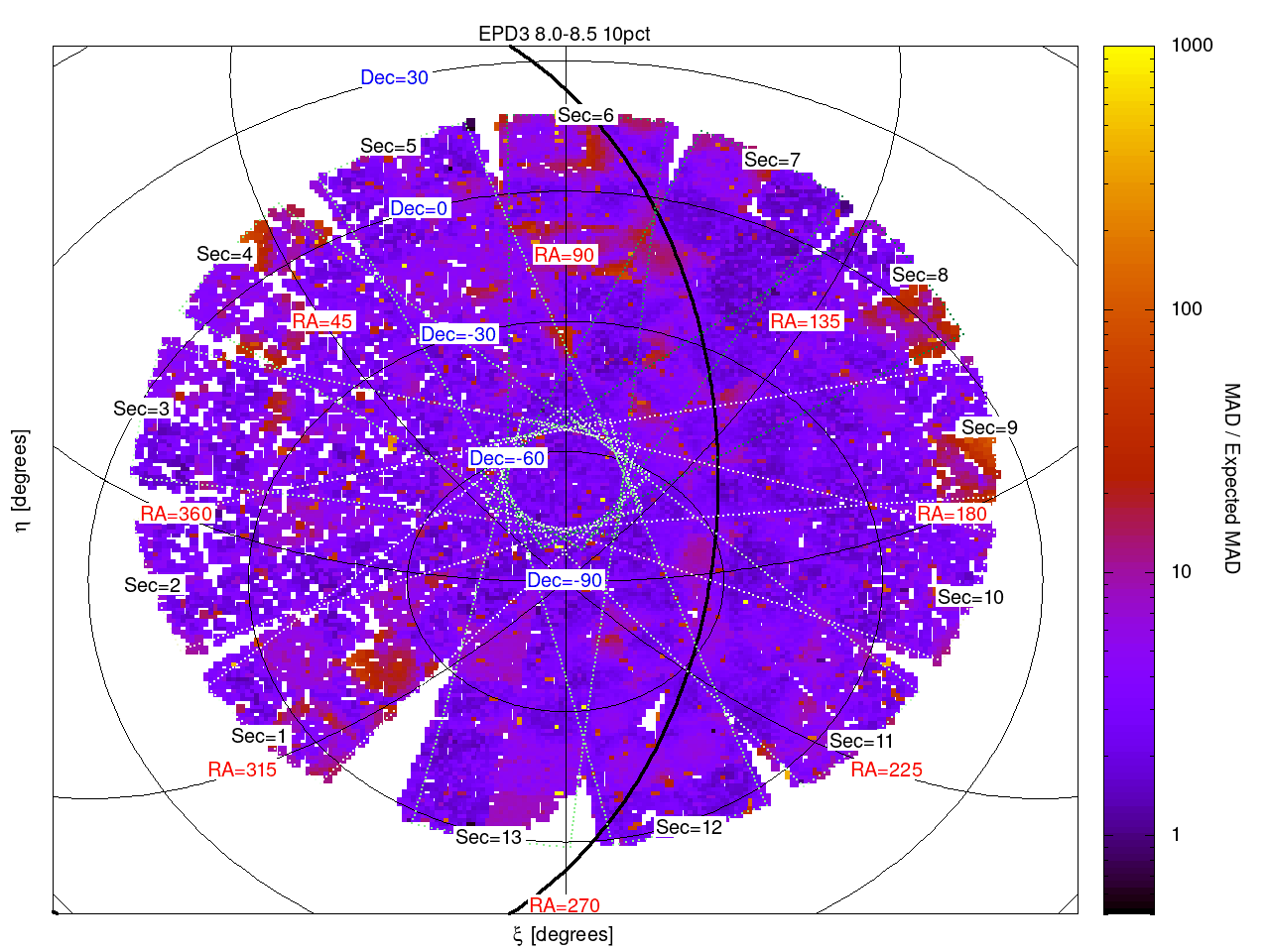}
\hfil
\includegraphics[width={0.5\linewidth}]{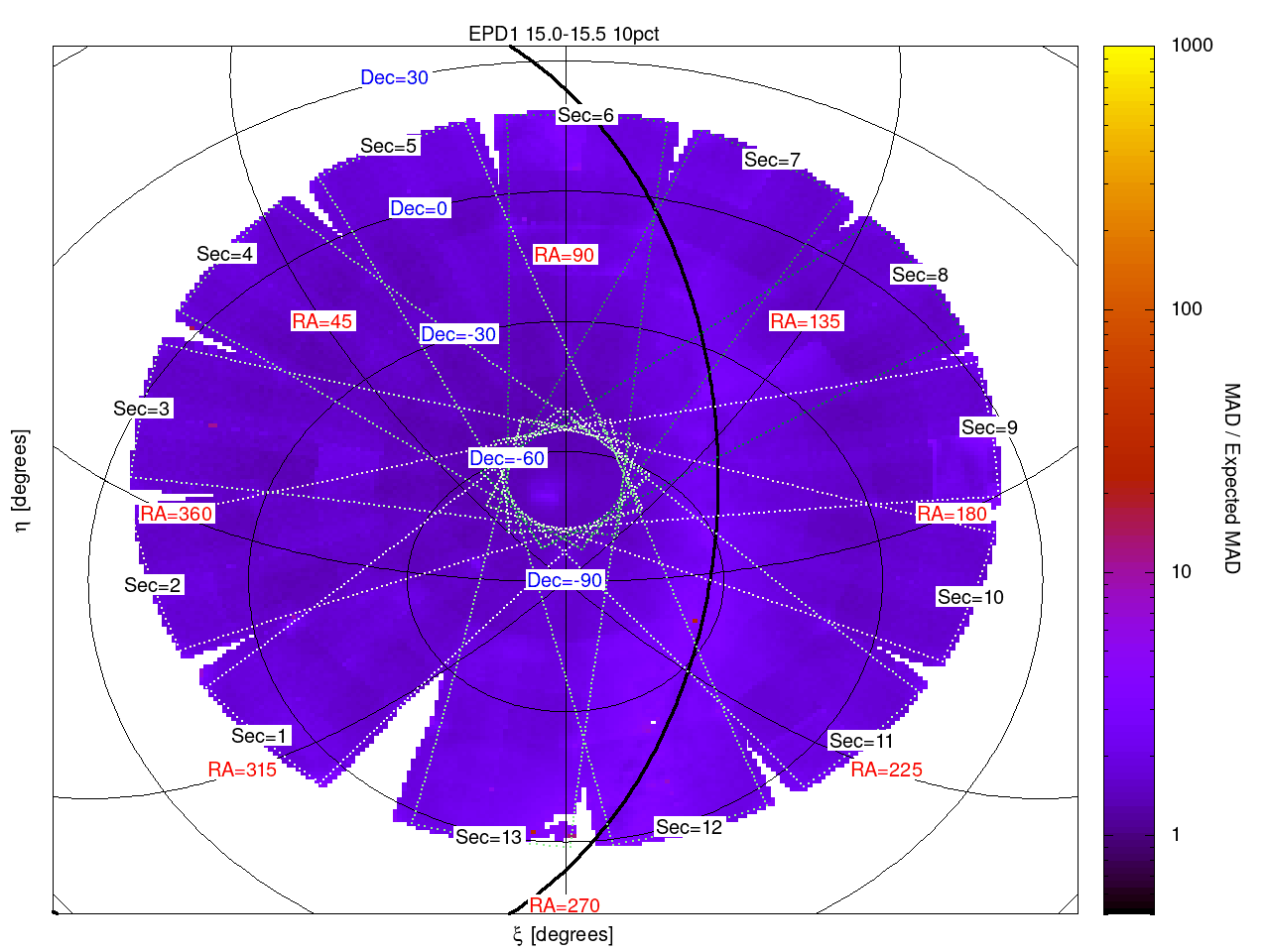}
}
{
\centering
\leavevmode
\includegraphics[width={0.5\linewidth}]{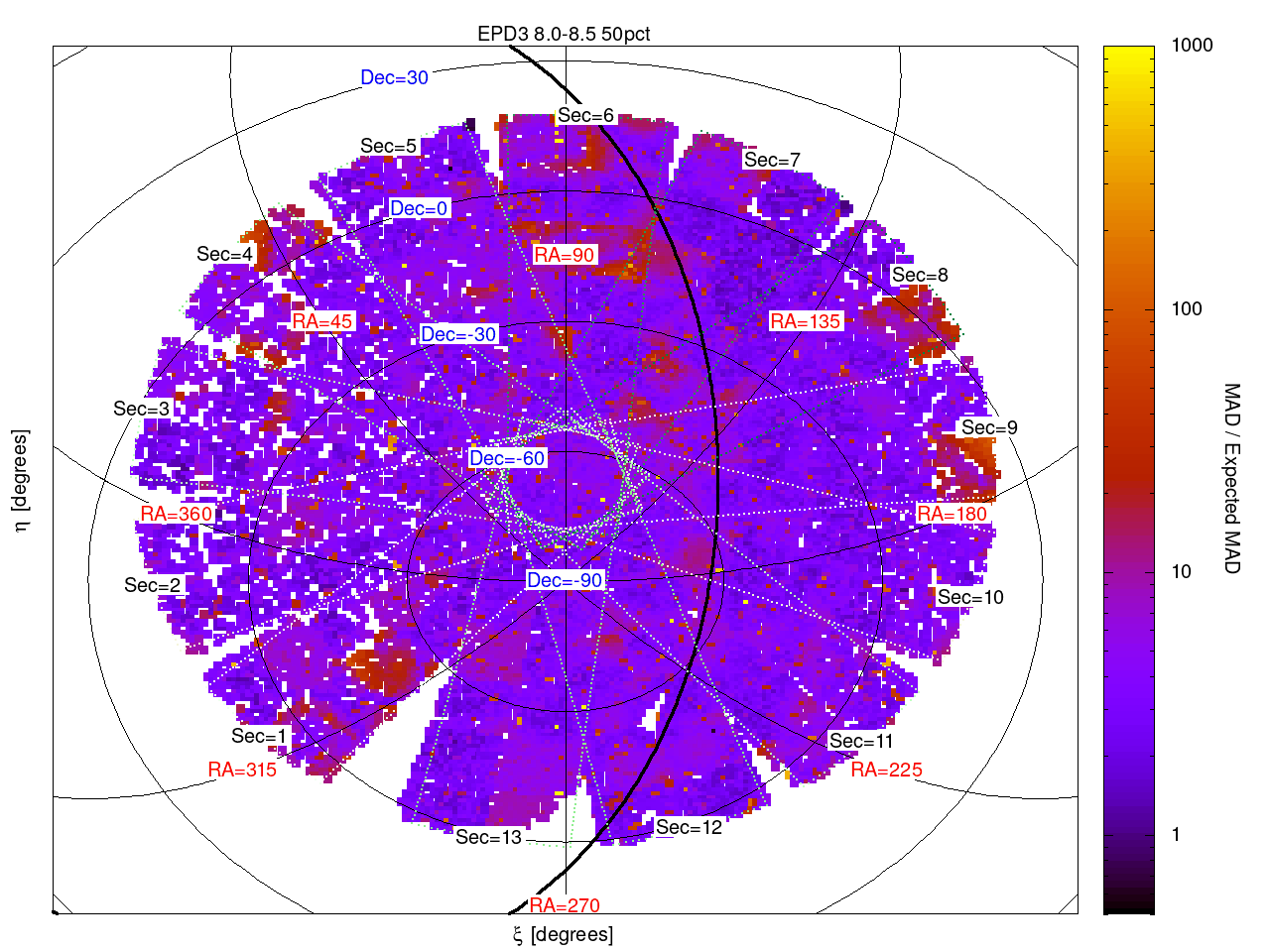}
\hfil
\includegraphics[width={0.5\linewidth}]{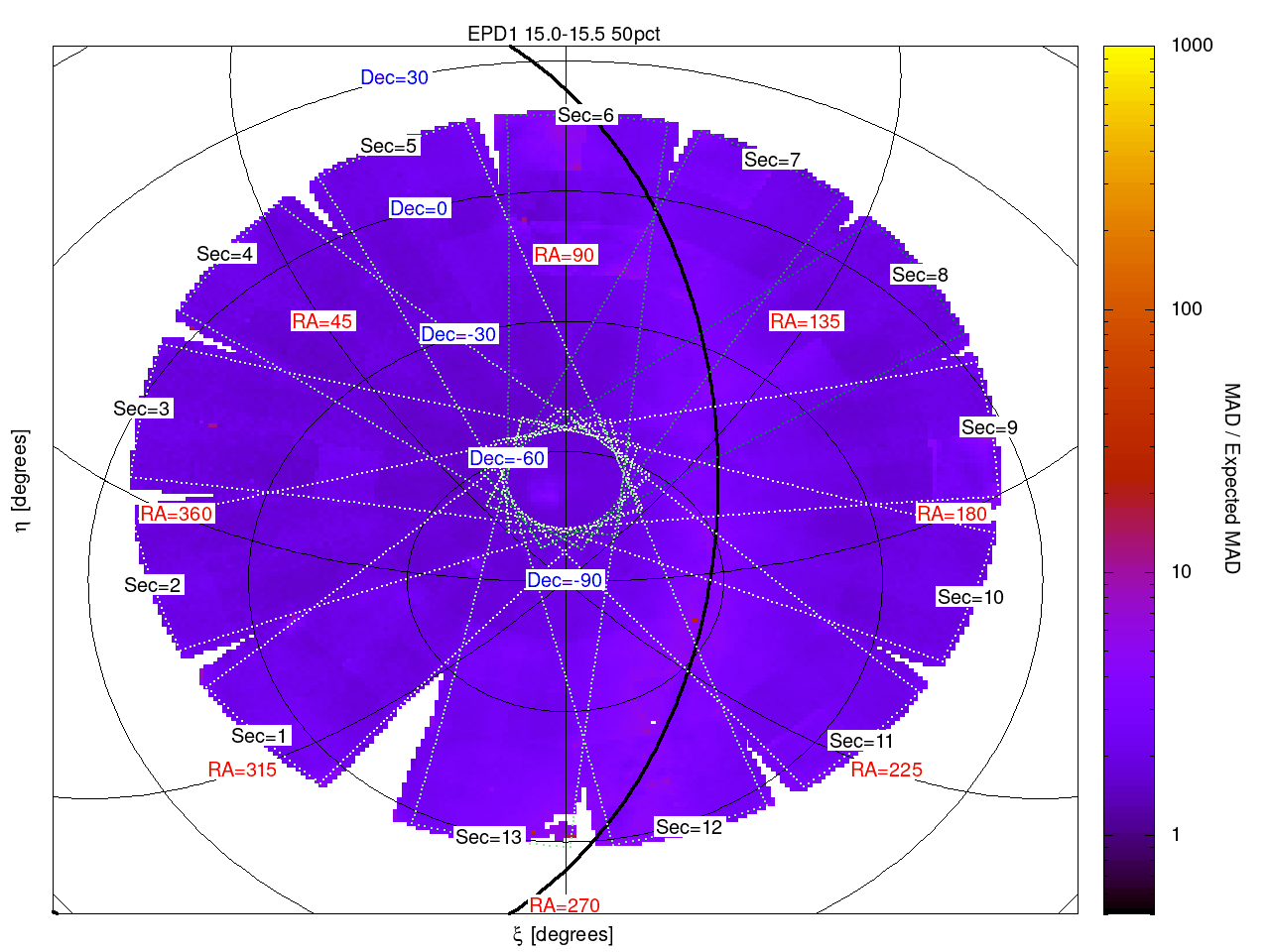}
}
{
\centering
\leavevmode
\includegraphics[width={0.5\linewidth}]{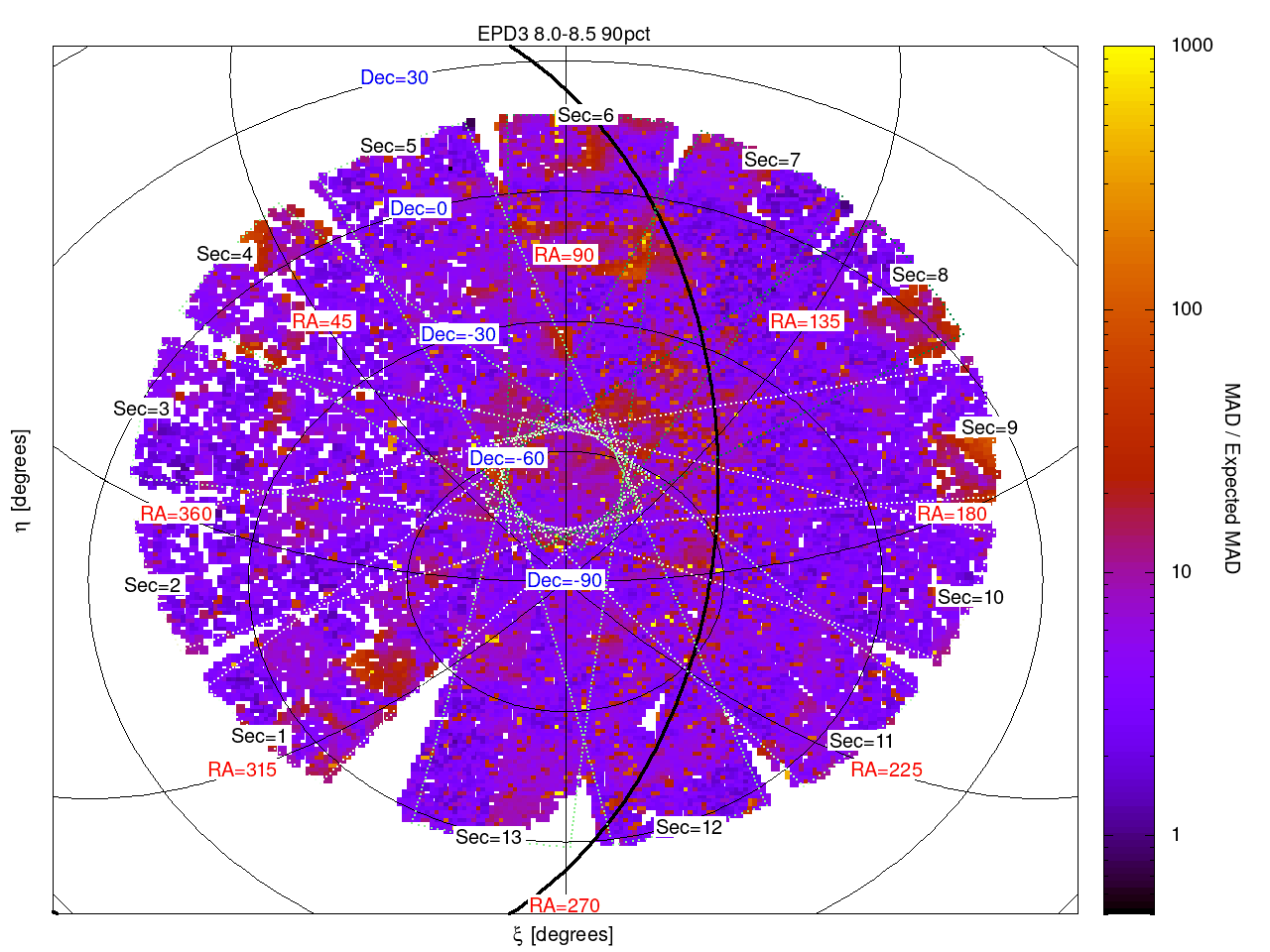}
\hfil
\includegraphics[width={0.5\linewidth}]{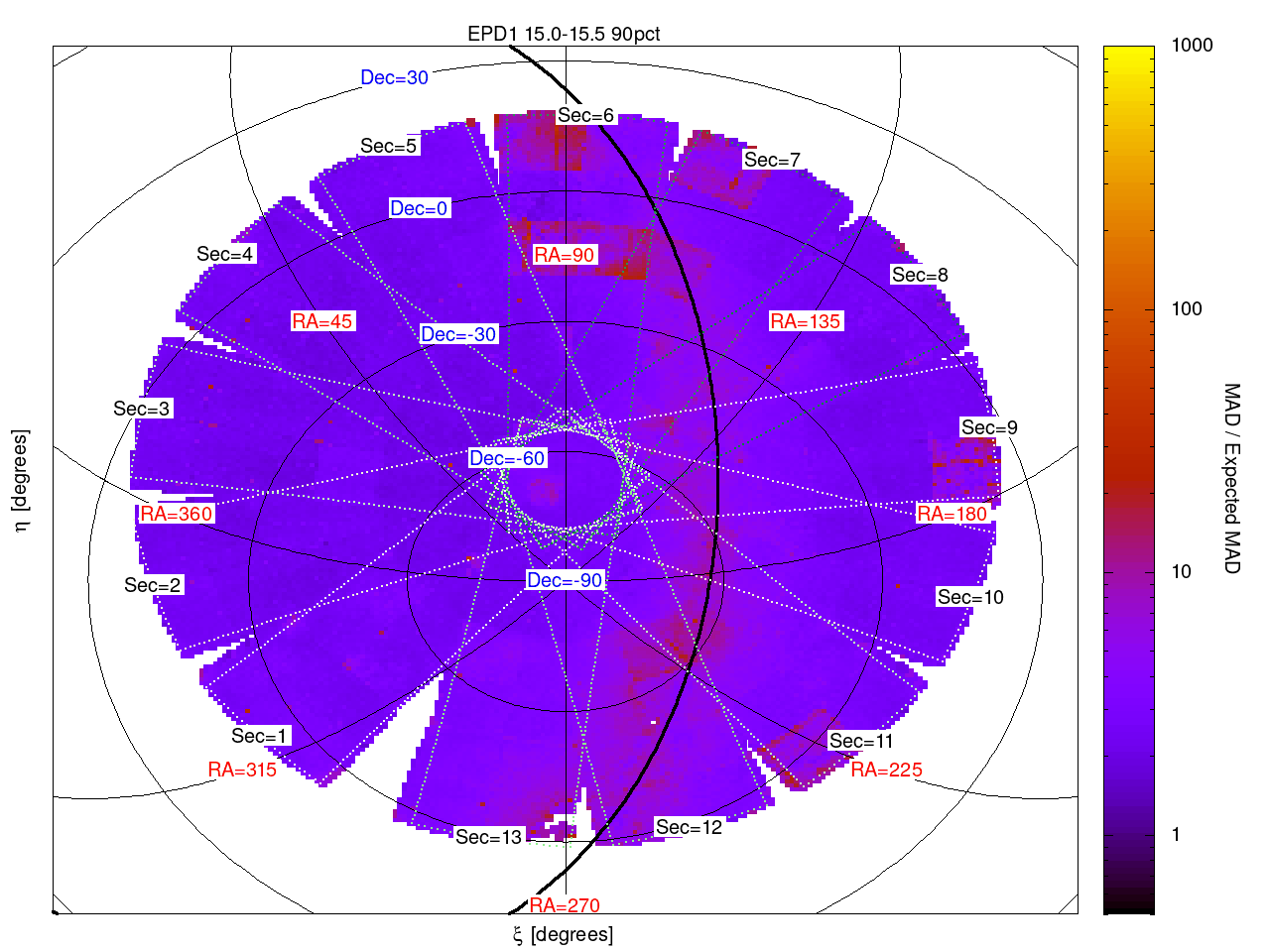}
}
\caption{Same as Figure~\ref{fig:madvspositionepd}, here we show the ratio of the MAD to the expected MAD.
\label{fig:madratiovspositionepd}}
\end{figure*}

\begin{figure*}[!ht]
{
\centering
\leavevmode
\includegraphics[width={0.5\linewidth}]{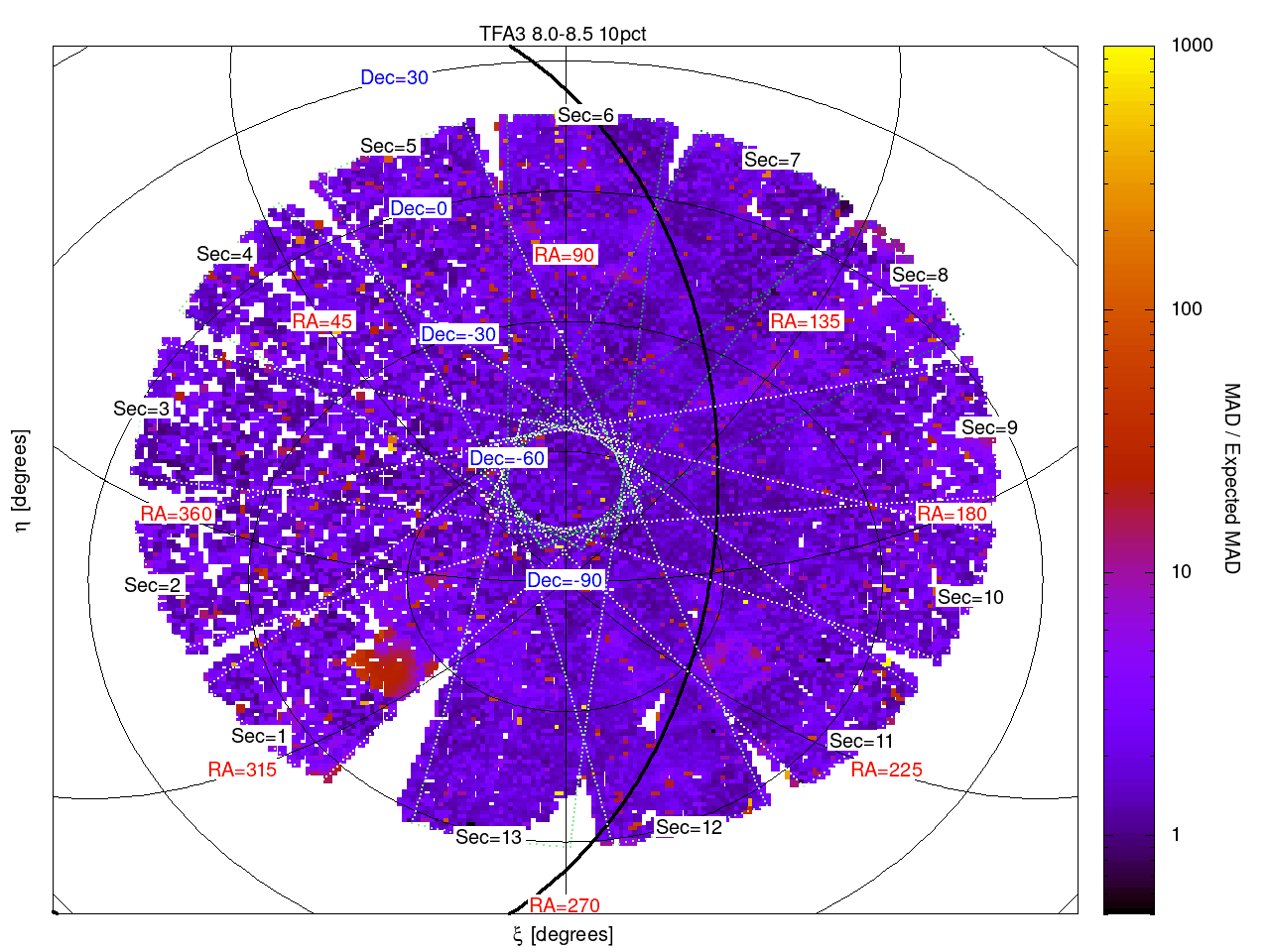}
\hfil
\includegraphics[width={0.5\linewidth}]{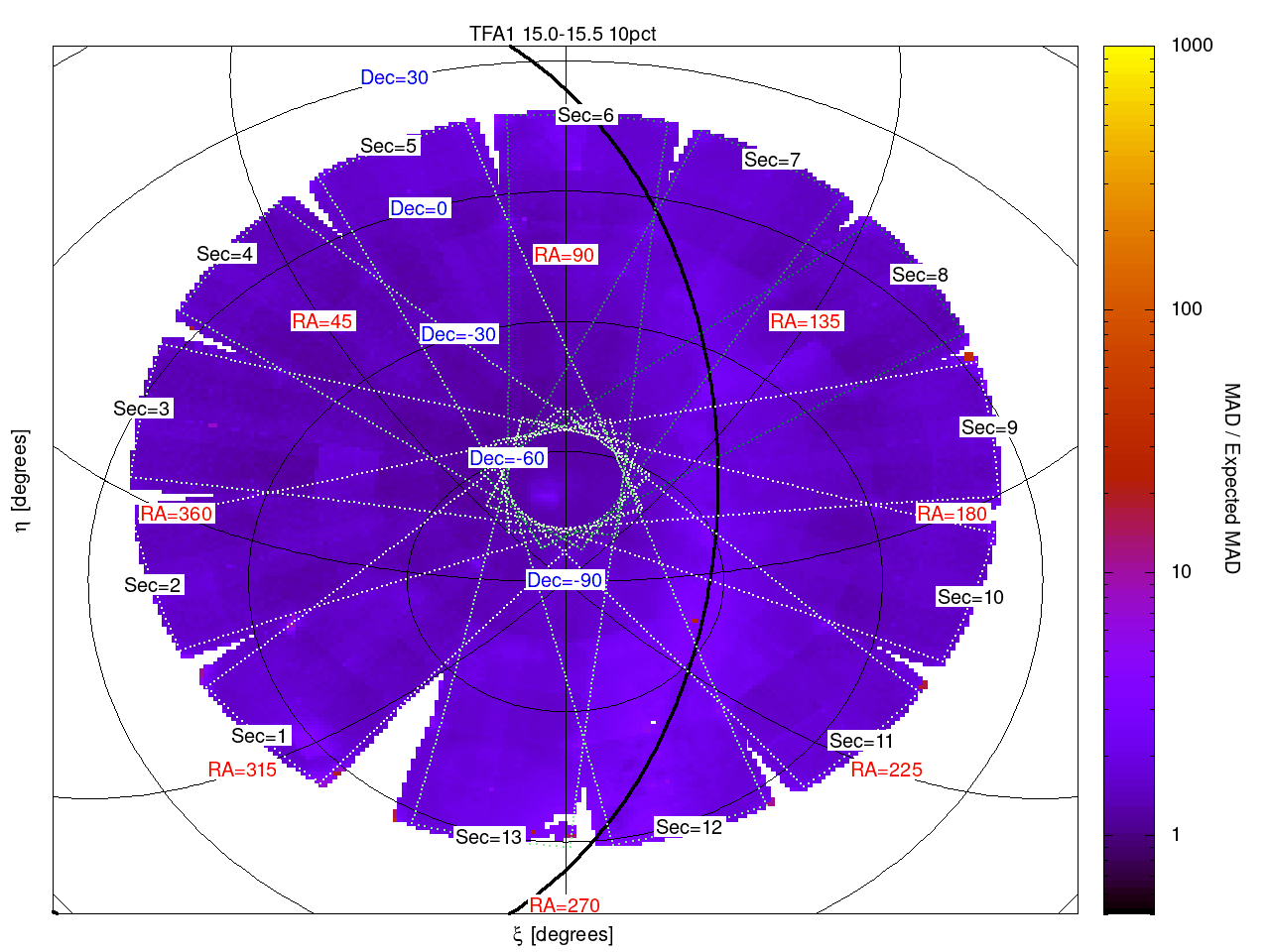}
}
{
\centering
\leavevmode
\includegraphics[width={0.5\linewidth}]{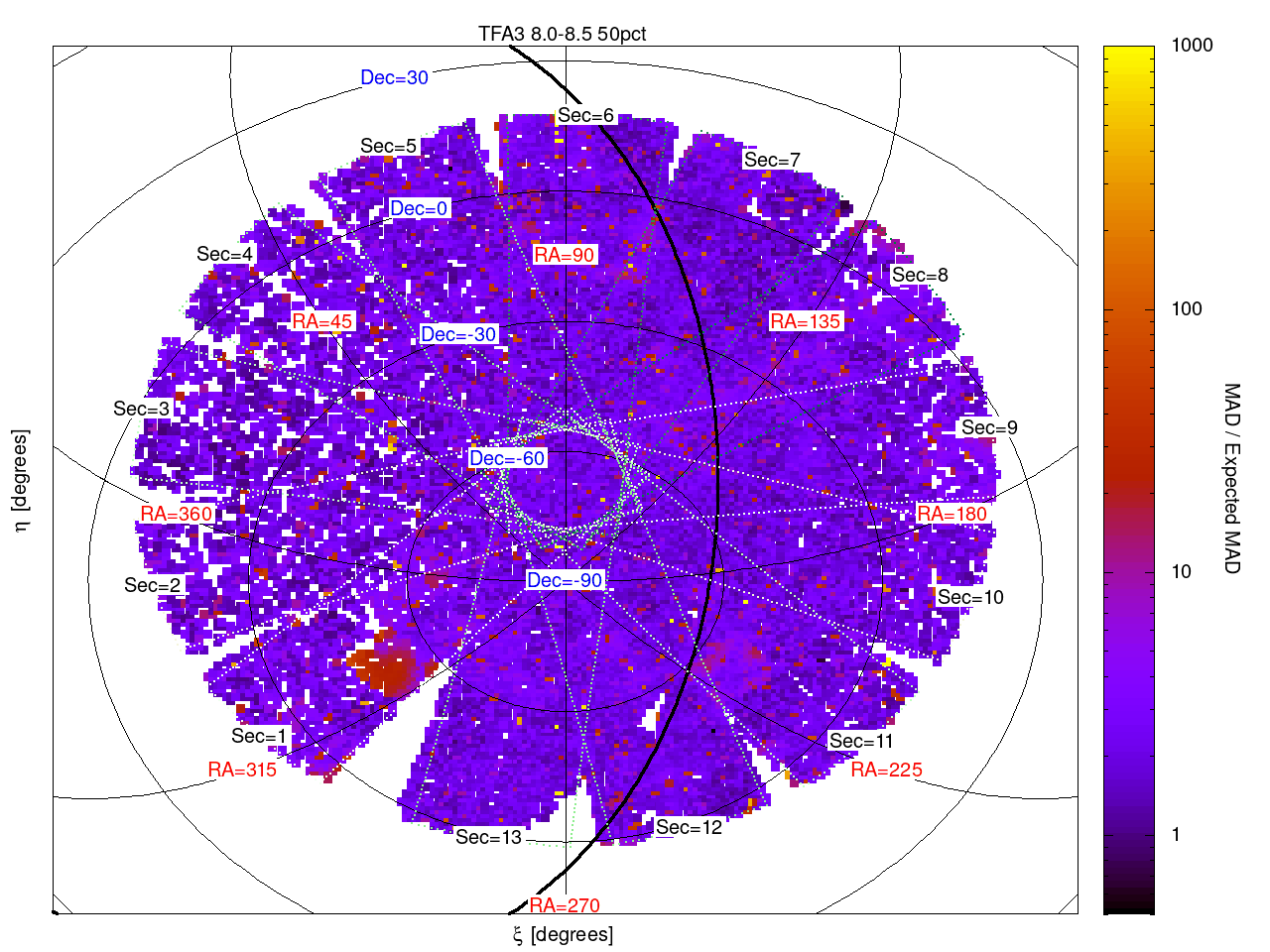}
\hfil
\includegraphics[width={0.5\linewidth}]{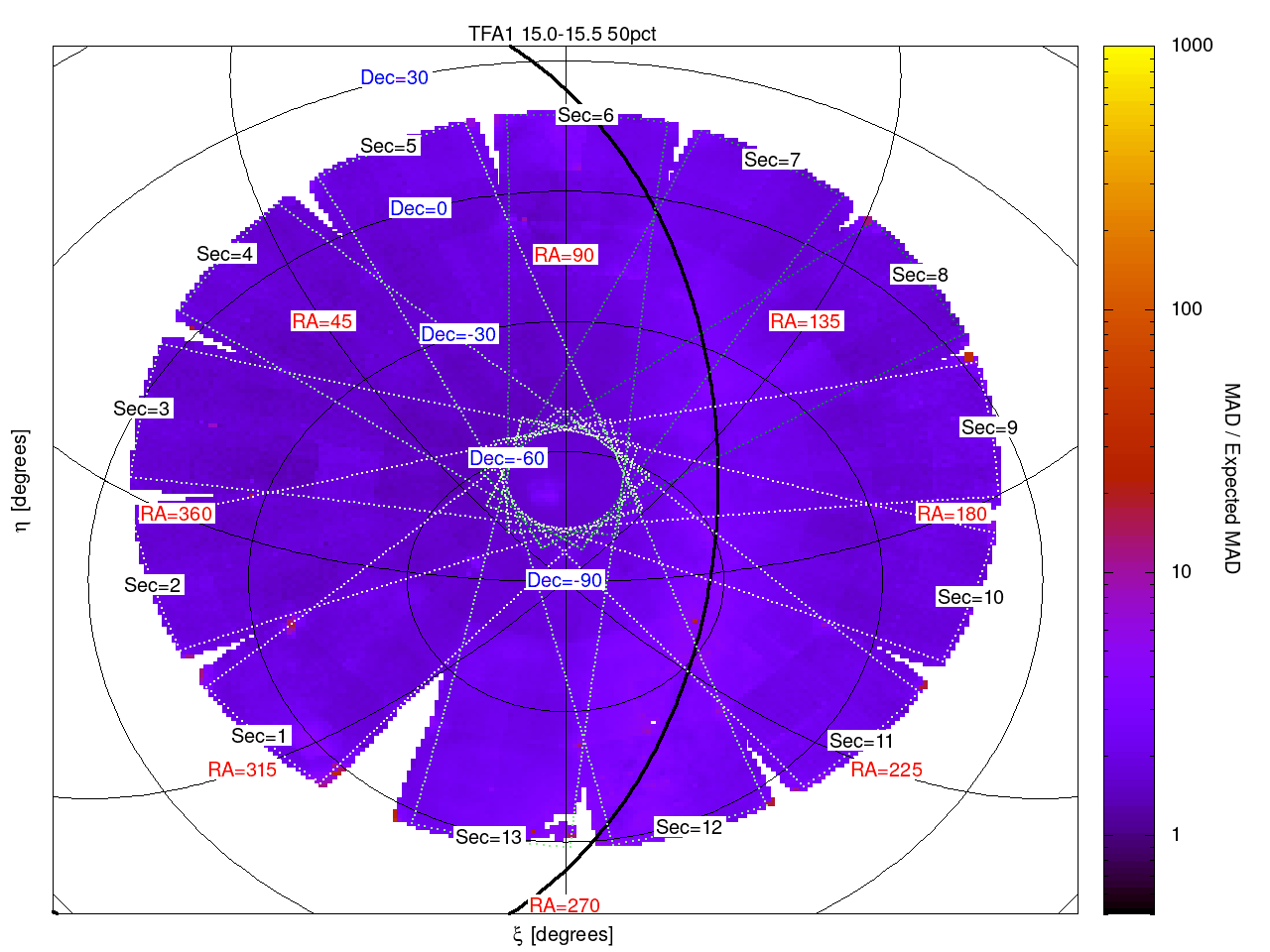}
}
{
\centering
\leavevmode
\includegraphics[width={0.5\linewidth}]{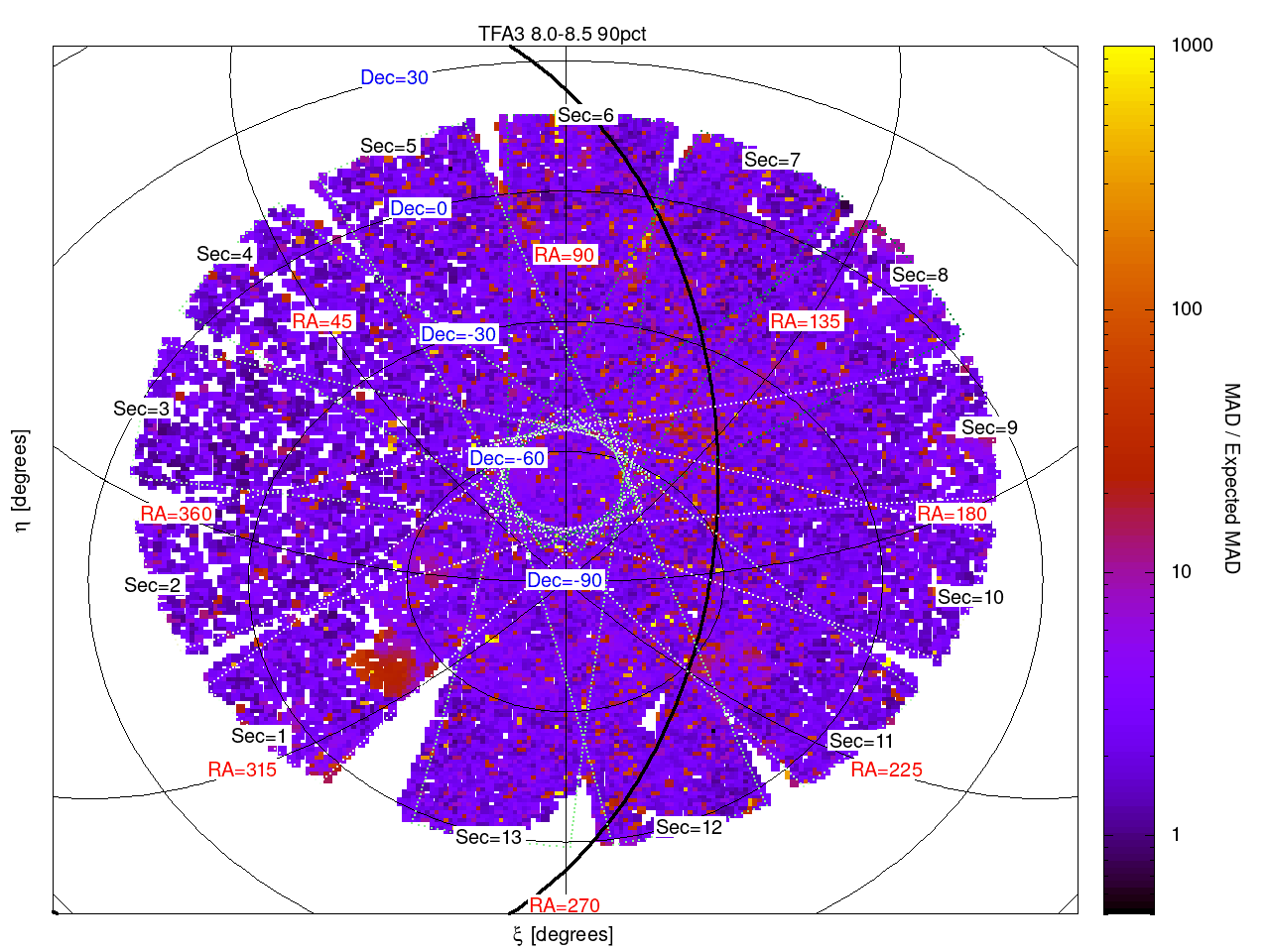}
\hfil
\includegraphics[width={0.5\linewidth}]{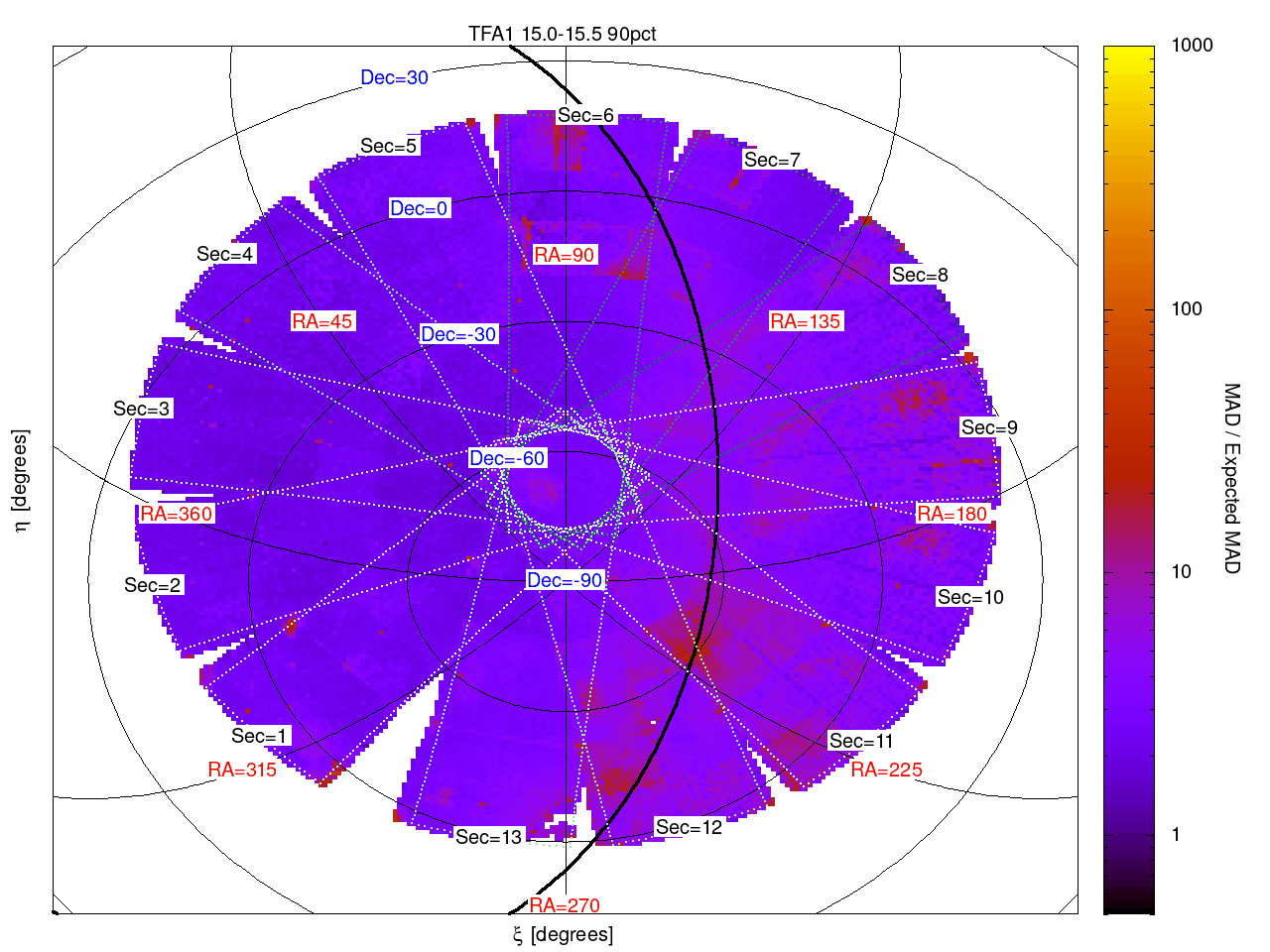}
}
\caption{Same as Figure~\ref{fig:madvspositiontfa}, here we show the ratio of the MAD to the expected MAD.
\label{fig:madratiovspositiontfa}}
\end{figure*}

Figures~\ref{fig:madratiovspositionirm},
\ref{fig:madratiovspositionepd}, and~\ref{fig:madratiovspositiontfa}
similarly show the ratio of the MAD to the expected MAD on the sky at
the 10th, median and 90th percentile levels for the raw, SEPD, and
TFA-detrended light curves, respectively.  For the faintest stars away
from the Galactic plane, the measured MAD is close to the expected MAD
before detrending, while for bright stars the pre-detrending MAD is
much higher than ($\gtrsim 100$ times) the expected MAD, but approaches the
expected MAD after TFA-detrending.

\subsection{Comparison to Other Reductions}
\label{sec:comparison}

We compare our light curves to the available FFI light curves produced
by four of the other projects listed in
Table~\ref{tab:ffiprojects}. Here we restrict the comparison to
projects that aim for a broad selection of stars, and do not compare
to the PATHOS or DIAMANTE projects that have released light curves
only for star cluster members and transiting planet hosts. We also do
not compare to the CDIPS project, since we are making direct use of
the CDIPS reduction pipeline. Here we do a thorough comparison for all
sectors, cameras, and CCDs to the SPOC-TESS and QLP projects. For the
TGLC and GSFC-ELEANOR-LITE projects, which have very large data
volumes, we selected 1\% of the light curves for comparison. To ensure a uniform sampling in magnitude and position on the sky, we made use of the batch light curve download scripts for these projects provided by MAST, downloading every hundredth file listed in the scripts.

\subsubsection{Precision Comparison}

\begin{figure*}[!ht]
{
\centering
\leavevmode
\includegraphics[width={0.5\linewidth}]{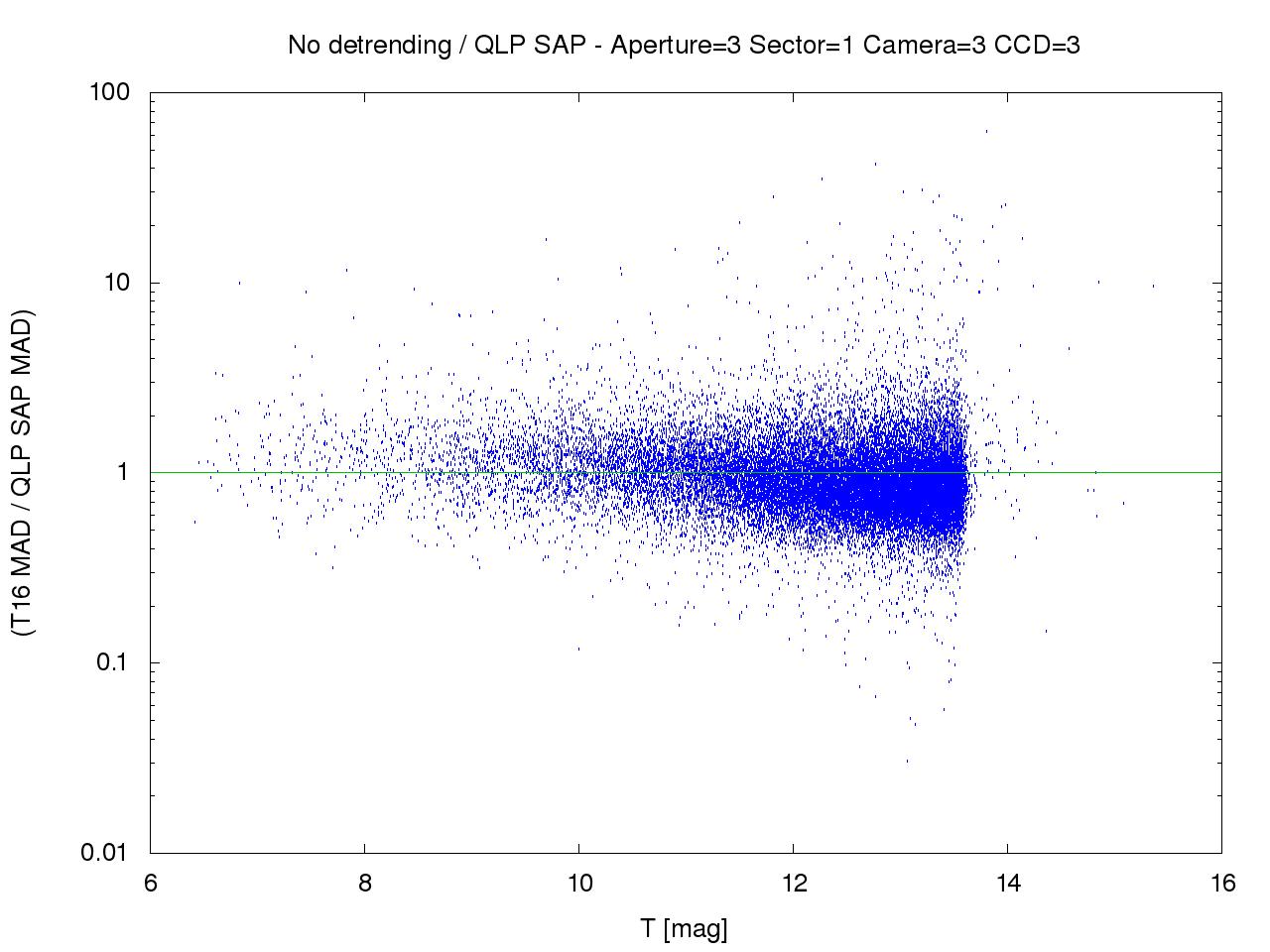}
\hfil
\includegraphics[width={0.5\linewidth}]{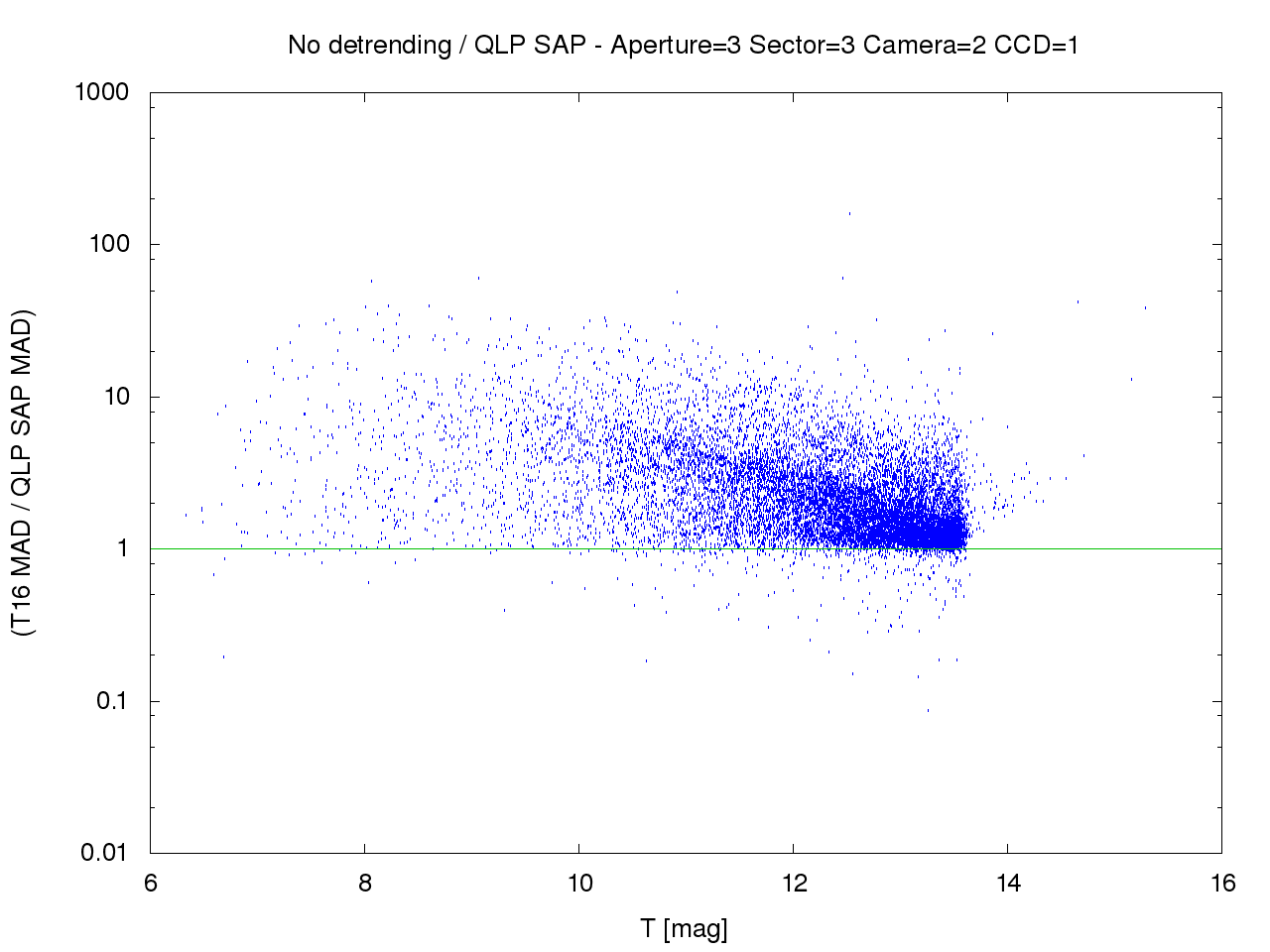}
}
{
\centering
\leavevmode
\includegraphics[width={0.5\linewidth}]{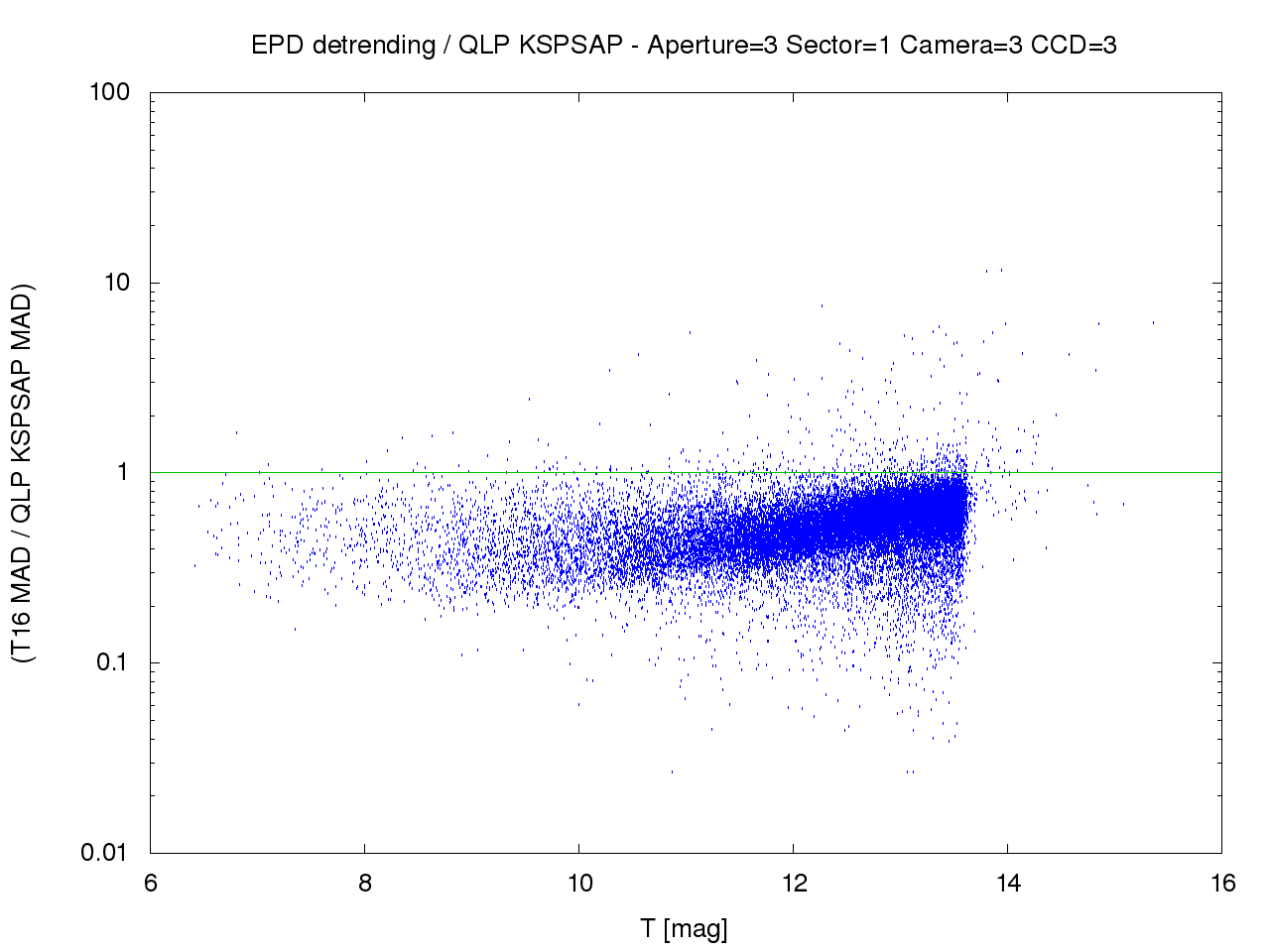}
\hfil
\includegraphics[width={0.5\linewidth}]{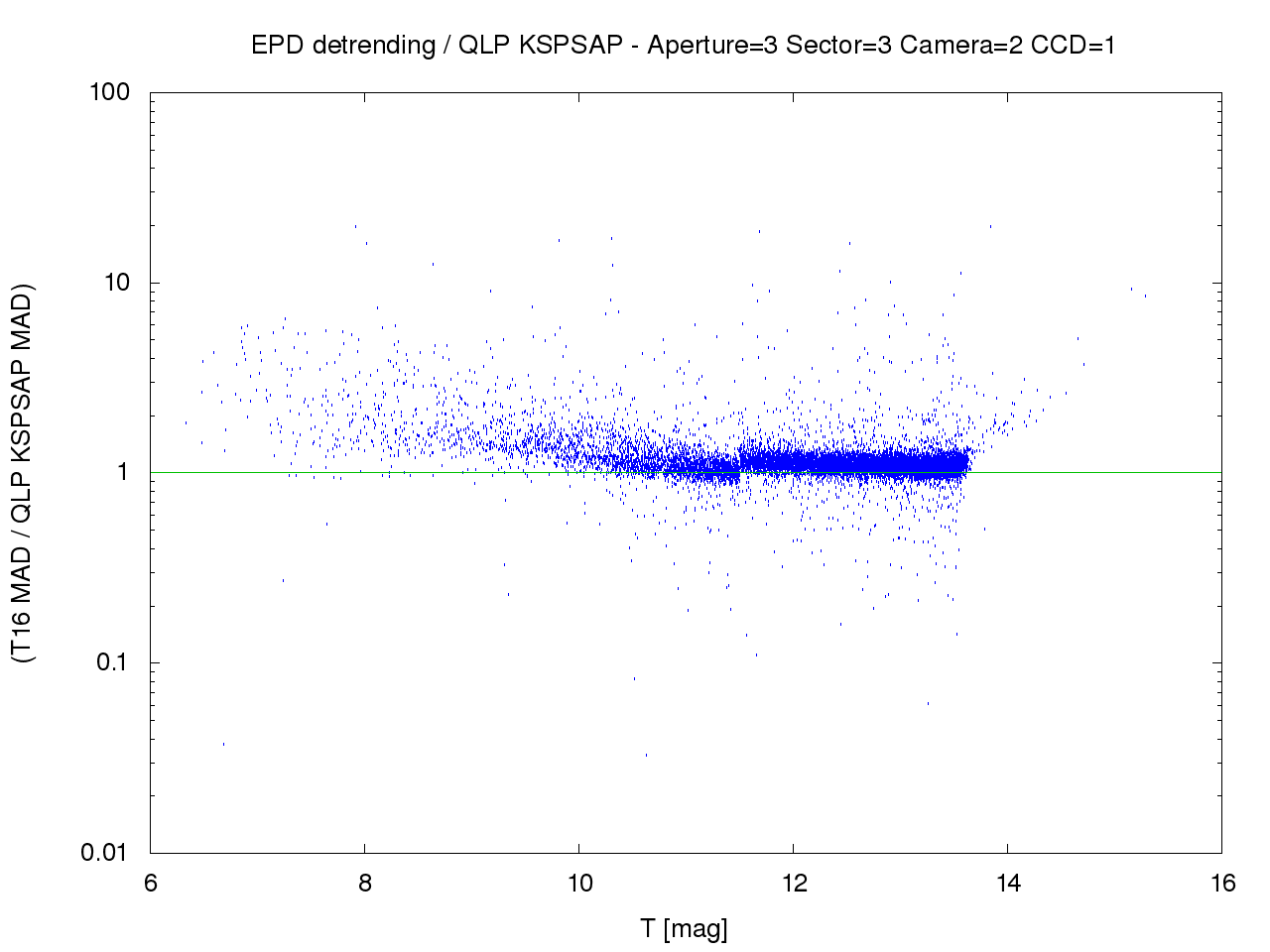}
}
{
\centering
\leavevmode
\includegraphics[width={0.5\linewidth}]{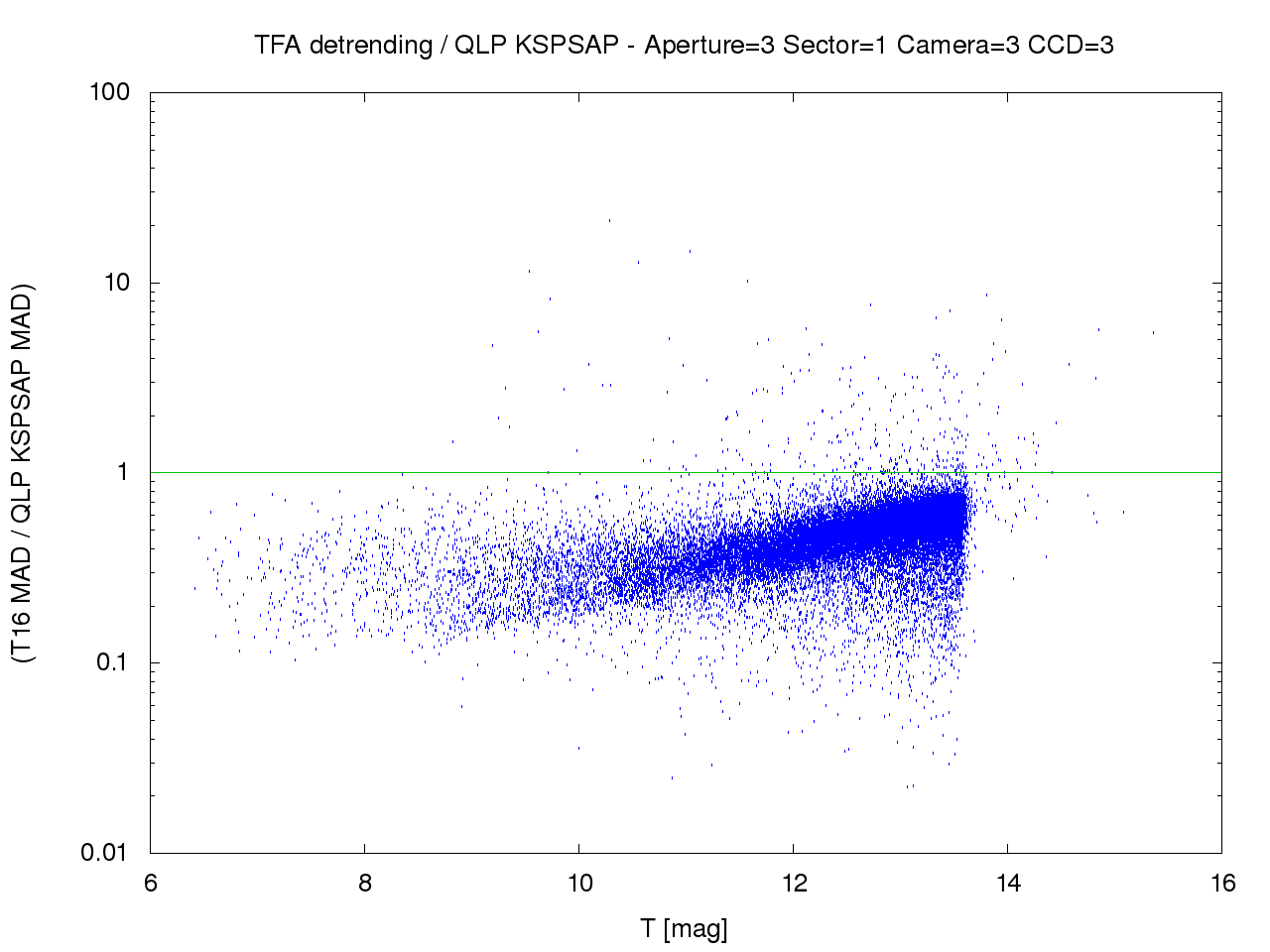}
\hfil
\includegraphics[width={0.5\linewidth}]{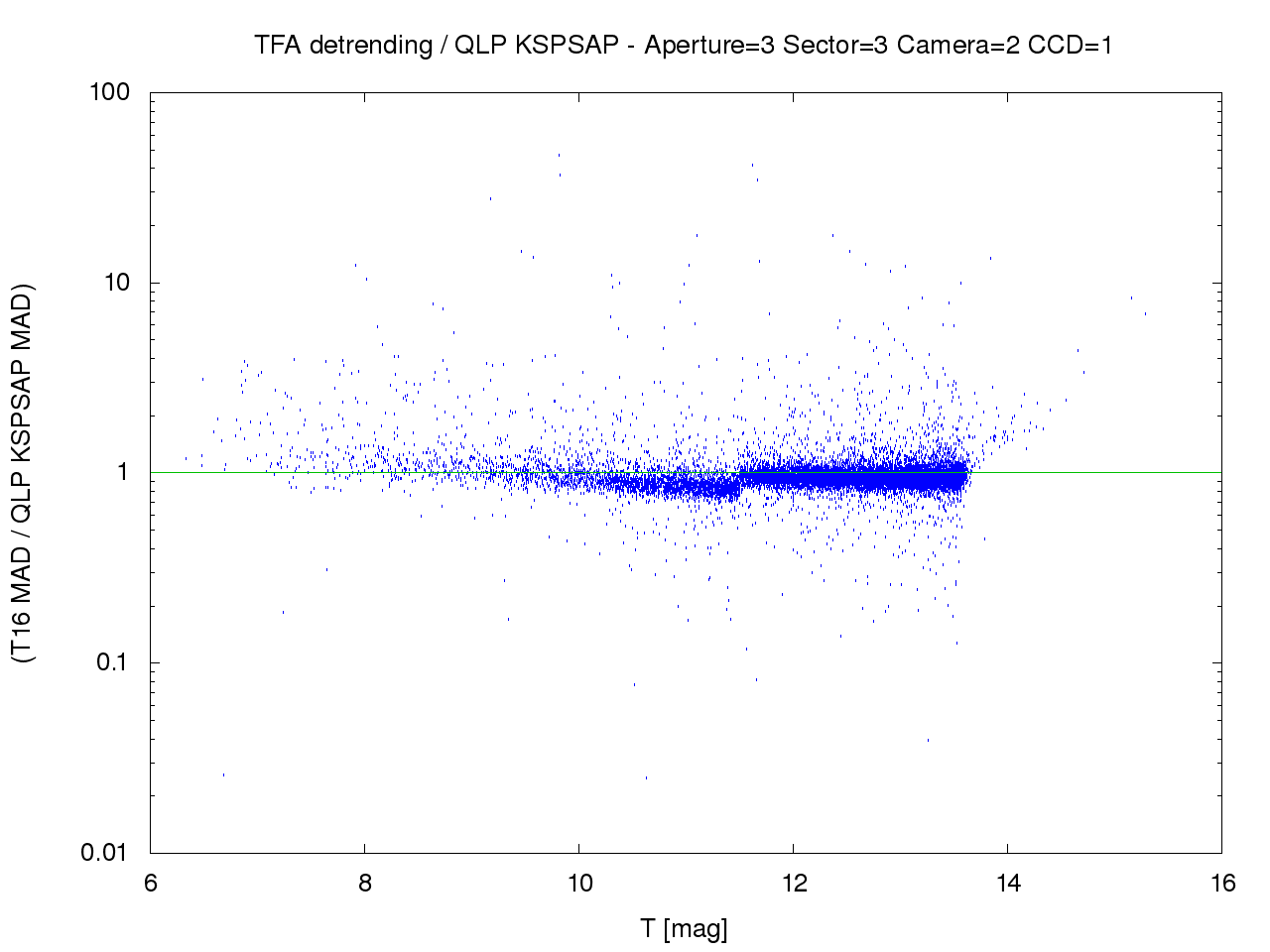}
}
\caption{The ratio of the T16 light curve MAD to the QLP light curve MAD, for a Sector/Camera/CCD where the T16 light curves have systematically lower MAD than QLP ({\em Left}), and for a Sector/Camera/CCD where the un-detrended QLP light curves have systematically lower MAD than the un-detrended T16 light curves ({\em right}). We show the comparison only for aperture 3 of the T16 light curves. For smaller apertures the precision of the T16 light curves tends to get worse for brighter stars and better for fainter stars, so the ratio shown here is lower at the bright end than it is for apertures 1 or 2, but higher at the faint end than it is for the other apertures. In the {\em top} row we compare the un-detrended T16 light curve precision to the un-detrended QLP light curve precision. In the {\em middle} row we compare the SEPD-detrended T16 light curve precision to the KSPSAP detrended QLP light curve precision. In the {\em bottom} row we compare the TFA-detrended T16 light curve precision to the KSPSAP detrended QLP light curve precision.
\label{fig:madratioqlp}}
\end{figure*}

\begin{figure*}[!ht]
{
\centering
\leavevmode
\includegraphics[width={0.5\linewidth}]{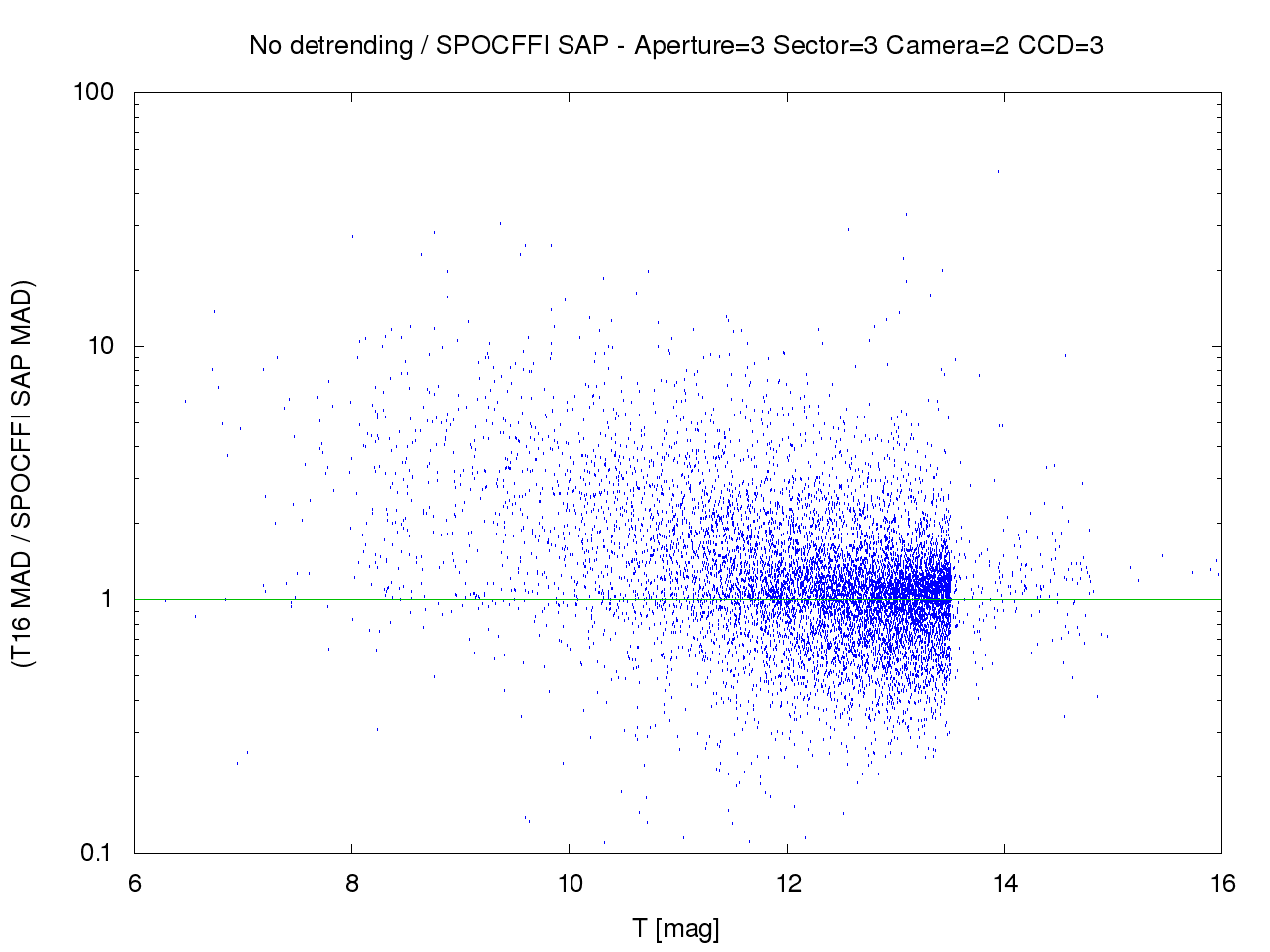}
\hfil
\includegraphics[width={0.5\linewidth}]{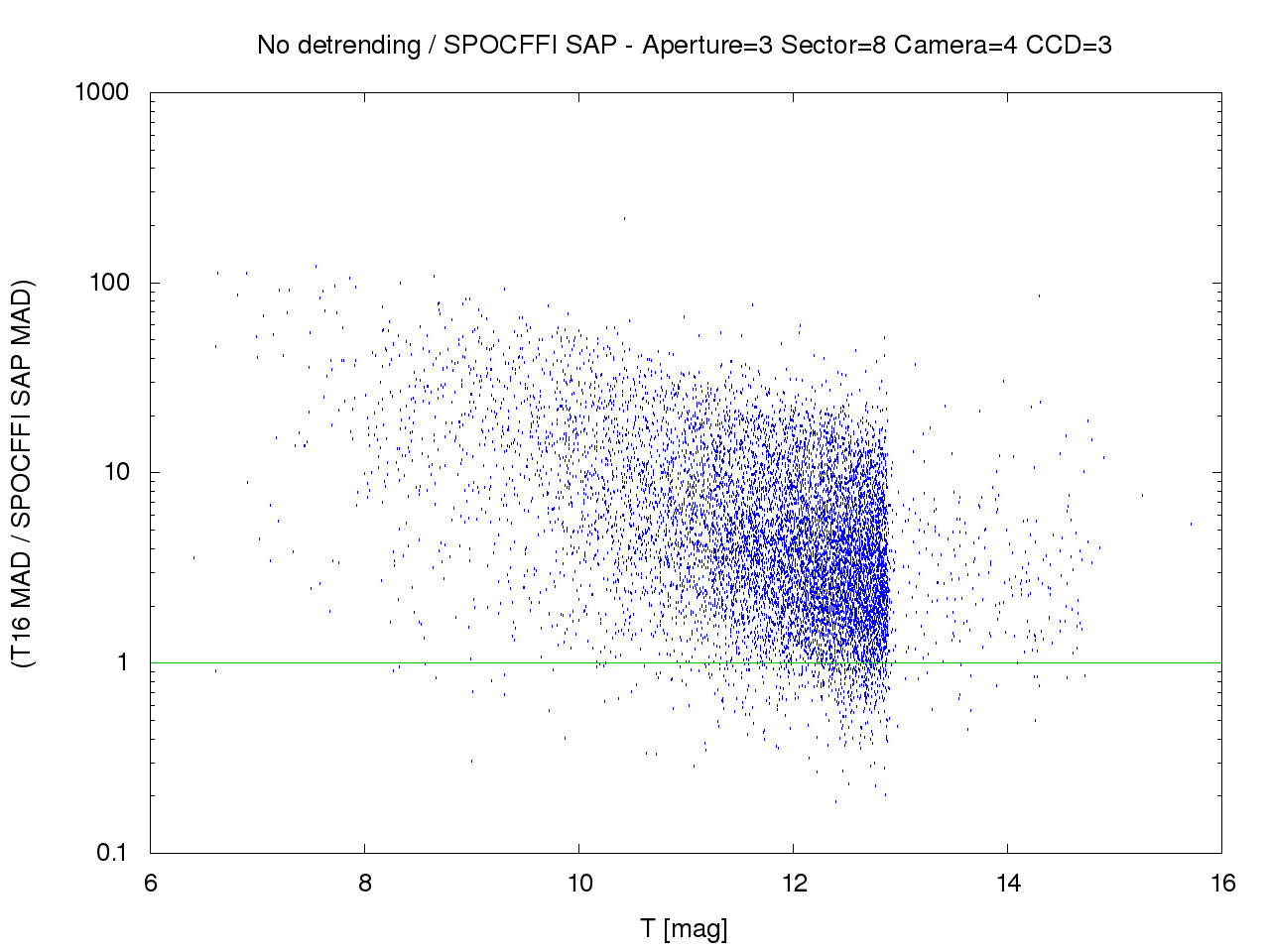}
}
{
\centering
\leavevmode
\includegraphics[width={0.5\linewidth}]{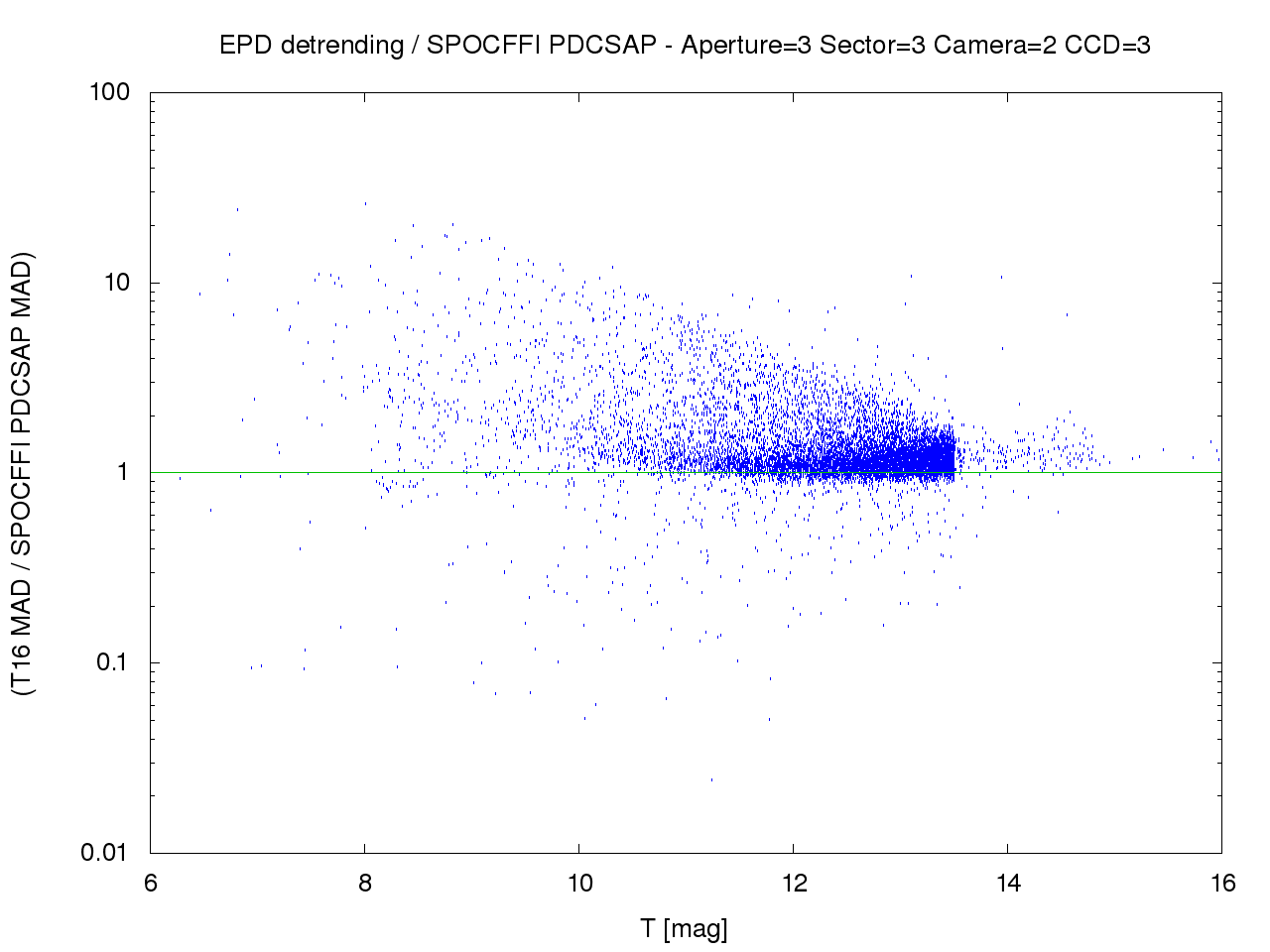}
\hfil
\includegraphics[width={0.5\linewidth}]{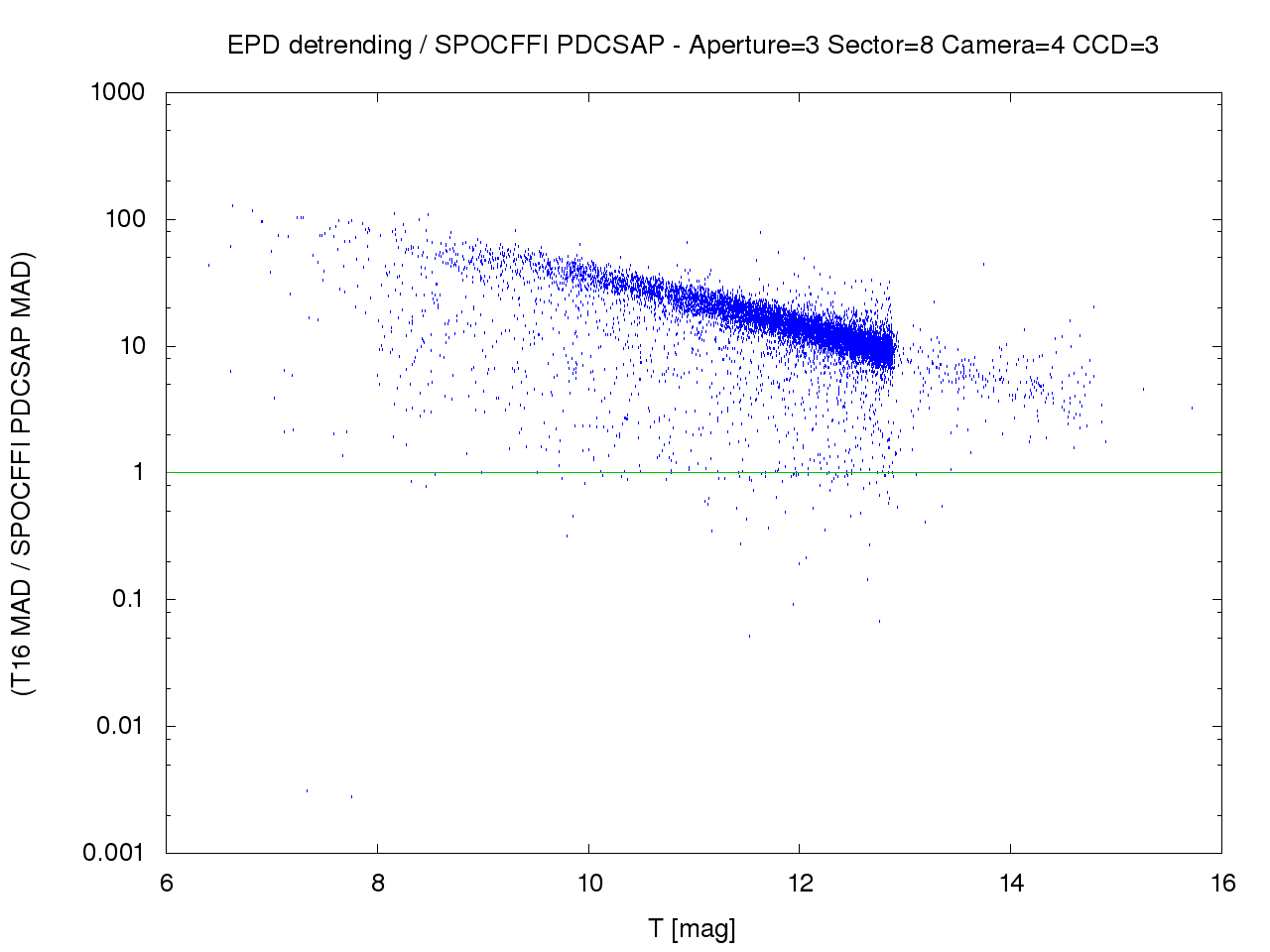}
}
{
\centering
\leavevmode
\includegraphics[width={0.5\linewidth}]{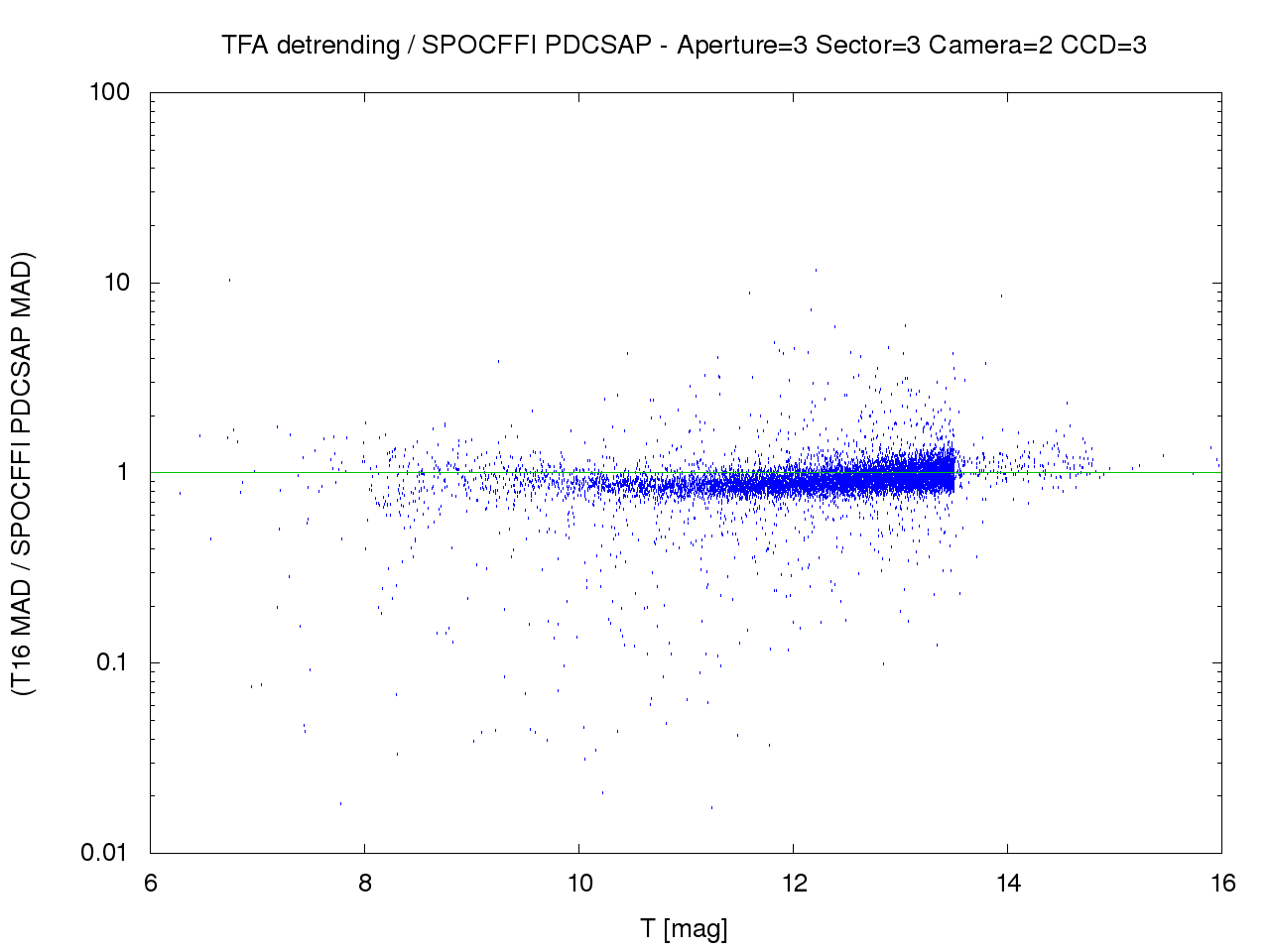}
\hfil
\includegraphics[width={0.5\linewidth}]{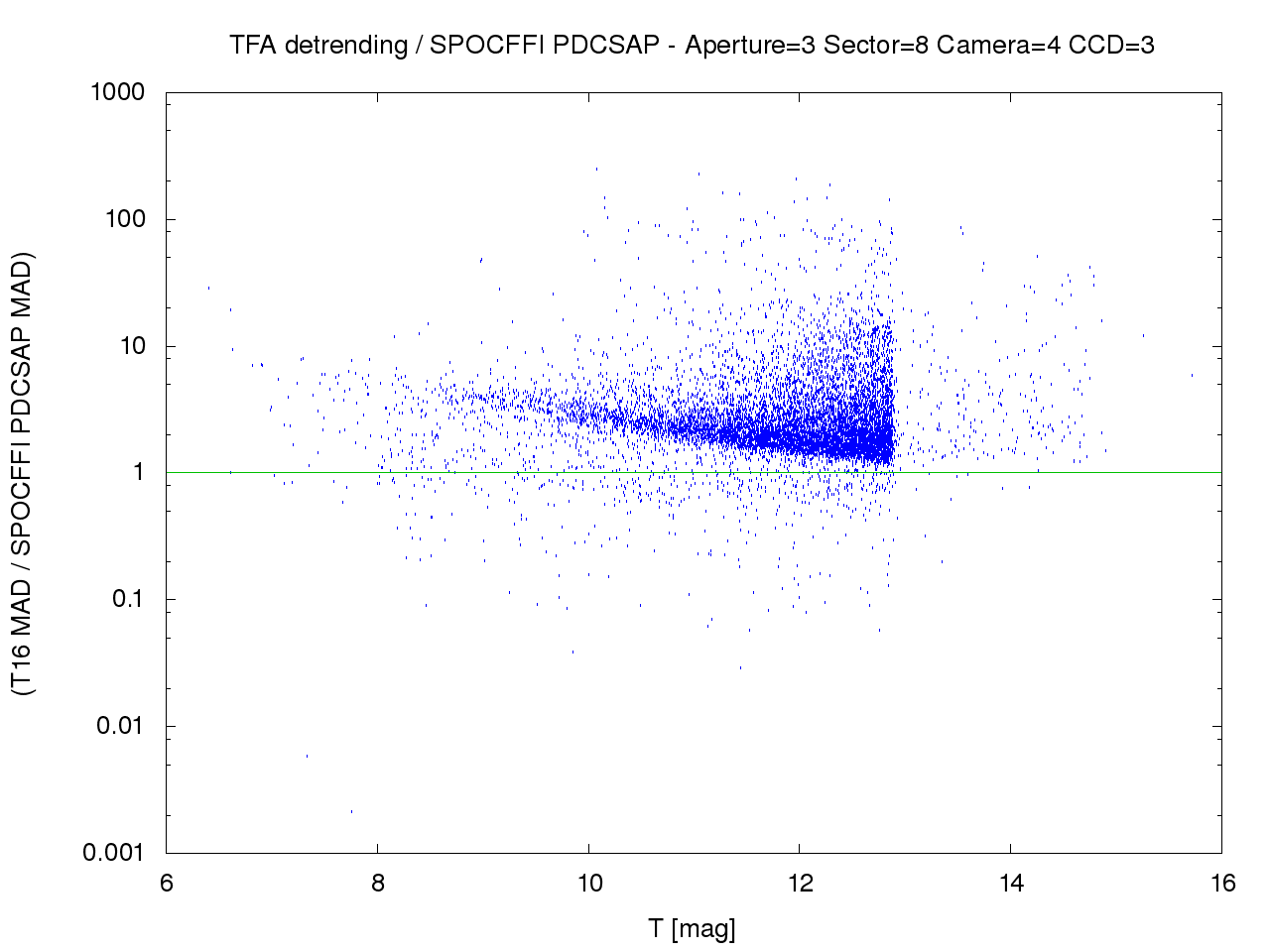}
}
\caption{Similar to Fig.~\ref{fig:madratioqlp}, here we compare the T16 light curve MAD to the SPOC-FFI light curve MAD. The Sector/Camera/CCD combinations shown here differ from those show in Fig.~\ref{fig:madratioqlp}.
\label{fig:madratiospoc}}
\end{figure*}

\begin{figure*}[!ht]
{
\centering
\leavevmode
\includegraphics[width={0.5\linewidth}]{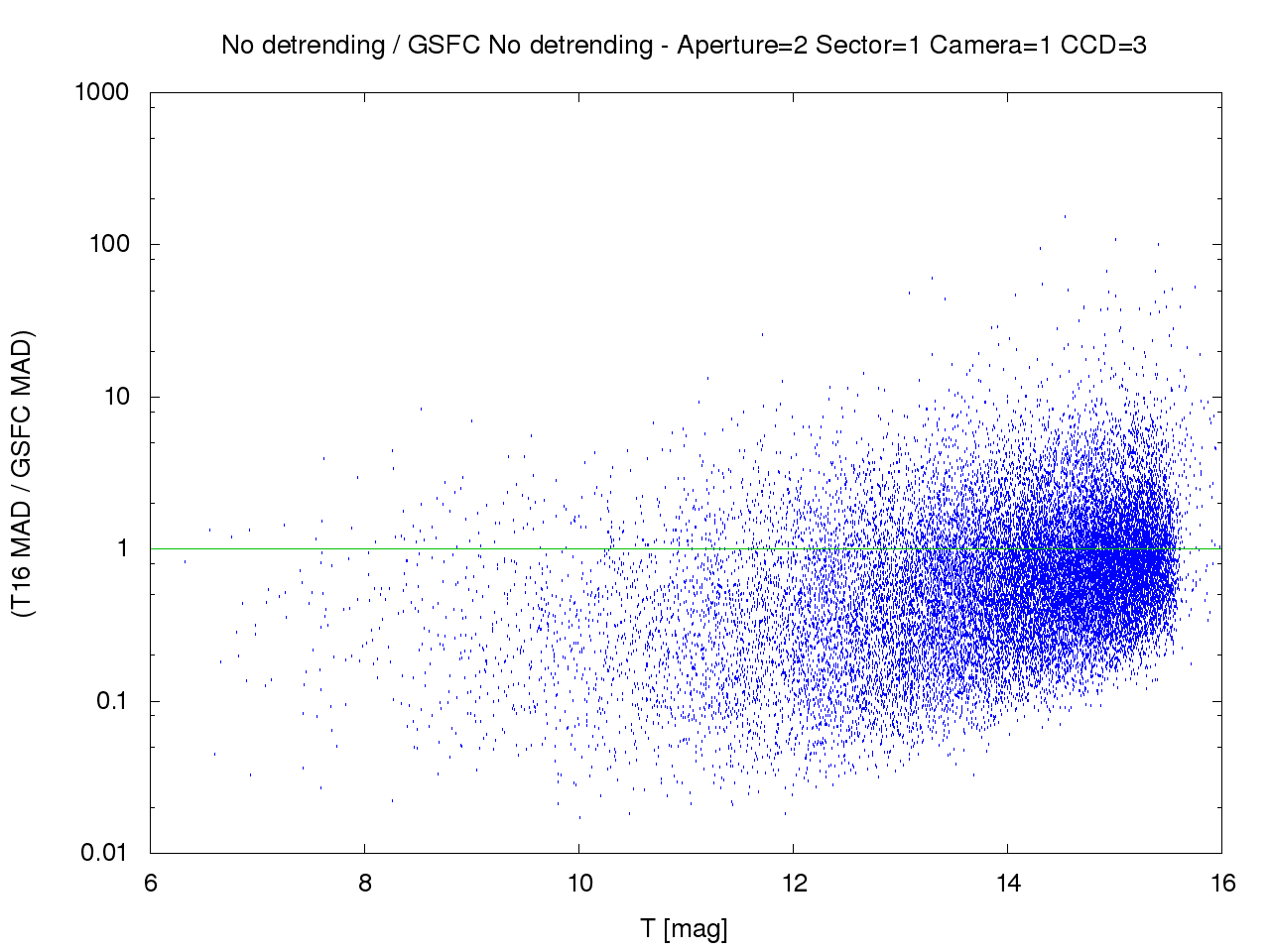}
\hfil
\includegraphics[width={0.5\linewidth}]{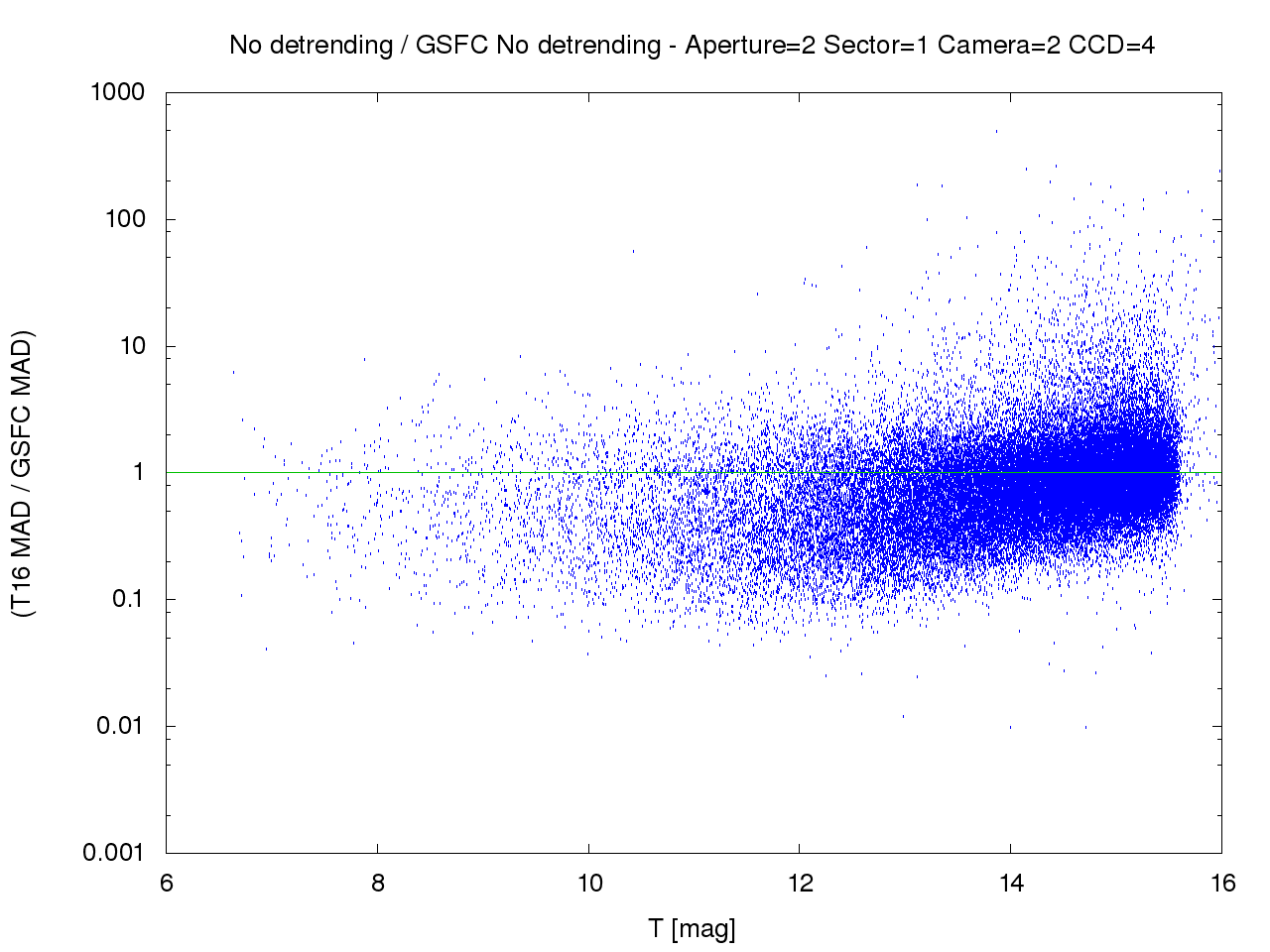}
}
{
\centering
\leavevmode
\includegraphics[width={0.5\linewidth}]{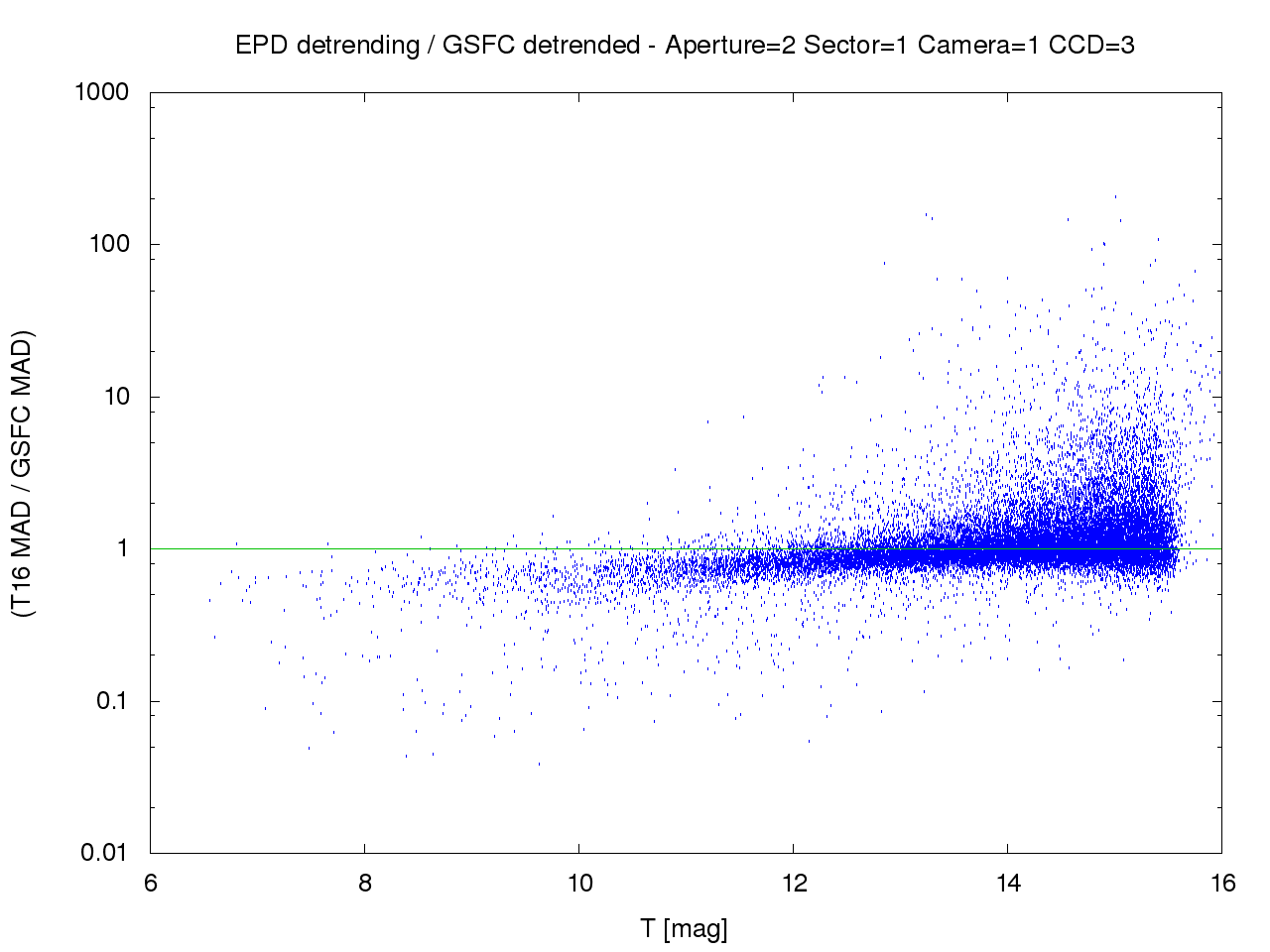}
\hfil
\includegraphics[width={0.5\linewidth}]{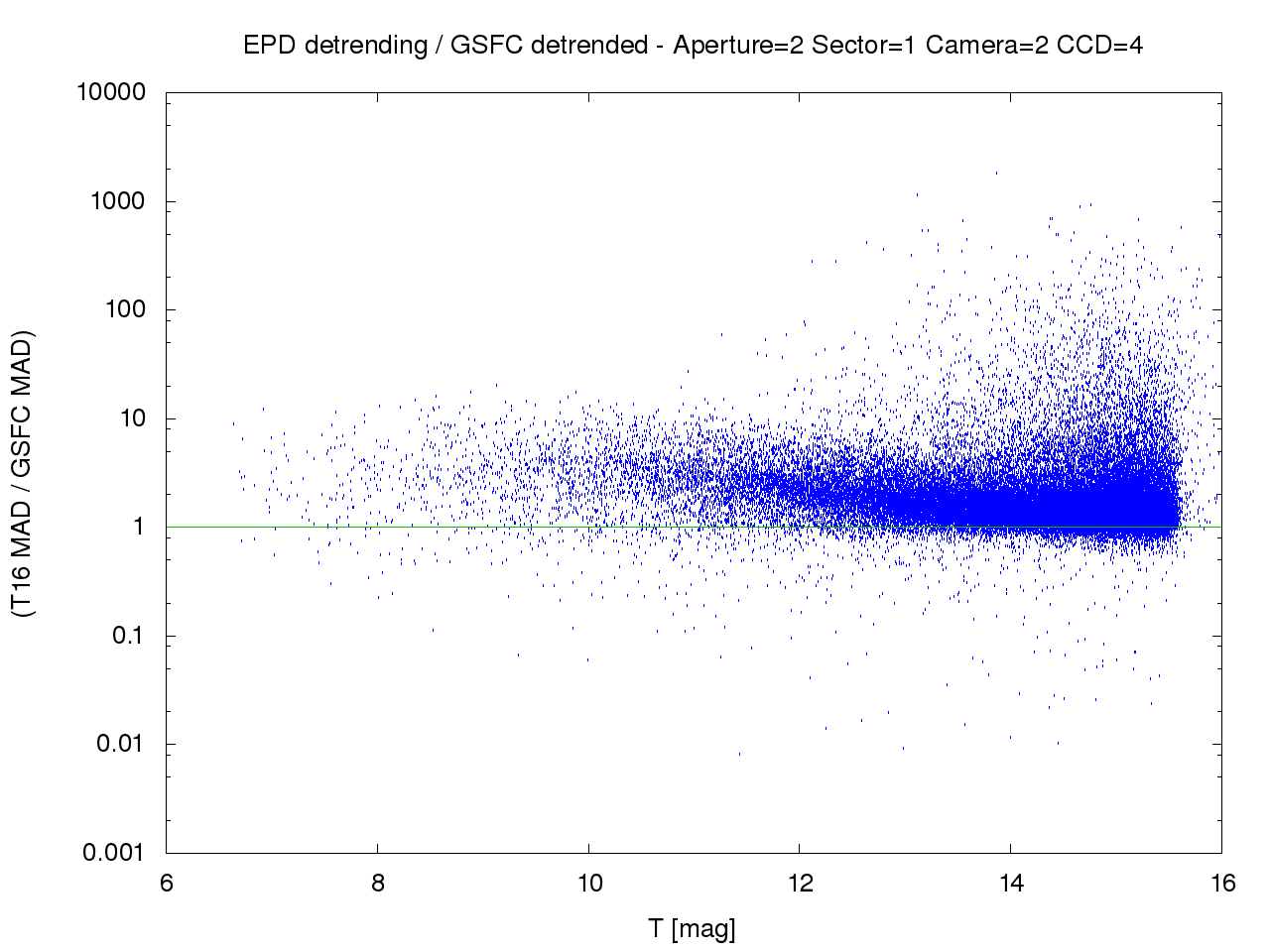}
}
{
\centering
\leavevmode
\includegraphics[width={0.5\linewidth}]{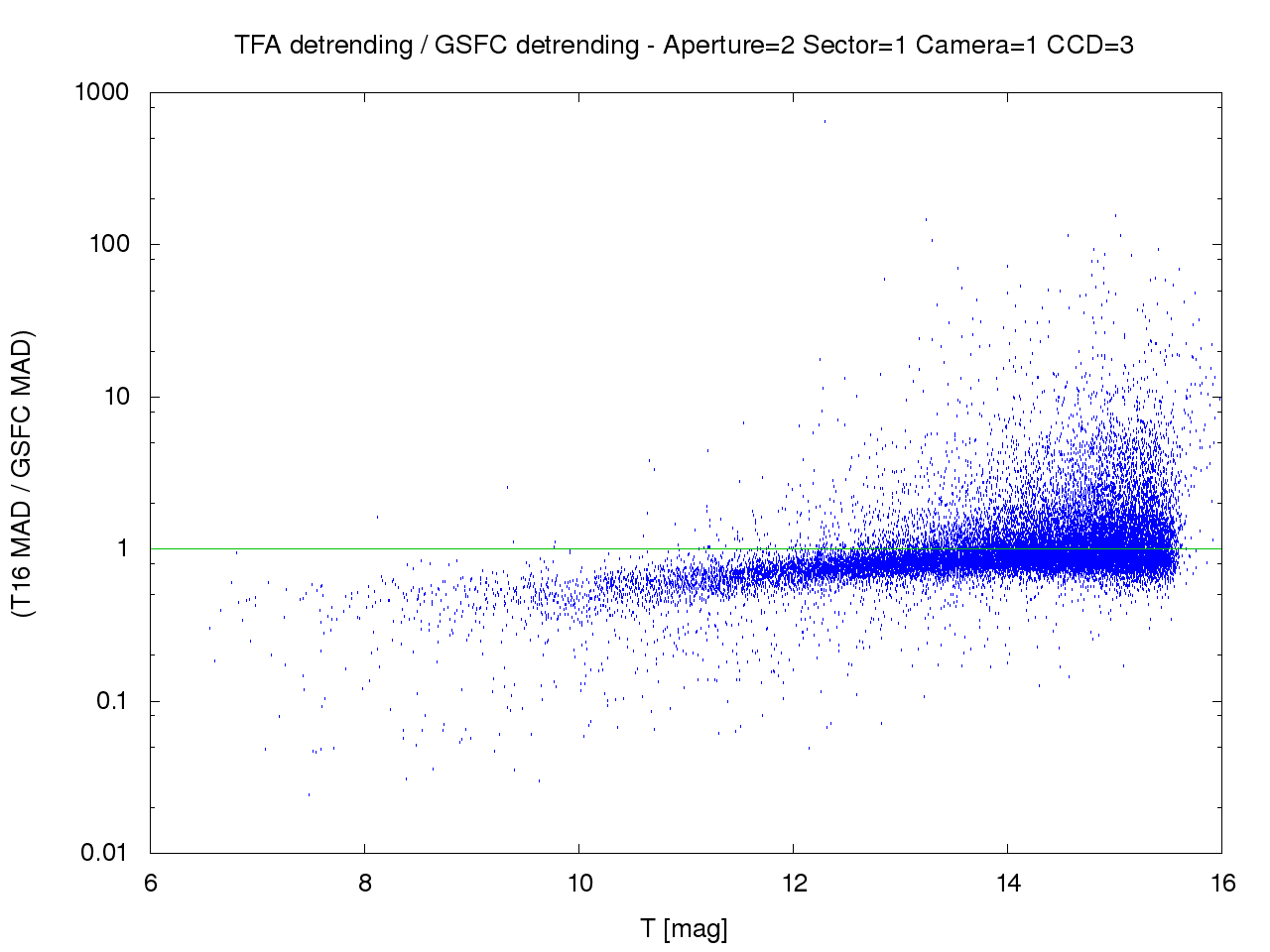}
\hfil
\includegraphics[width={0.5\linewidth}]{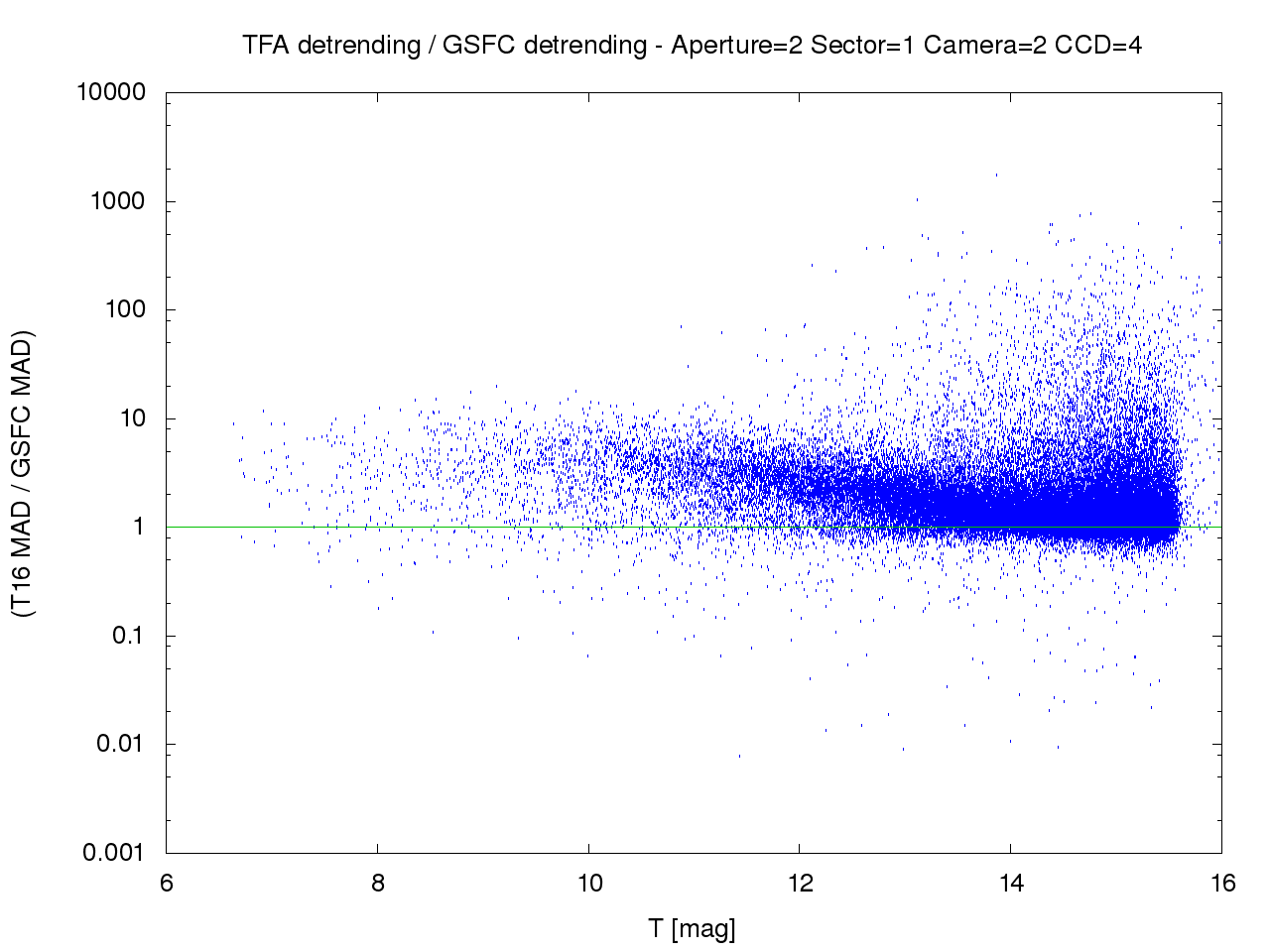}
}
\caption{Similar to Fig.~\ref{fig:madratioqlp}, here we compare the T16 light curve MAD to the GSFC-ELEANOR-LITE light curve MAD. The Sector/Camera/CCD combinations shown here differ from those show in Fig.~\ref{fig:madratioqlp}. We also use aperture 2 for T16 rather than aperture 3.
\label{fig:madratiogsfc}}
\end{figure*}

\begin{figure*}[!ht]
{
\centering
\leavevmode
\includegraphics[width={0.5\linewidth}]{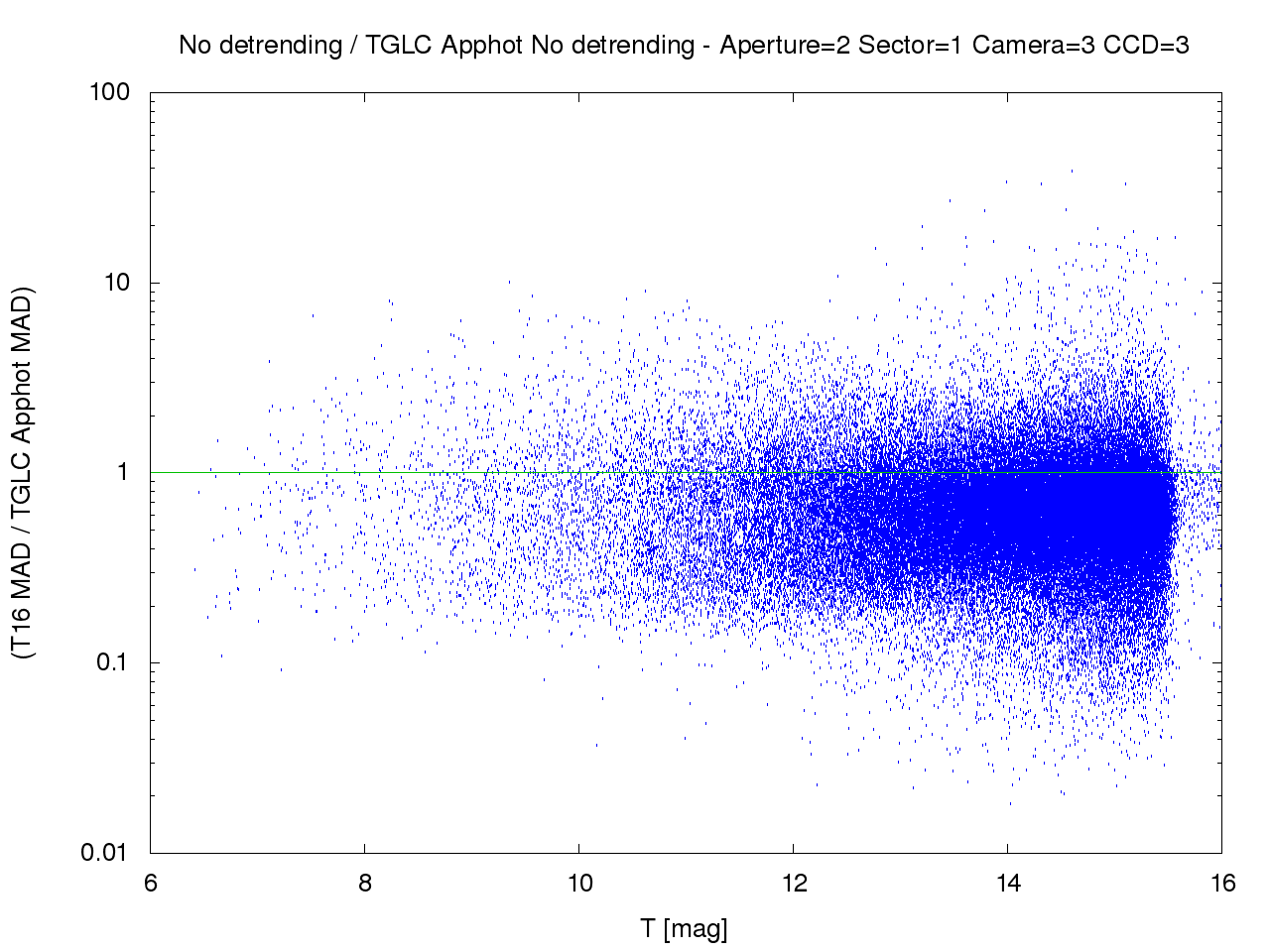}
\hfil
\includegraphics[width={0.5\linewidth}]{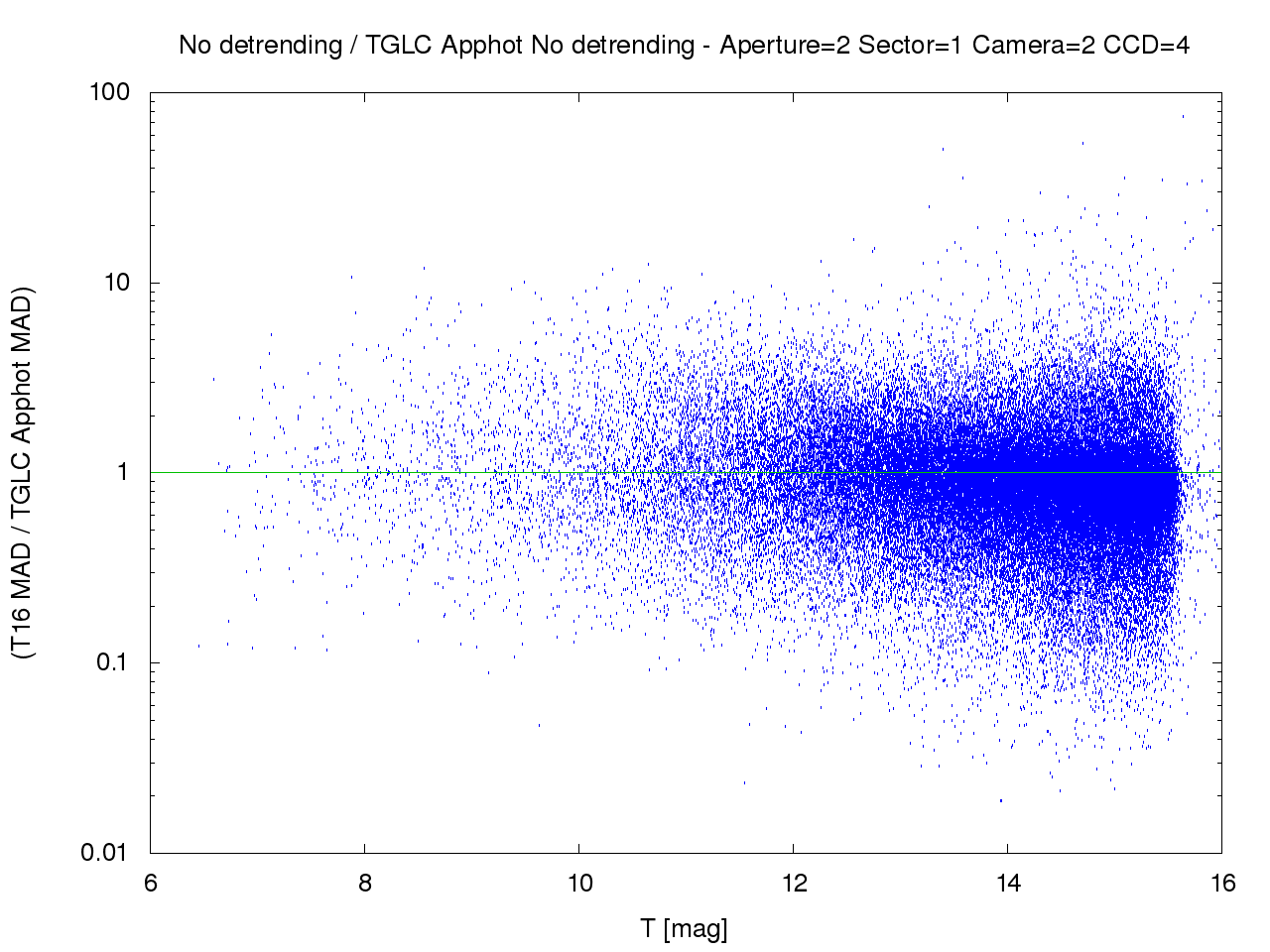}
}
{
\centering
\leavevmode
\includegraphics[width={0.5\linewidth}]{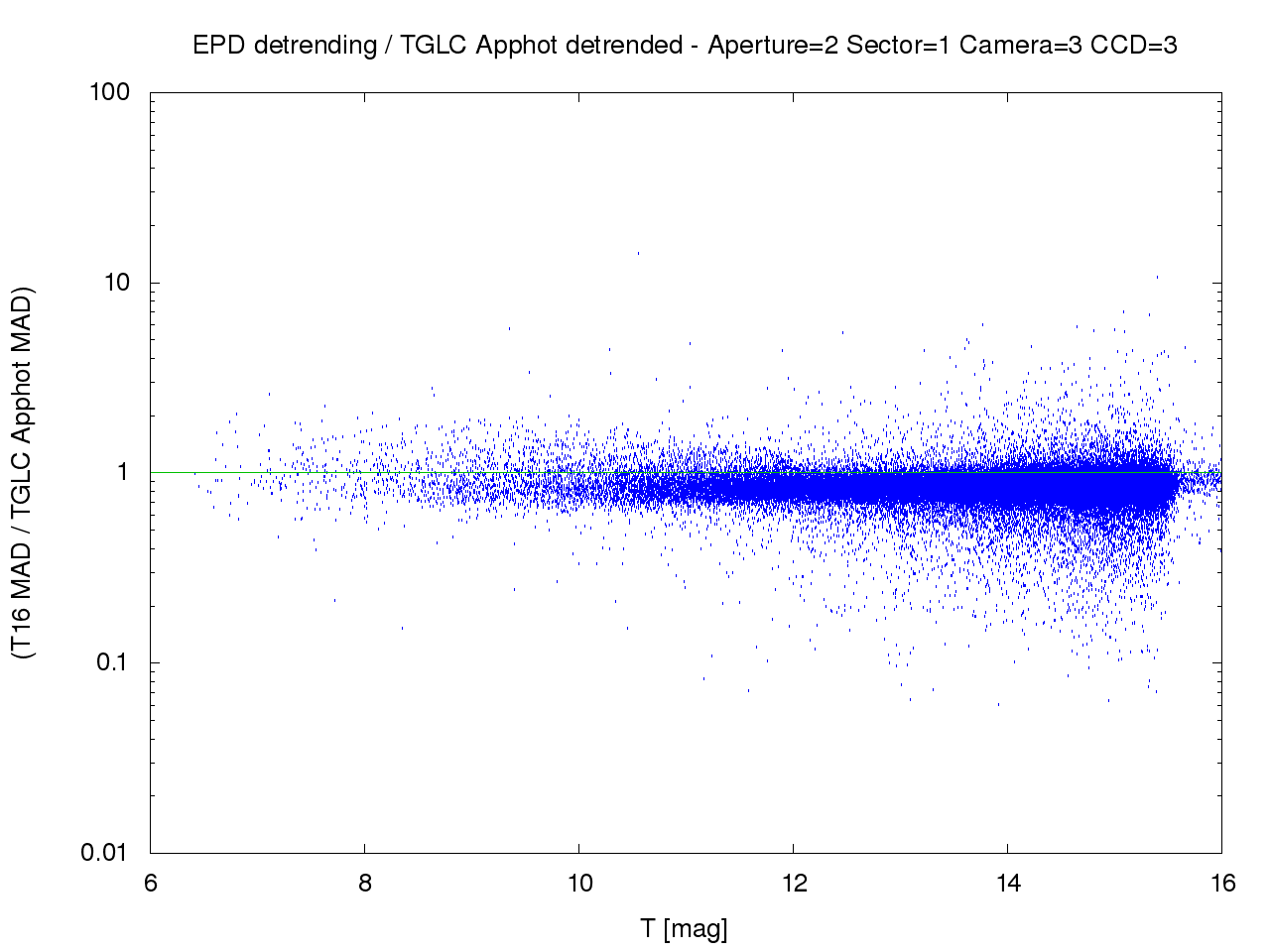}
\hfil
\includegraphics[width={0.5\linewidth}]{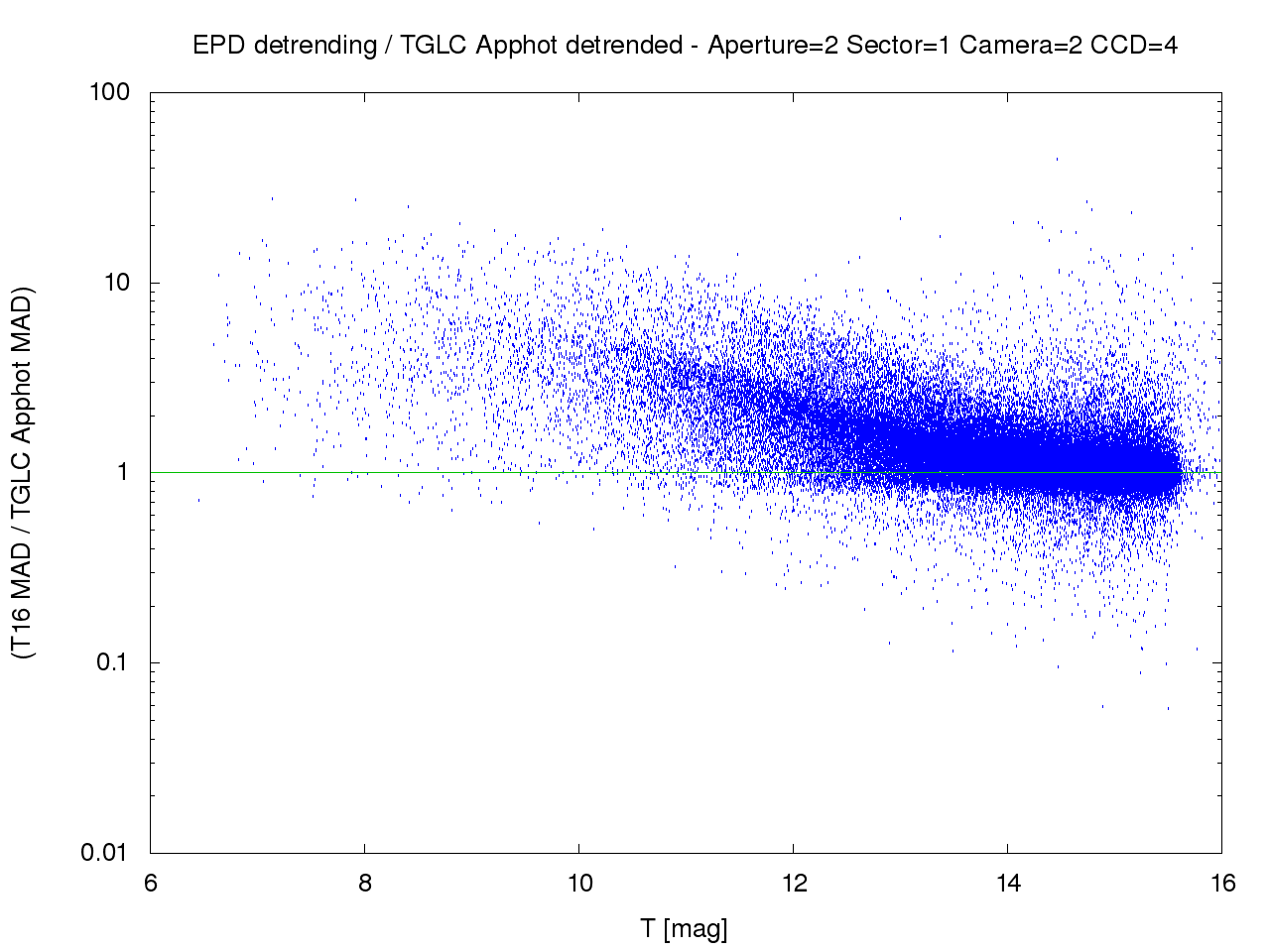}
}
{
\centering
\leavevmode
\includegraphics[width={0.5\linewidth}]{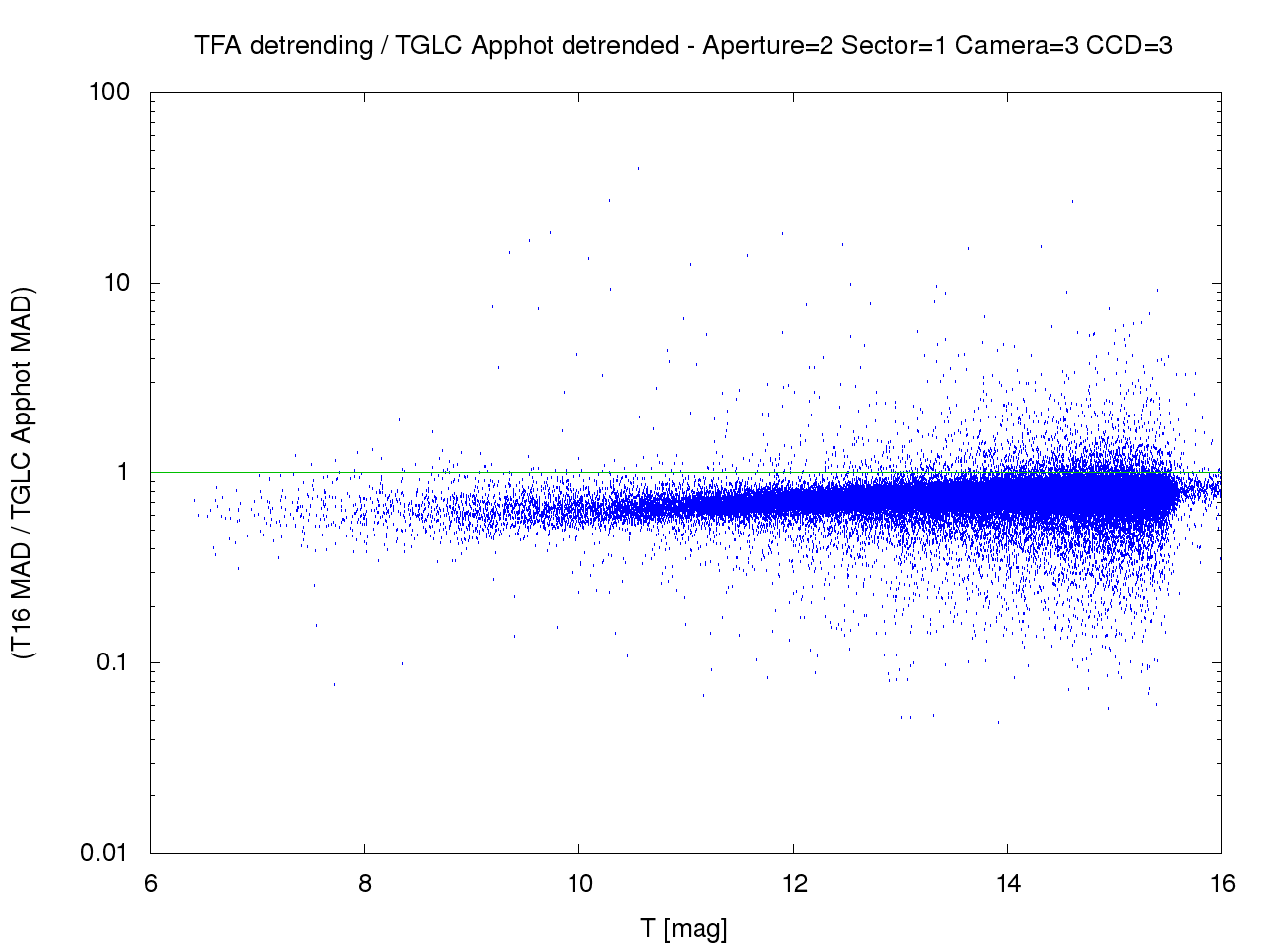}
\hfil
\includegraphics[width={0.5\linewidth}]{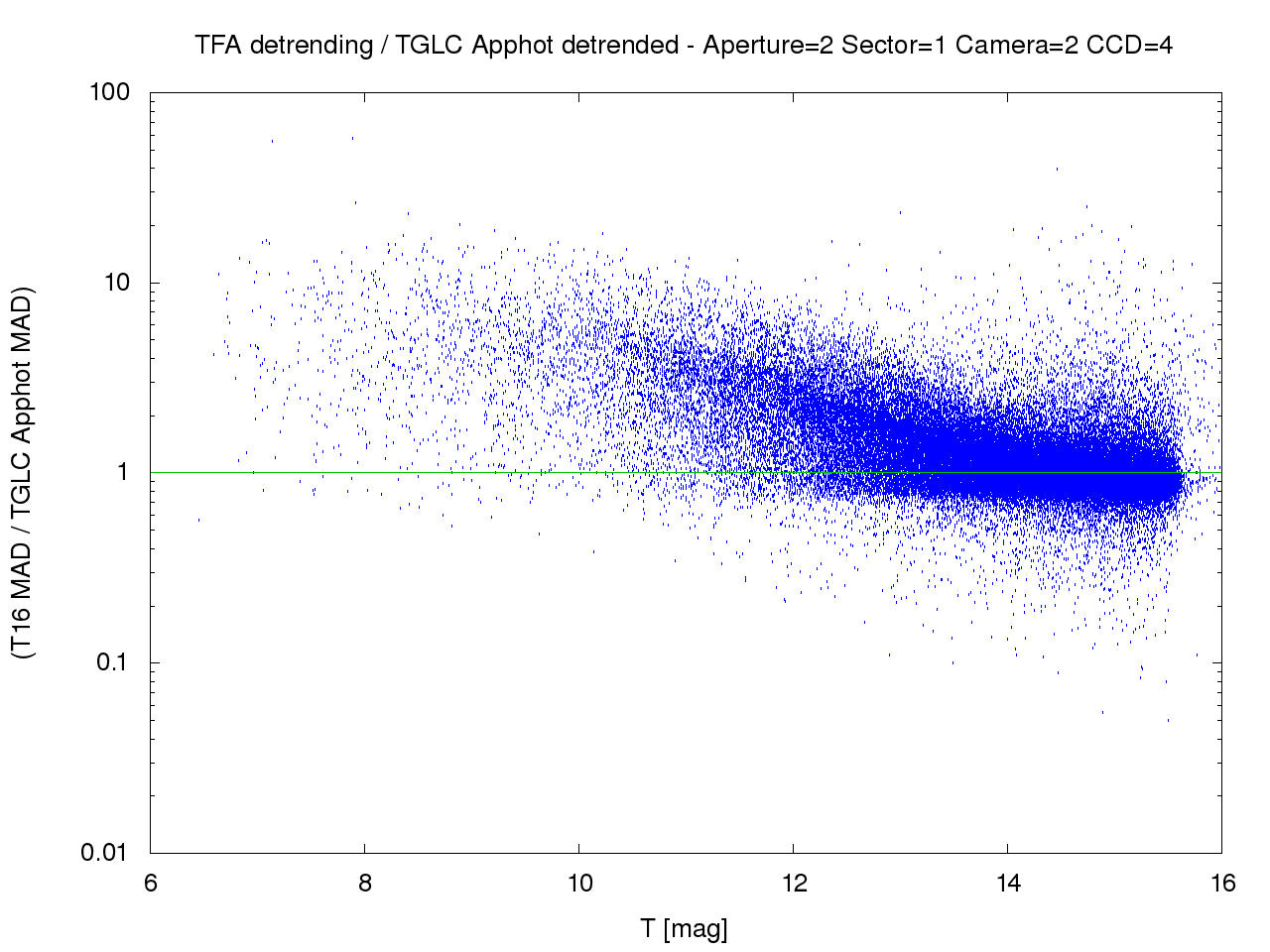}
}
\caption{Similar to Fig.~\ref{fig:madratioqlp}, here we compare the T16 light curve MAD to the TGLC aperture photometry light curve MAD. Note that the TGLC aperture photometry has systematically better precision than the PSF photometry for sector 1, so we restrict the comparison shown here to the aperture photometry. The Sector/Camera/CCD combinations shown here differ from those show in Fig.~\ref{fig:madratioqlp}. We also use aperture 2 for T16 rather than aperture 3.
\label{fig:madratiotglc}}
\end{figure*}

To perform the comparison we obtained the light curves from each of
these projects from MAST. We then compute the MAD for each light
curve in all available methods and states of trend-filtering. For the
QLP light curves we compute this for both the simple aperture
photometry (SAP column in the available FITS files) and Kepler-Spline
simple aperture photometry (KSPSAP column) time series, while for
SPOC-TESS we compute this for both the SAP and pre-search data
conditioning simple aperture photometry (PDCSAP column) time series
data. For TGLC we compute the MAD for the PSF-fitting photometry
(psf\_flux column), aperture photometry (aperture\_flux), calibrated
PSF-fitting photometry (cal\_psf\_flux), and calibrated aperture
photometry (cal\_aper\_flux) time series. For GSFC-ELEANOR-LITE we
use the raw flux (RAW\_FLUX) and corrected flux (CORRECTED\_FLUX) time
series. In all cases we exclude any data point that is flagged as
being potentially affected by some instrumental artifact.

Note that because each project applied a different detrending method,
the precision comparison that we are performing here should not be
interpreted as a pure comparison of the different photometry methods
or of the different detrending methods, but is rather a comparison of
the different combined photometry+detrending methods. The goal here is
to compare the precision of the different light curves that are
readily available to the public, and not to compare the individual
stages of the reduction methods.

Figure~\ref{fig:madratioqlp} compares the T16 light curve precision to
the QLP precision, Figure~\ref{fig:madratiospoc} shows this comparison
for the SPOC-FFI light curves, Figure~\ref{fig:madratiogsfc} shows the
comparison for the GSFC-ELEANOR-LITE light curves, and
Figure~\ref{fig:madratiotglc} shows the comparison for the TGLC light
curves. A figure set that shows these comparisons for every {\em
  TESS} sector, camera, and CCD combination, for each of the three
photometric apertures that we employ, for the IRM, EPD and TFA light
curves from T16, and for all of the different photometric time series
provided on MAST is available at \url{https://doi.org/10.5281/zenodo.14278698}. In this article we show a small subset of these
different comparison plots.

\begin{deluxetable*}{lrrrrrrrrrr}
\tablewidth{0pc}
\tabletypesize{\tiny}
\tablecaption{
    Fraction of light curves with T16 MAD lower than MAD from other projects by sector
    \label{tab:precisioncomparison}
}
\tablehead{
    \multicolumn{1}{c}{Sector} &
    \multicolumn{1}{c}{QLP\tablenotemark{a}} &
    \multicolumn{1}{c}{QLP\tablenotemark{b}} &
    \multicolumn{1}{c}{SPOC-FFI\tablenotemark{a}} &
    \multicolumn{1}{c}{SPOC-FFI\tablenotemark{b}} &
    \multicolumn{1}{c}{GSFC\tablenotemark{a}\tablenotemark{c}} &
    \multicolumn{1}{c}{GSFC\tablenotemark{b}\tablenotemark{c}} &
    \multicolumn{1}{c}{TGLC\tablenotemark{a}\tablenotemark{c}} &
    \multicolumn{1}{c}{TGLC\tablenotemark{a}\tablenotemark{c}} &
    \multicolumn{1}{c}{TGLC\tablenotemark{b}\tablenotemark{c}} &
    \multicolumn{1}{c}{TGLC\tablenotemark{b}\tablenotemark{c}} \\
    \multicolumn{1}{c}{} &
    \multicolumn{1}{c}{SAP} &
    \multicolumn{1}{c}{KSPSAP} &
    \multicolumn{1}{c}{SAP} &
    \multicolumn{1}{c}{PDCSAP} &
    \multicolumn{1}{c}{Raw} &
    \multicolumn{1}{c}{Corrected} &
    \multicolumn{1}{c}{Aperture} &
    \multicolumn{1}{c}{PSF} &
    \multicolumn{1}{c}{Cal.\ Aperture} &
    \multicolumn{1}{c}{Cal.\ PSF}
}
\startdata
1 & 60.6\% & 95.5\% & 38.5\% & 62.9\% & 66.8\% & 47.1\% & 82.1\% & 75.4\% & 88.8\% & 80.3\% \\
2 & 11.7\% & 82.7\% & 32.8\% & 75.0\% & 58.1\% & 45.7\% & 56.1\% & 67.0\% & 84.1\% & 77.9\% \\
3 & 17.3\% & 87.1\% & 39.2\% & 84.1\% & 83.0\% & 79.2\% & 79.2\% & 83.5\% & 95.3\% & 89.8\% \\
4 & 8.6\% & 75.3\% & 54.0\% & 76.7\% & 71.0\% & 55.3\% & 76.9\% & 81.2\% & 84.7\% & 80.8\% \\
5 & 20.7\% & 76.0\% & 40.5\% & 70.6\% & 67.2\% & 36.2\% & 67.6\% & 76.8\% & 78.0\% & 72.4\% \\
6 & 25.1\% & 78.1\% & 41.4\% & 78.4\% & 51.4\% & 23.3\% & 65.4\% & 69.0\% & 81.3\% & 74.8\% \\
7 & 30.1\% & 86.4\% & 42.8\% & 82.8\% & 55.4\% & 20.2\% & 72.6\% & 71.1\% & 84.9\% & 76.4\% \\
8 & 14.6\% & 54.8\% & 43.2\% & 56.8\% & 58.7\% & 38.1\% & 70.4\% & 70.0\% & 53.2\% & 52.3\% \\
9 & 28.0\% & 73.1\% & 46.6\% & 63.9\% & 55.5\% & 34.5\% & 62.9\% & 62.6\% & 68.1\% & 61.9\% \\
10 & 29.6\% & 85.1\% & 44.6\% & 78.8\% & 57.4\% & 46.8\% & 59.7\% & 57.4\% & 81.0\% & 74.0\% \\
11 & 23.7\% & 65.9\% & 31.4\% & 51.1\% & 51.4\% & 41.6\% & 57.5\% & 51.4\% & 50.7\% & 48.2\% \\
12 & 42.7\% & 83.6\% & 36.0\% & 59.0\% & 57.4\% & 42.0\% & 85.5\% & 75.6\% & 75.7\% & 64.0\% \\
13 & 33.1\% & 86.3\% & 28.4\% & 71.7\% & 67.7\% & 51.2\% & 85.6\% & 73.6\% & 89.9\% & 75.4\% \\
{\bf Total} & {\bf 28.8\%} & {\bf 78.0\%} & {\bf 40.0\%} & {\bf 69.8\%} & {\bf 62.3\%} & {\bf 43.1\%} & {\bf 79.5\%} & {\bf 73.7\%} & {\bf 85.6\%} & {\bf 77.3\%} \\
\enddata
\tablenotetext{a}{Comparing to the Undetrended T16 Photometry}
\tablenotetext{b}{Comparing to the TFA-detrended T16 Photometry}
\tablenotetext{c}{For GSFC and TGLC the comparison is done on a randomly selected subset of the light curves.}
\end{deluxetable*}

Table~\ref{tab:precisioncomparison} compares the precision of the T16
light curves to that of the QLP, SPOC-TESS, GSFC and TGLC light curves
by listing the fraction of light curves by Sector for which
the T16 light curves has a lower MAD than the comparison light
curves. For SPOC-TESS and QLP we perform this comparison for all available light curves, while for GSFC and TGLC, for which the volume of data on MAST is very large, we randomly selected 1\% of the available GSFC and TGLC light curves per sector for comparison. 

When comparing un-detrended light curves, we find that QLP and
SPOC-TESS both tend to produce light curves with lower MAD than
T16. We find that 28.8\% of un-detrended T16 light curves have a lower
MAD than the corresponding un-detrended SAP light curves from QLP, and
40.0\% of un-detrended T16 light curves have a lower MAD than the
un-detrended SAP light curves from SPOC-TESS.  Here it is important to
note that the Cycle 1 QLP light curves are not automatically corrected for
contamination from neighboring stars. This can lead to an artificially
low MAD in the magnitude light curves of faint stars blended with
brighter stars. Such blending is a common occurrence given the
21\arcsec\,pixel$^{-1}$ plate scale of {\em TESS}. For T16 and
SPOC-TESS the contamination flux is effectively subtracted from the
light curves.  When comparing the detrended light curves, we find that
a majority of the T16 light curves have lower MAD than the
corresponding QLP and SPOC-TESS light curves, with 78\% of T16
detrended light curves having a lower MAD than the KSPSAP light curves
from QLP, and 69.8\% of T16 detrended light curves having a lower MAD
than the PDCSAP light curves from QLP.

We find that 62.3\% of the undetrended T16 light curves
have lower MAD than their corresponding raw GSFC light curves, but
this drops to 43.1\% for the detrended T16 light curves compared to
the corrected GSFC light curves. We find that 79.5\% and 73.7\% of the
undetrended T16 light curves have lower MAD than the TGLC aperture and
psf photometry light curves, respectively, while 85.6\% and 77.3\% of
the detrended T16 light curves have lower MAD than the respective
calibrated TGLC light curves.

\subsubsection{Example Light Curve Comparison}
\label{sec:examples}

\paragraph{Transiting Planet Systems}
\label{sec:teps}

As a check of the light curve quality, we identified {\em
  TESS} Objects of Interest (TOIs) listed on the NASA Exoplanet
Archive\footnote{as of the date 2023 Jan 25; note that TOIs identified after this date are largely longer period candidates that have no, or few, transits in Cycle I, so we chose not update the analysis to include those objects.} for which we have produced a
light curve. We used the {\sc VARTOOLS} program to perform a Box Least
Squares \citep[BLS;][]{kovacs:2002:BLS} search for periodic transit
signals in these light curves.

We searched the TFA-detrended light curves in all three available
photometric apertures, treating the light curves from each sector,
camera and CCD independently (i.e., we do not combine all light curves
for a given target, but instead search the individual single sector
light curves).  In each case we search for periods between 0.2\,days
and 30\,days, scanning 100,000 frequencies with 1000 phase bins per
frequency. At each frequency we search a range of transit durations
from half the expected duration for a zero-impact-parameter transiting
planet on a circular orbit around a solar-density star, to twice the
expected duration for such a system. We identify the top five peaks in
the periodogram for each run of BLS, fitting a trapezoid transit model
at each identified peak.

We consider the transit signal to be recovered for the purposes of
this exercise if one of the identified periods from at least one of
the BLS runs for a given target has a period that is within 1\% of the
listed period, or an integer harmonic of the listed period (we
consider the second through tenth harmonics).  We note that because we
are not combining the light curves from multiple sectors, cameras or
CCDs, and we are not optimizing the BLS search on a star-by-star
basis, we do not expect this search to recover all of the
TOIs that might be recoverable within our data. We also note that many
of the TOIs were identified based on {\em TESS} observations obtained
after Cycle 1, so we do not expect to recover these signals within the
light curves presented here. Based on the above selection method, we
recover 2639 out of the 2971 TOIs for which we have produced a light
curve.  We conclude that our T16 light curves are of sufficient
quality to recover the large majority of known transiting planet
signals from {\em TESS}, even when running a non-optimized, single-sector search.

%% TBD - Show some example phase-folded transit light curves. Some targets to consider:
% 650778499868023296 - high SNR
% 5519728867449179264 - consider for median SNR
% 3167323052618369408 - consider for median SNR - not great for T16
% 4768613025228556416 - consider for median SNR
% 4674216245427964416 - consider for lower SNR - this is a multi planet system
% 2920348242527963904 - consider for lower SNR
% 3210444215030339584 - consider for lower SNR

\begin{figure}[!ht]
{
\centering
\leavevmode
\includegraphics[width=1.0\linewidth]{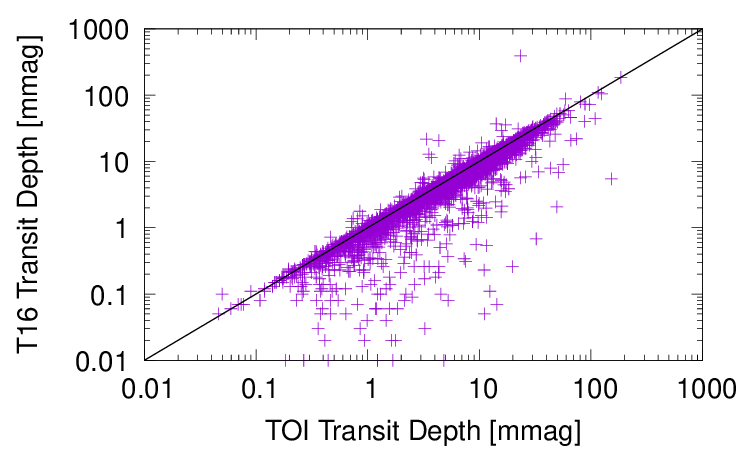}
}
\caption{Comparison between the transit depth measured from a non-optimized BLS analysis of the TFA-detrended T16 light curves to the listed TOI transit depth for known TOIs for which we have produced a light curve. We only show this comparison for those objects for which the BLS transit ephemeris is consistent with the reported TOI ephemeris. The solid line shows equal transit depths.
\label{fig:blsdepthcompare}}
\end{figure}

Figure~\ref{fig:blsdepthcompare} compares the transit depth from the
TFA-detrended T16 light curves to the reported TOI transit depth. For
simplicity the T16 transit depth is measured by performing a BLS
transit search on each light curve at a period fixed to the reported
TOI period, and we show only those systems for which the transit epoch
recovered in the T16 light curve is consistent with the reported TOI
ephemeris.  We find that the median ratio of the T16 transit depth to
the reported TOI transit depth is 91\%. The modest reduction in depth
may be due to over-filtering by SEPD+TFA (these are being run without
simultaneously fitting for any astrophysical variability signal), the
simple BLS search finding a slightly longer transit duration or
slightly offset transit epoch compared to the TOI values, and/or the
use of a box-shape transit model rather than the limb-darkened
model. We conclude that while there may be a slight distortion and
damping of transit-like variable signals in our TFA-detrended light
curves, overall the image subtraction light curves have variability
amplitudes for transit systems that are in line with expectations.

There are a number of cases where the T16 transit depth is significantly lower than the listed TOI value. We find 79 cases (corresponding to 2.5\% of the sample) where the T16 depth is less than 10\% of the TOI depth (counting only examples where there is at least one point in transit in the T16 light curve).  There does not appear to be a single, simple explanation of these. The TOIs with low T16 transit depths do not have systematically fewer points in transit, systematically longer periods, or systematically different transit durations, or stellar apparent magnitudes than TOIs where the T16 transit depths match well to the TOI values. They also do not appear to be clustered in any particular region of the sky. We do find that 25\% of the objects with exceptionally low T16 transit depths have been set aside as false positives or marked as likely eclipsing binaries or blended eclipsing binaries, which is higher than the 15\% of the general sample of TOIs. One case (TOI~2556.01) has exceptionally strong TTVs that smear out the transits when folded at fixed period. In some cases the estimated reference flux based on applying our transformations to the Gaia DR2 photometry appears to be too high, yielding a transit depth that is too low in the image subtraction light curves.

\begin{figure*}[!ht]
{
\centering
\leavevmode
\includegraphics[width={0.31\linewidth}]{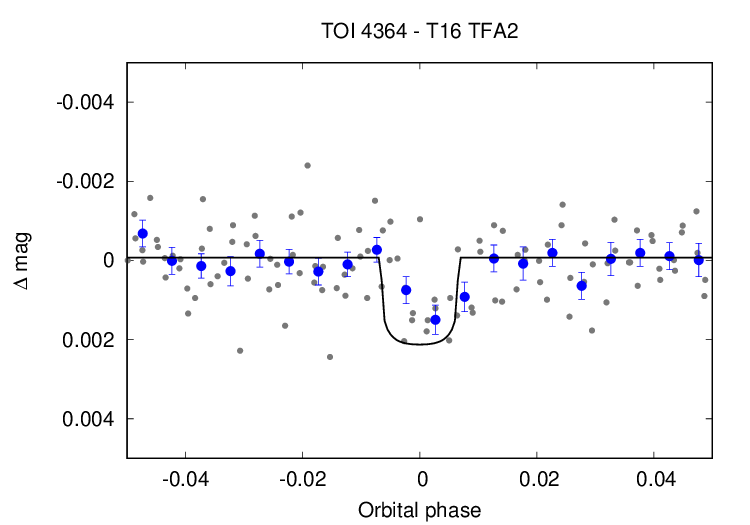}
\hfil
\includegraphics[width={0.31\linewidth}]{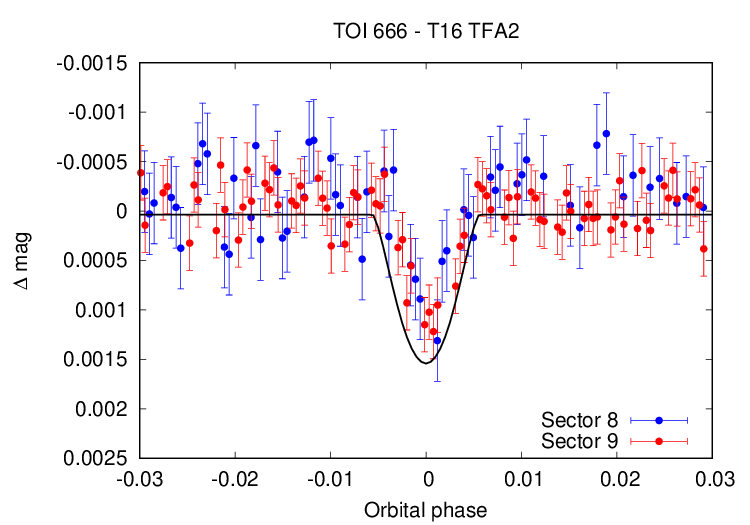}
\hfil
\includegraphics[width={0.31\linewidth}]{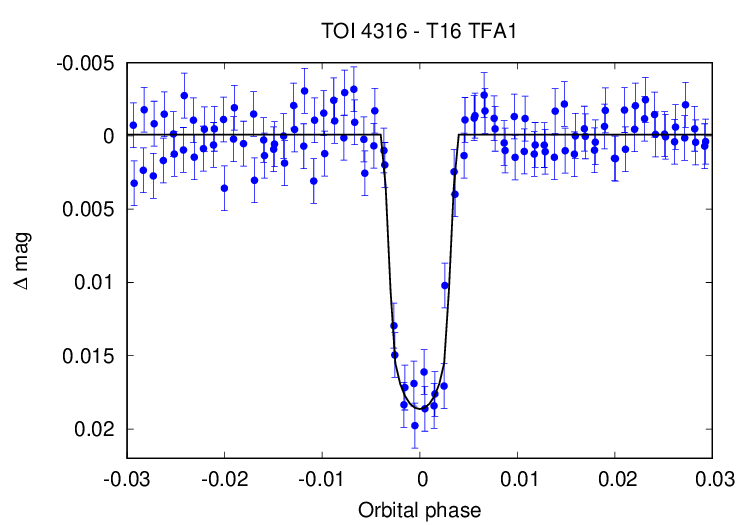}
}
{
\centering
\leavevmode
\includegraphics[width={0.31\linewidth}]{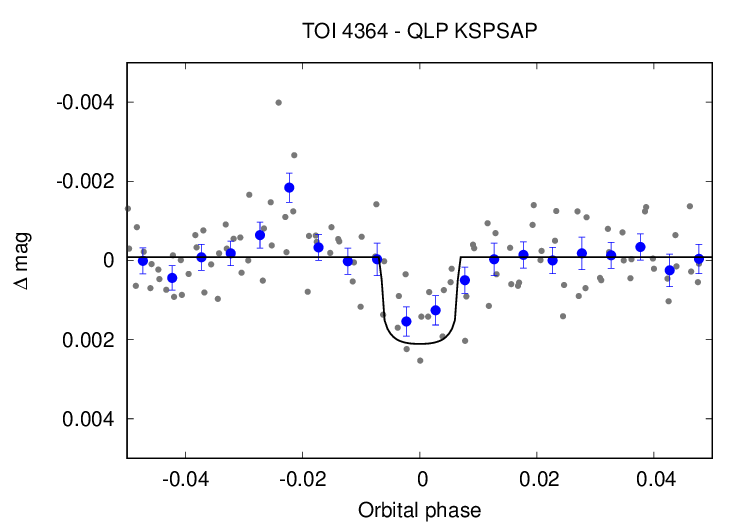}
\hfil
\includegraphics[width={0.31\linewidth}]{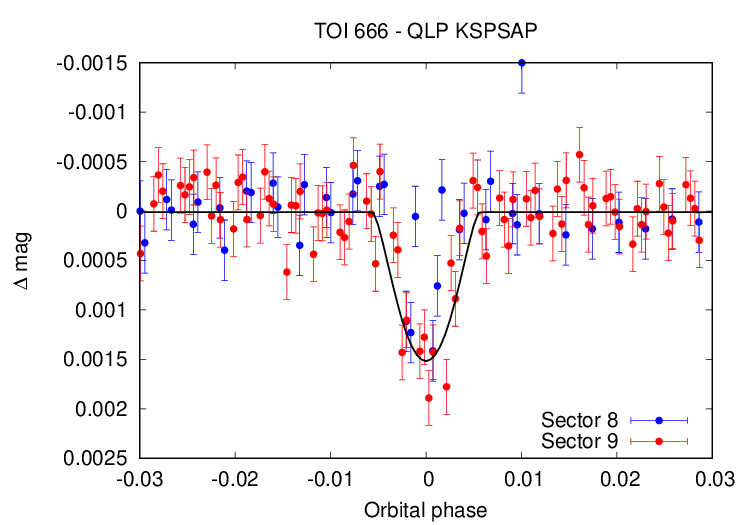}
\hfil
\includegraphics[width={0.31\linewidth}]{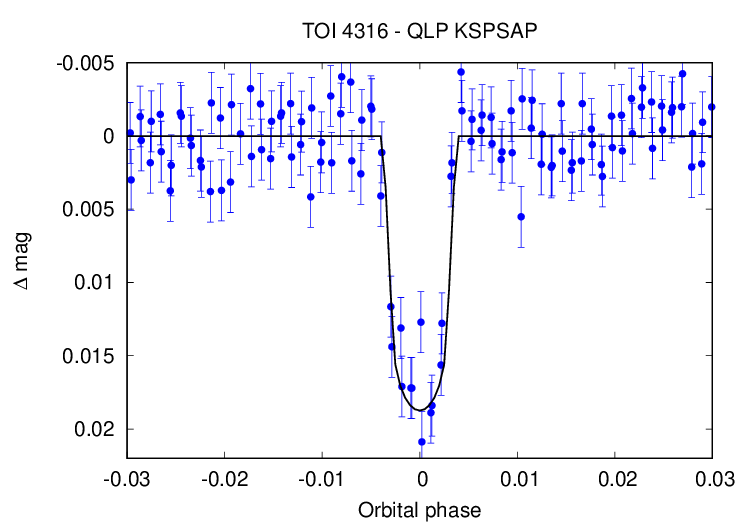}
}
{
\centering
\leavevmode
\includegraphics[width={0.31\linewidth}]{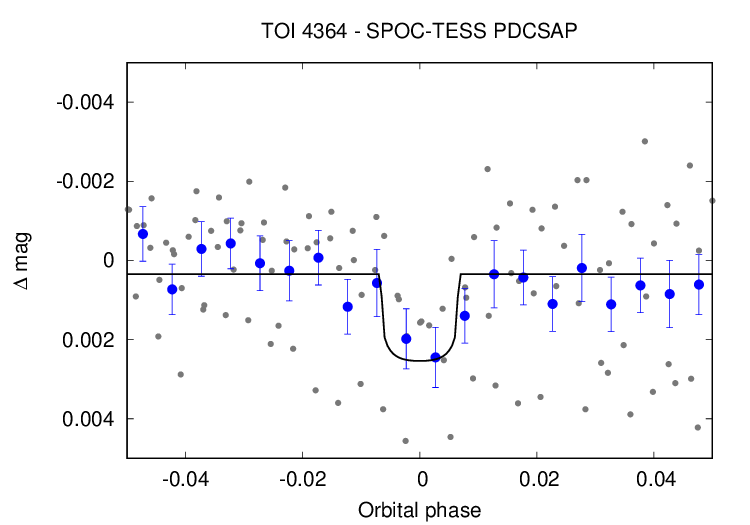}
\hfil
\includegraphics[width={0.31\linewidth}]{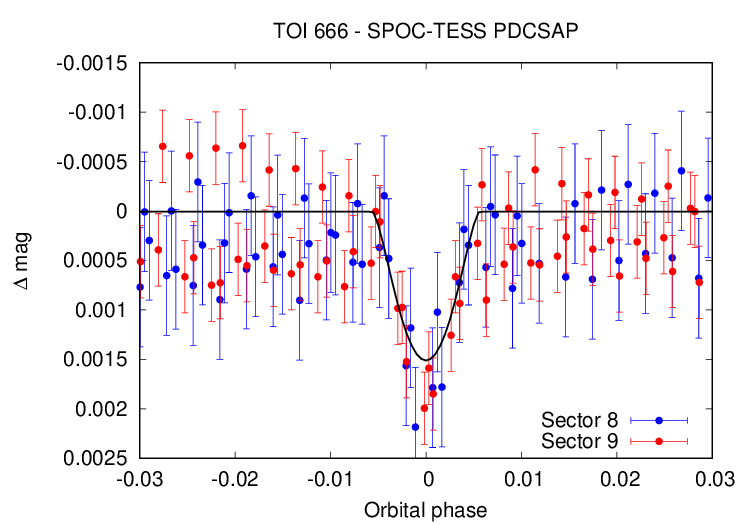}
\hfil
\includegraphics[width={0.31\linewidth}]{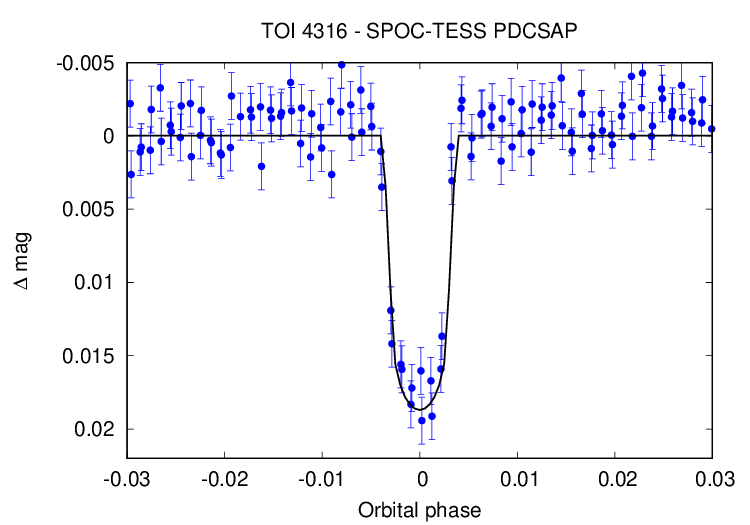}
}
{
\centering
\leavevmode
\includegraphics[width={0.31\linewidth}]{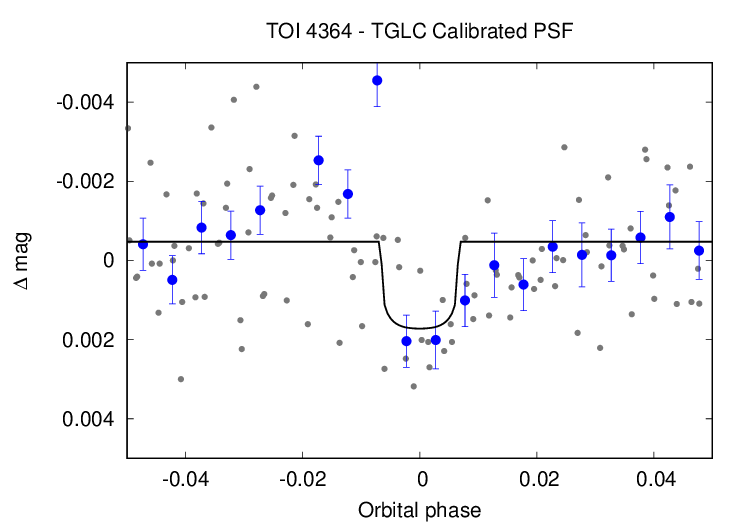}
\hfil
\includegraphics[width={0.31\linewidth}]{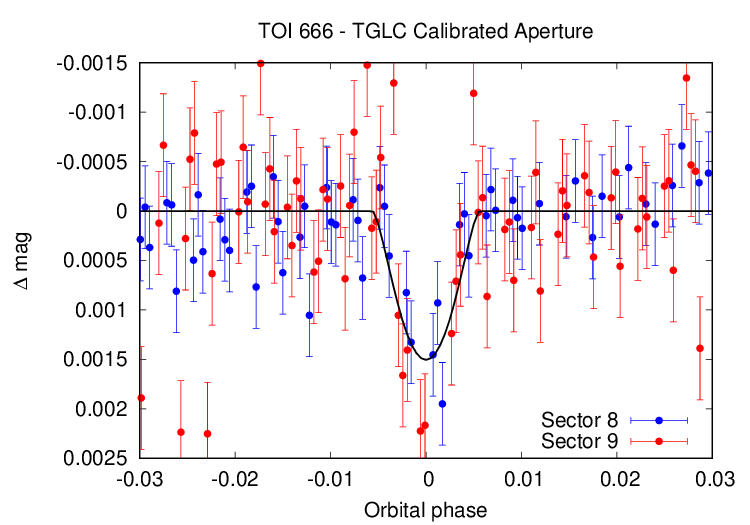}
\hfil
\includegraphics[width={0.31\linewidth}]{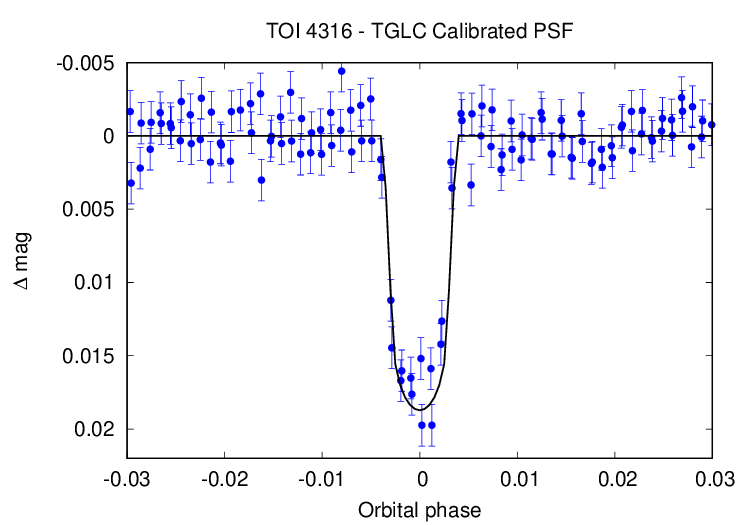}
}
{
\centering
\leavevmode
\includegraphics[width={0.31\linewidth}]{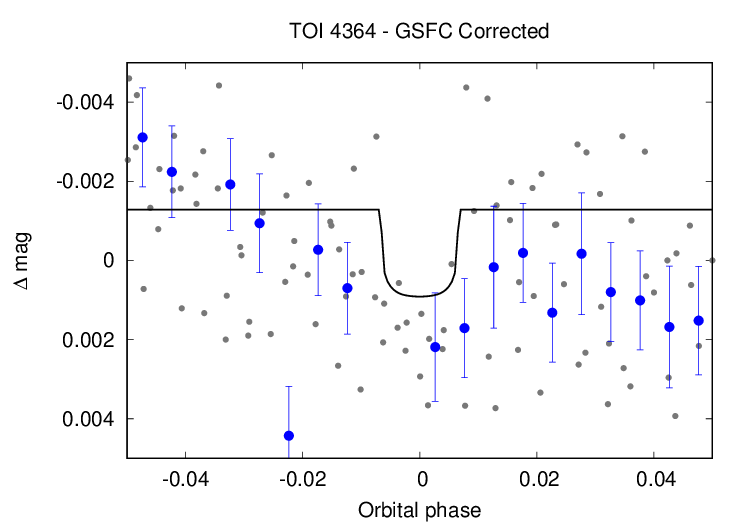}
\hfil
\includegraphics[width={0.31\linewidth}]{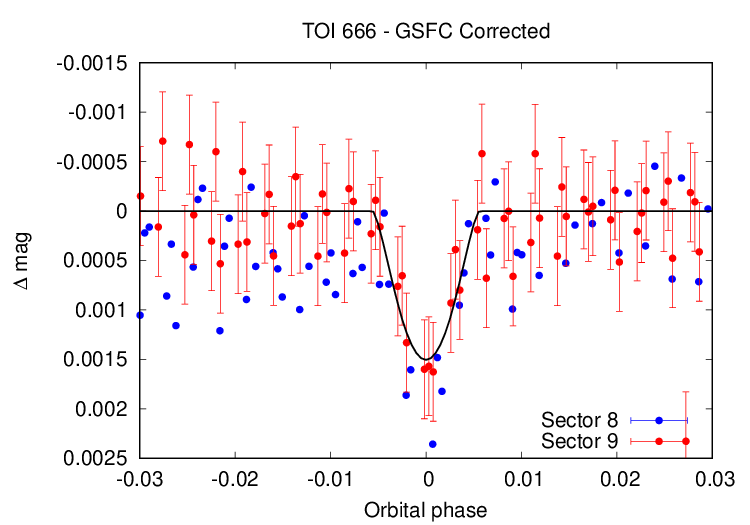}
\hfil
\includegraphics[width={0.31\linewidth}]{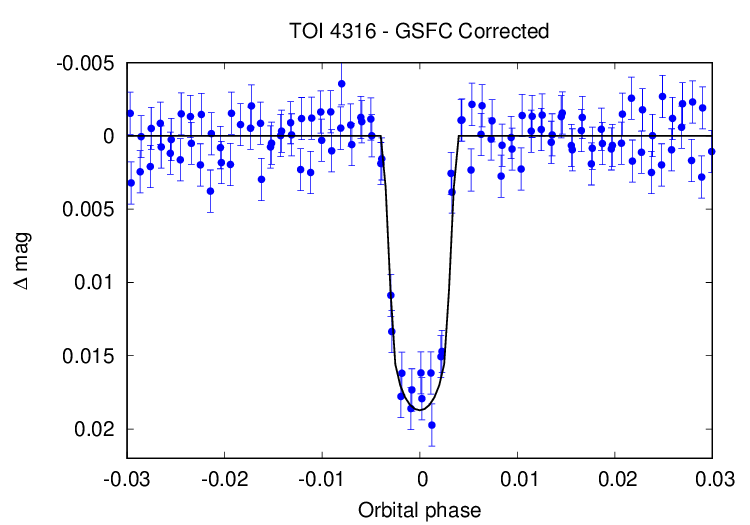}
}
\caption{Phase-folded detrended light curves from the T16 project ({\em top row}), QLP project ({\em second row}), SPOC-FFI project ({\em third row}), TGLC project ({\em fourth row}), and GSFC-ELEANOR-LITE project ({\em bottom row}) for three representative TOIs: TOI~4364 ({\em left column}), TOI~666 ({\em middle column}) and TOI~4316 ({\em right column}). For TOI~4364 we show the individual light curve observations as gray circles, while blue circles show the phase-binned values. For TOI~666 individual observations from Sectors 8 and 9 are shown using blue and red circles, respectively. For TOI~4316 we show only the individual observations using blue circles. For all three systems we overplot a \citet{mandel:2002} transit model calculated using planetary system parameters taken from the ExoFOP-TESS website. For the T16 light curves the photometric aperture displayed is indicated above each panel. We show the calibrated aperture photometry TGLC light curve for TOI~666 and the calibrated PSF photometry TGLC light curves for the other two TOIs.
\label{fig:toilccompare}}
\end{figure*}

\begin{figure*}[!ht]
{
\centering
\leavevmode
\includegraphics[width=0.5\linewidth]{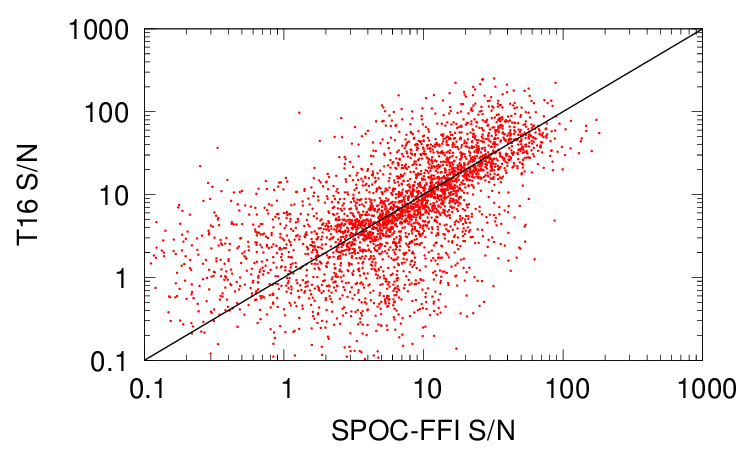}
\hfil
\includegraphics[width=0.5\linewidth]{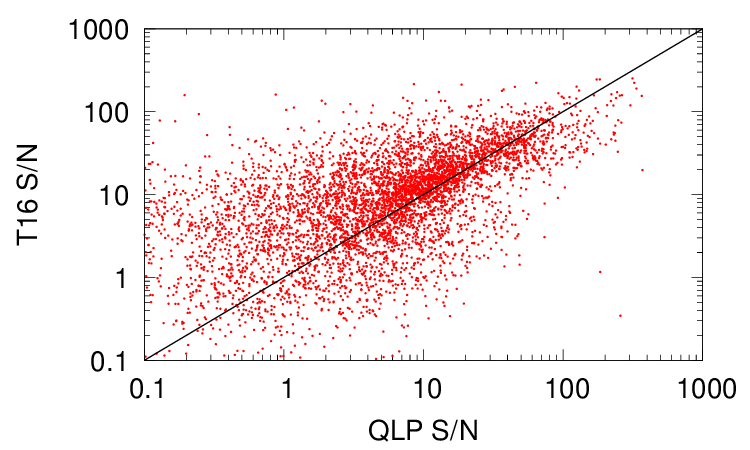}
}
{
\centering
\leavevmode
\includegraphics[width=0.5\linewidth]{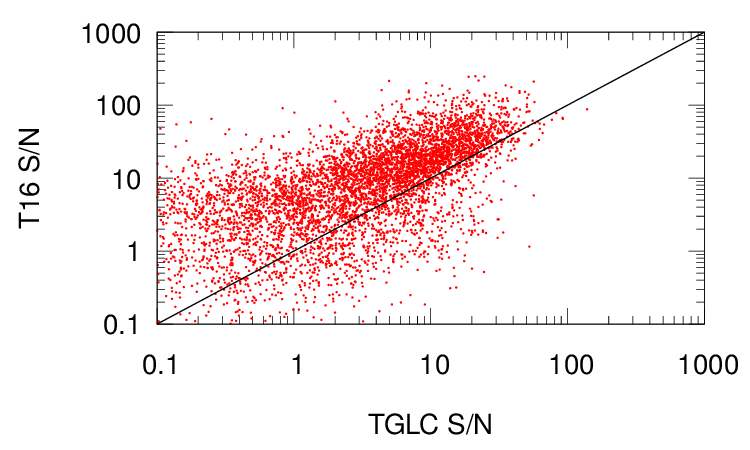}
\hfil
\includegraphics[width=0.5\linewidth]{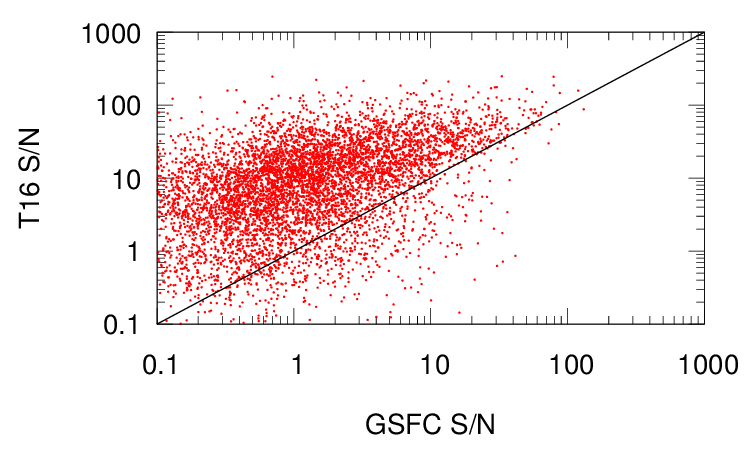}
}
\caption{Comparison between S/N$_{\rm pink}$ measured from the T16
  TFA-detrended light curves for known TOIs to S/N$_{\rm pink}$ measured from the
  SPOC-FFI light curves ({\em top left}), the QLP light curves ({\em top right}), the TGLC light curve ({\em bottom left}), and the GSFC-ELEANOR-LITE light curves ({\em bottom right}), for the same objects. The solid line shows where the two
  measures are equal. We find that S/N$_{\rm pink}$ is higher in the
  T16 light curves for 44.9\%, 62.8\%, 76.6\% and 85.5\% of TOIs compared to the SPOC, QLP, TGLC and GSFC-ELEANOR-LITE light curves, respectively.
\label{fig:blsspocsncompare}}
\end{figure*}

Figure~\ref{fig:toilccompare} compares the QLP, SPOC-FFI, TGLC, GSFC-ELEANOR-LITE, and T16
light curves for three representative TOIs (TOI~4364, TOI~666 and
TOI~4316) for which light curves are available from all three
projects. In all cases we show the phase-folded detrended light curves
available on MAST from each project, and over-plot the same
\citet{mandel:2002} transit model evaluated for the fixed planetary
system parameters taken from the ExoFOP-TESS
webpage\footnote{\url{https://exofop.ipac.caltech.edu/tess/index.php}}. The
three systems selected for display have transit S/N in the T16 light
curves that is near the 10th, median and 90th percentile among all
known TOIs with transits recoverable in the T16 light curves
(Section~\ref{sec:teps}). An extended figure set showing the T16 phase-folded light curves for all of the TOIs investigated is available at \url{https://doi.org/10.5281/zenodo.14278698}.

Figure~\ref{fig:blsspocsncompare} compares the S/N$_{\rm pink}$
measured from the T16 light curves for the TOIs to this same quantity
measured from the SPOC, QLP, TGLC and GSFC light curves. Here
S/N$_{\rm pink}$ is the signal-to-pink-noise ratio calculated by the
{\em -BLSFixPerDurTc} command in {\sc vartools}, and is defined in
\citet{hartman:2016:vartools}. We find that for 44.9\% of cases, the
T16 S/N$_{\rm pink}$ is greater than the SPOC-FFI S/N$_{\rm pink}$, for 62.8\% of cases it is greater than the QLP S/N$_{\rm
  pink}$, for 85.5\% of cases it is greater than the corrected GSFC S/N$_{\rm pink}$, and for 76.5\% of cases it is greater than the calibrated PSF-photometry TGLC S/N$_{\rm pink}$. We conclude that while the SPOC-FFI light curves may be
somewhat better for transit recovery than the TFA-detrended T16 light
curves, the T16 light curves are generally better for this purpose than
the QLP, TGLC, or GSFC light curves. A significant advantage of T16 over either SPOC-FFI or QLP is that fainter targets with $T > 13.5$\,mag are
included in the T16 light curve release.

\paragraph{Pulsating Variable Stars}
\label{sec:cepheids}

\begin{figure*}[!ht]
{
\centering
\leavevmode
\includegraphics[width={0.31\linewidth}]{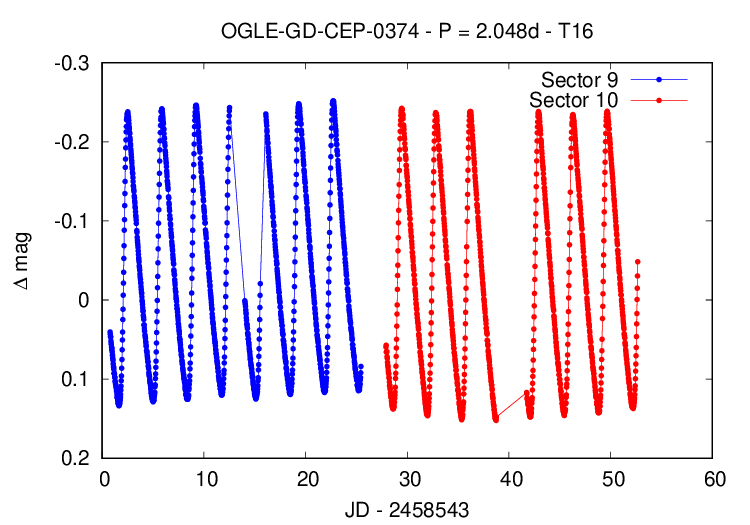}
\hfil
\includegraphics[width={0.31\linewidth}]{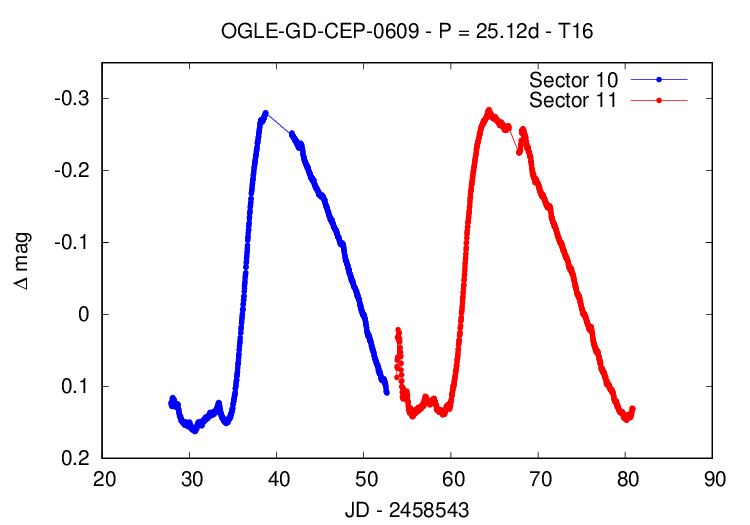}
\hfil
\includegraphics[width={0.31\linewidth}]{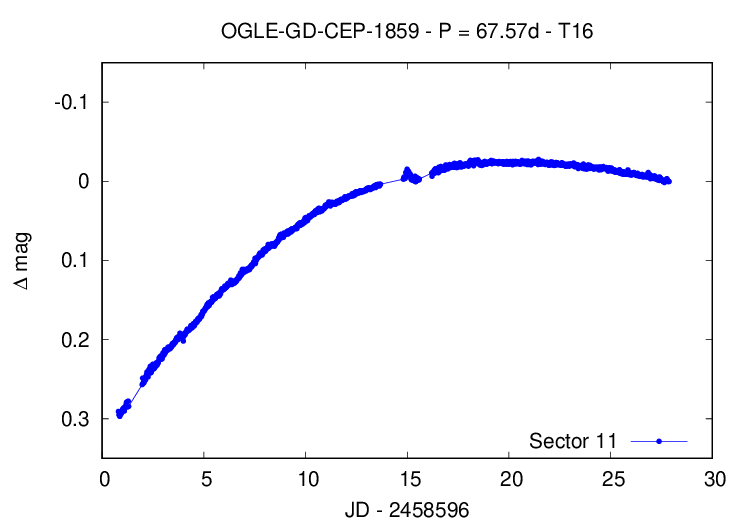}
}
{
\centering
\leavevmode
\includegraphics[width={0.31\linewidth}]{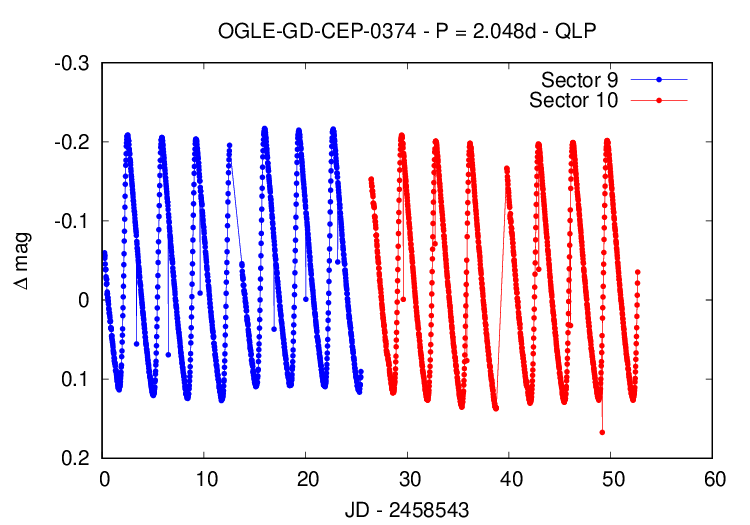}
\hfil
\includegraphics[width={0.31\linewidth}]{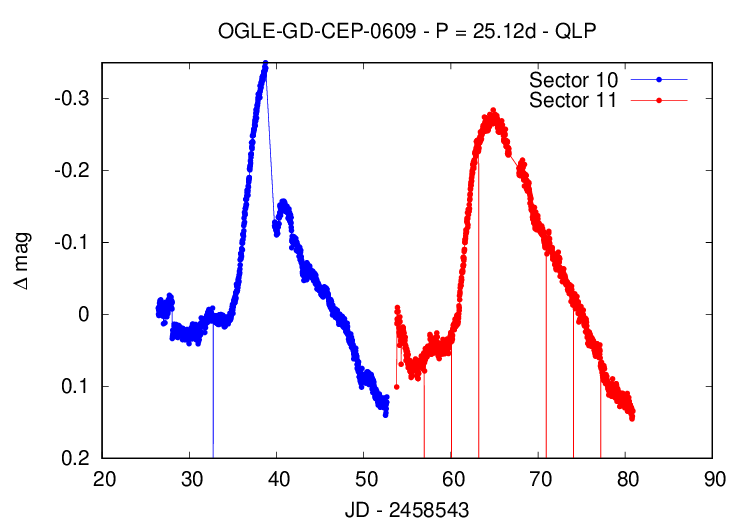}
\hfil
\includegraphics[width={0.31\linewidth}]{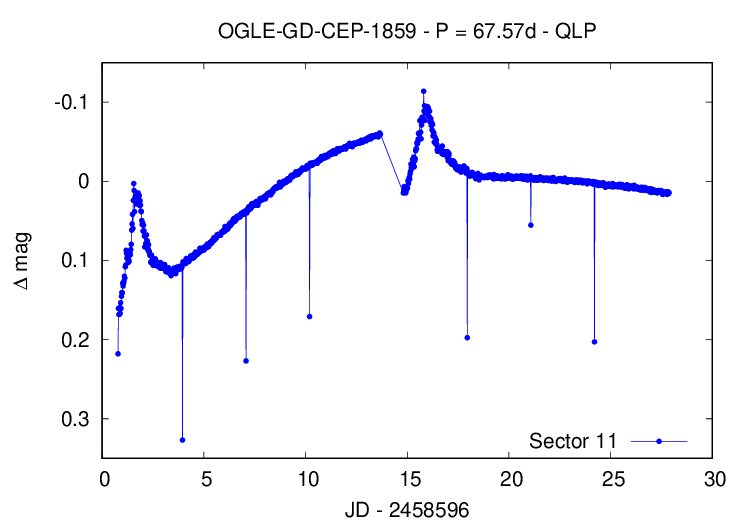}
}
\caption{T16 ({\em top row}) and QLP ({\em bottom row}) light curves for three representative Cepheid variable stars taken from the OGLE catalog of variable stars in the Galactic plane \citep{udalski:2015,udalski:2018,soszynski:2020}. The three stars shown here were selected to have a range of periods, as indicated in the title of each panel. For both the T16 and QLP light curves we show the un-detrended data (IRM magnitudes for T16, and SAP fluxes converted to magnitudes for QLP). SPOC-FFI light curves were not available for these three objects. While instrumental systematic variations are present in the T16 data, these artifacts are generally much less pronounced in the un-detrended T16 light curves than they are in the QLP SAP light curves.
\label{fig:cepheidlccompare}}
\end{figure*}

To evaluate the performance of the T16 reduction process on longer
period variable sources, we inspected light curves for large amplitude
Cepheid variable stars from the OGLE survey
\citep{udalski:2015,udalski:2018,soszynski:2020}. These stars are
restricted to the Galactic plane. Figure~\ref{fig:cepheidlccompare} compares the T16 light
curves for three representative Cepheids chosen to span a range of
periods to the QLP light curves. Here we compare the un-detrended
light curves from both projects, as both projects implement a
detrending process that would distort large amplitude, longer period
variable stars, such as these. We find that the T16 light curves,
which include an effective ensemble magnitude correction through the
image subtraction process, perform much better than the QLP SAP light
curves for these stars. This is especially pronounced for the longer
period variables, such as OGLE-GD-CEP-1859, with a period of
67.57\,days, that shows a slow variation in time in the T16 light curve
over the course of Sector 11, but for which the variation is
indistinguishable from other systematic artifacts in the QLP SAP light
curve. We found much less overlap for variable stars like these with
the SPOC-TESS project, due to the more restrictive selection of stars
for which SPOC-TESS generated light curves.

\begin{figure*}[!ht]
{
\centering
\leavevmode
\includegraphics[width=\linewidth]{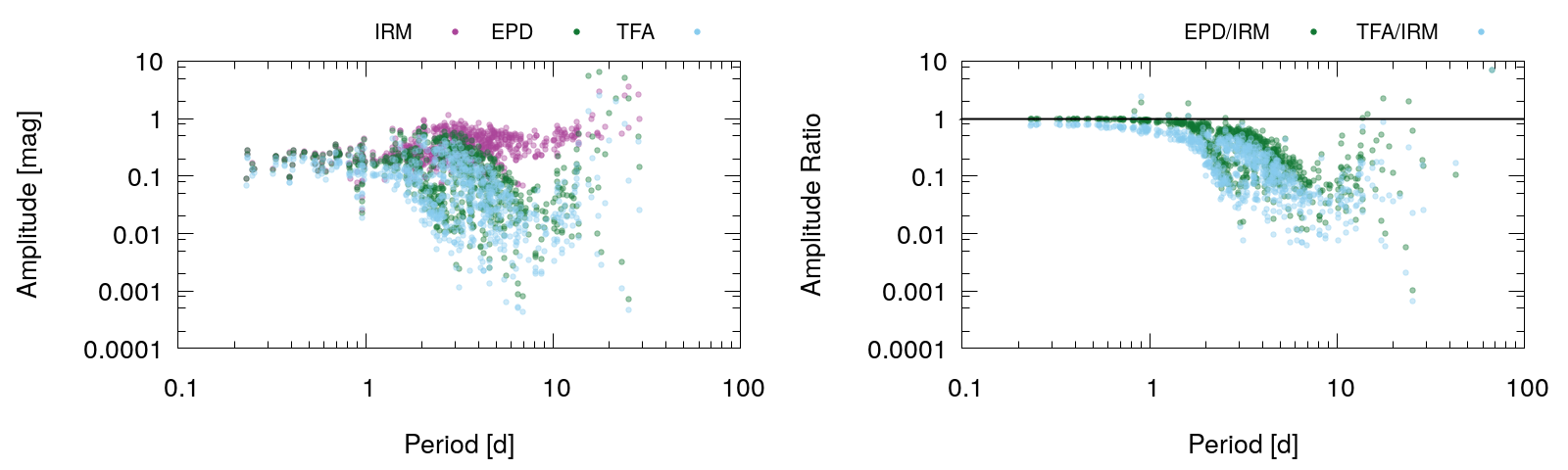}
}
\caption{{\em Left:} Peak-to-peak amplitude vs.\ period for Cepheid variables from the OGLE catalog measured from the T16 undetrended IRM light curves, as well as the EPD- and TFA-detrended light curves. A few example light curves are shown in Fig.~\ref{fig:cepheidlccompare}. {\em Right:} The ratio of the EPD to IRM and TFA to IRM amplitudes as a function of period for the same Cepheid variables. For periods greater than $\sim 1$\,day the EPD- and TFA-detrended light curves have significantly lower amplitudes than the undetrended IRM light curves, with the amplitude of the signal suppressed by as much as a factor of $\sim 100$ and some cases.
\label{fig:cepheidampchange}}
\end{figure*}

We also use the OGLE Cepheid variables to assess the impact of our detrending methods on the amplitudes of variable stars. Fig.~\ref{fig:cepheidampchange} compares the peak-to-peak amplitudes measured from the undetrended IRM light curves, and from the EPD- and TFA-detrended light curves for Cepheids from this catalog, as a function of period. The amplitudes are measured by fitting a 10-harmonic Fourier series to the T16 light curve, fixing the period to the value listed in the OGLE catalog. The results shown are for the smallest aperture photometry, and are similar to the results for the medium and largest apertures. We also show the ratios of the EPD to IRM and the TFA to IRM amplitudes as a function of period. For periods less than $\sim 1$\,day, the EPD amplitudes are generally consistent with the IRM amplitudes, while the TFA amplitudes show a slight reduction that is uncorrelated with the period. For periods longer than 1\,day, the EPD and TFA amplitudes drop precipitously, and can be as low as $\sim$1\% of the IRM amplitudes.

\paragraph{Eclipsing Binary Stars}

\begin{figure*}[!ht]
{
\centering
\leavevmode
\includegraphics[width={0.31\linewidth}]{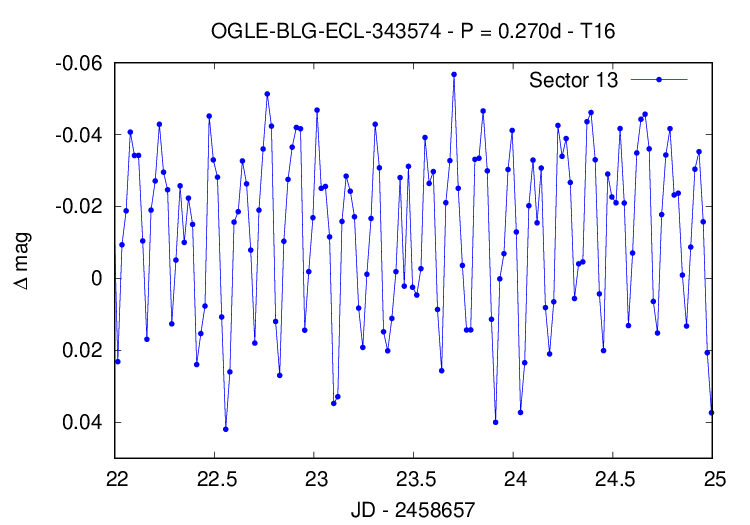}
\hfil
\includegraphics[width={0.31\linewidth}]{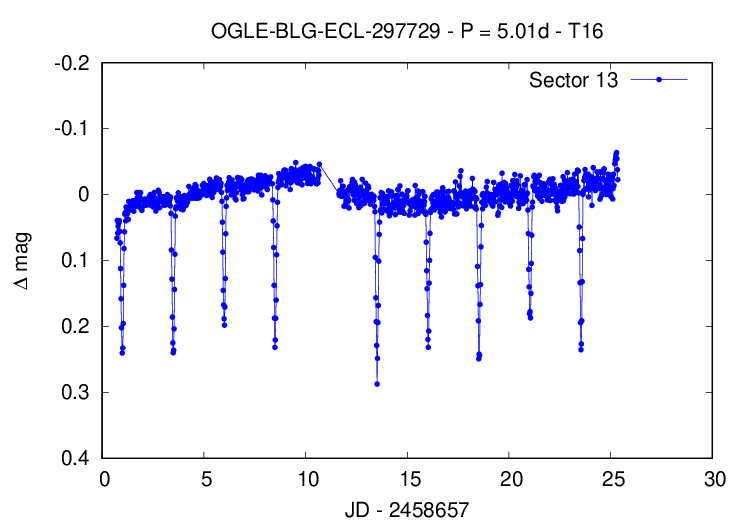}
\hfil
\includegraphics[width={0.31\linewidth}]{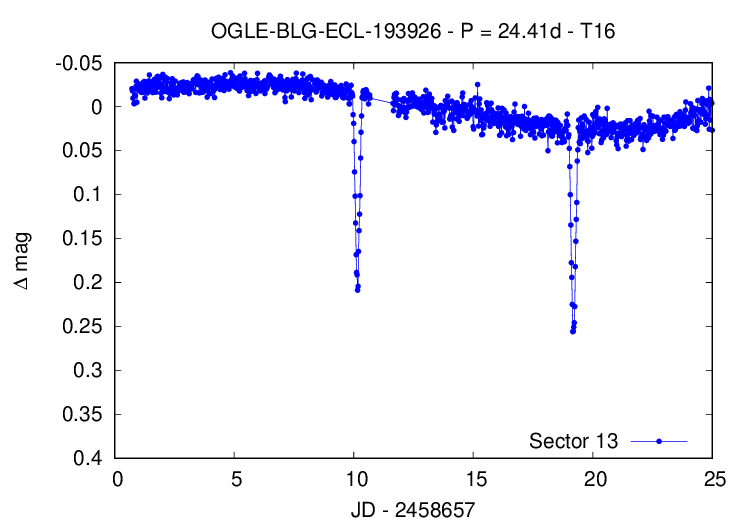}
}
{
\centering
\leavevmode
\includegraphics[width={0.31\linewidth}]{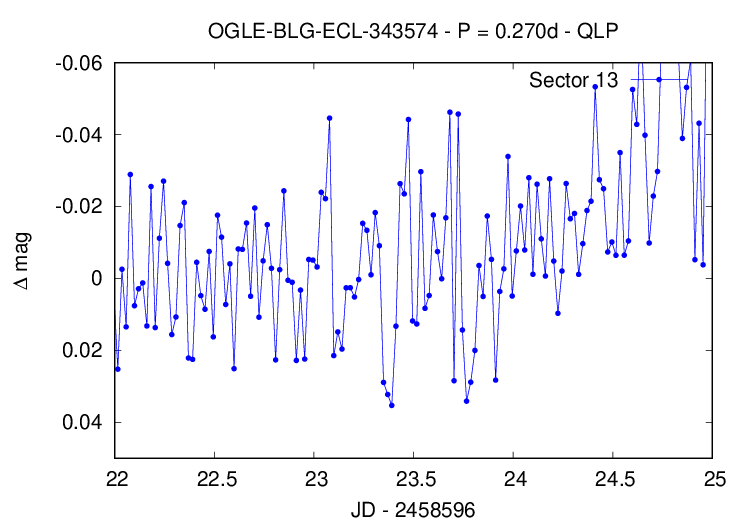}
\hfil
\includegraphics[width={0.31\linewidth}]{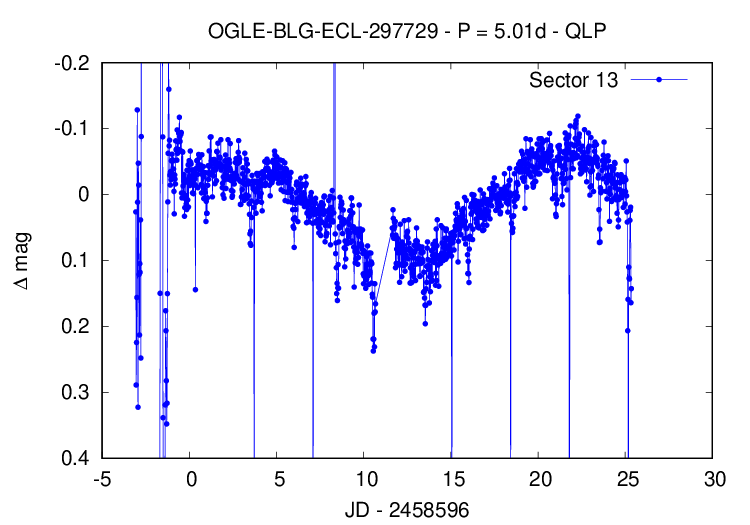}
\hfil
\includegraphics[width={0.31\linewidth}]{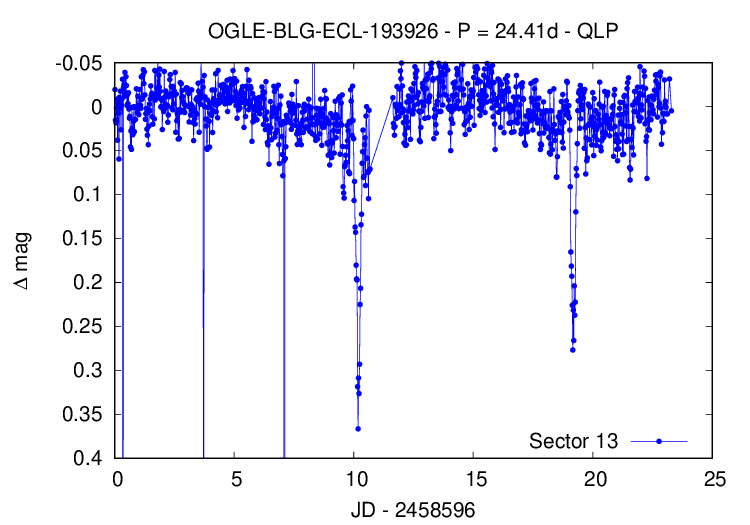}
}
\caption{Similar to Fig.~\ref{fig:cepheidlccompare}, here we compare the T16 ({\em top row}) and QLP ({\em bottom row}) light curves for three representative Eclipsing binary stars taken from the OGLE catalog of variable stars in the Galactic bulge \citep{soszynski:2016}. For OGLE-BLG-ECL-343574 we show only 3 days of data to make the individual eclipses visible. As for the Cepheids, we find that the un-detrended T16 light curves have much lower instrumental systematic artifacts compared to the un-detrended QLP light curves for these deep eclipsing binary stars in the highly crowded Galactic bulge.
\label{fig:eblccompare}}
\end{figure*}

We also inspected the light curves of eclipsing binary stars taken
from the OGLE catalog of eclipsing binaries in the Galactic bulge
\citep{soszynski:2016}. Again we find overlap for most of these
objects only with the QLP light curves, and we show a comparison for
three representative cases in Figure~\ref{fig:eblccompare}. As for
Cepheids, we find that the undetrended T16 light curves provide more
reliable measurements for these large amplitude variable stars than
the un-detrended QLP SAP light curves.

\section{Conclusion}
\label{sec:conclusion}

In this paper we have introduced the T16 project, which is producing
image subtraction light curves for all stars with $T < 16$\,mag from
the {\em TESS} Full-Frame Images. We have produced \nlctotal\ light
curves for \nstarstotal\ stars with $T < 16$\,mag from the {\em TESS}
Cycle 1 Full-Frame Images, making use of the image subtraction carried
out by the CDIPS project. Both raw, and detrended light curves are
made available at \url{http://doi.org/10.17909/8nxx-tw70}. Additional {\em TESS} Cycles are being processed, and the data will be released at this same location.

We compared our light curves to those from the QLP, SPOC-FFI, GSFC and
TGLC projects. We find that the detrended T16 light curves have lower
MAD for a majority of stars when compared to the detrended QLP,
SPOC-FFI and TGLC light curves, whereas 43.1\% of detrended T16 light
curves have lower MAD than the detrended GSFC light curves. We find
that 2639 out of 2971 TOIs have transit signals detectable in the
Cycle 1 T16 light curves, and that the transit depth and S/N is
generally comparable to that seen in the light curves from other
projects.

We find that undetrended T16 light curves may be especially
useful for studying large amplitude, and long time-duration variable
stars. Such variations are often distorted in the light curves
produced by other projects.

Finally, we note that a significant contribution of the T16 project is
in simply providing light curves that are produced using a different
method from other projects. The choice of which publicly available
light curve to use will vary by the star and science goals of the
investigation. We have found that all of the different projects
produce the highest precision light curve available at least some of
the time, and it is important to note that precision is not the only
basis on which one might select a light curve. The more choices that
are available for a given source, the more likely it is that there
will be a high quality, scientifically useful light curve available.

\acknowledgements 

The authors thank the anonymous referee for numerous helpful comments that improved the quality of the work.
This project received funding from the NASA Astrophysics Data Analysis Program (ADAP) grant 80NSSC22K0409.
We acknowledge the use of TESS High Level Science Products (HLSP) produced by the Quick-Look Pipeline (QLP) at the TESS Science Office at MIT, which are publicly available from the Mikulski Archive for Space Telescopes (MAST). Funding for the TESS mission is provided by NASA's Science Mission directorate. We thank the data scientists at STScI for their assistance in making data from this project available on MAST.

\facilities{TESS}

\software{FITSH \citep{pal:2012}, VARTOOLS \citep{hartman:2016:vartools}, Astropy \citep{astropy:2013, astropy:2018, astropy:2022}, Pandas \citep{mckinney2010data}}

\bibliographystyle{aasjournal}

\end{document}